\documentclass[aps,a4paper,twocolumn,20pt,superscriptaddress,longbibliography,floatfix]{revtex4-1}

\usepackage{graphicx}
\usepackage{epsfig}
\usepackage{amsfonts,amsmath,amssymb,amsbsy,amscd,latexsym}
\usepackage{color}
\usepackage{bm}

\newcommand{\bc}{\begin{center}}
\newcommand{\ec}{\end{center}}
\newcommand{\be}{\begin{equation}}
\newcommand{\ee}{\end{equation}}
\newcommand{\ba}{\begin{eqnarray*}}
\newcommand{\ea}{\end{eqnarray*}}
\newcommand{\bna}{\begin{eqnarray}}
\newcommand{\ena}{\end{eqnarray}}

\newcommand{\mpaa}{\begin{minipage}[t]{7.5cm}}
\newcommand{\mpea}{\end{minipage}}

\definecolor{pdcolor}{rgb}{1,0.5,0}

\definecolor{pdblue}{rgb}{0,0,1}

\definecolor{rkgreen}{rgb}{0,1,0}

\begin{document}

\title{Search efficiency of discrete fractional Brownian motion
  in a random distribution of targets}

\author{S.\ Mohsen J.\ Khadem}
\email{jebreiilkhadem@tu-berlin.de}
\affiliation{Institute for Theoretical Physics, Technical University of Berlin,
  Hardenbergstr.\ 36, D-10623 Berlin, Germany}

\author{Sabine H.L.\ Klapp}
\email{sabine.klapp@tu-berlin.de}
\affiliation{Institute for Theoretical Physics, Technical University of Berlin,
  Hardenbergstr.\ 36, D-10623 Berlin, Germany}
  
\author{Rainer Klages}
  \email{r.klages@qmul.ac.uk}
\affiliation{Institute for Theoretical Physics, Technical University of Berlin,
  Hardenbergstr.\ 36, D-10623 Berlin, Germany}
  \affiliation{Queen Mary University of London, School of Mathematical
    Sciences, Mile End Road, London E1 4NS, United Kingdom}

  \date{\today}

\begin{abstract}
Efficiency of search for randomly distributed targets is a prominent
problem in many branches of the sciences. For the stochastic process
of L\'evy walks, a specific range of optimal efficiencies was
suggested under variation of search intrinsic and extrinsic
environmental parameters. In this article, we study fractional
Brownian motion as a search process, which under parameter variation
generates all three basic types of diffusion, from sub- to normal to
superdiffusion. In contrast to L\'evy walks, fractional Brownian
motion defines a Gaussian stochastic process with power law memory
yielding anti-persistent, respectively persistent motion. Computer
simulations of search by time-discrete fractional Brownian motion in a
uniformly random distribution of targets show that maximising search
efficiencies sensitively depends on the definition of efficiency, the
variation of both intrinsic and extrinsic parameters, the perception
of targets, the type of targets, whether to detect only one or many of
them, and the choice of boundary conditions. In our simulations we
find that different search scenarios favour different modes of motion
for optimising search success, defying a universality across all
search situations. Some of our numerical results are explained by a
simple analytical model. Having demonstrated that search by fractional
Brownian motion is a truly complex process, we propose an over-arching
conceptual framework based on classifying different search
scenarios. This approach incorporates search optimisation by L\'evy
walks as a special case.

\end{abstract}

\maketitle

\section{introduction}

Finding randomly located objects is a challenge for every human being,
be it the search for mushrooms \cite{CBV15}, for lost keys
\cite{BLMV11}, or for food \cite{RWGMMP14}. Within the context of
modern society, attempts to solve this problem fuelled the development
of operations research, which aims at optimising tasks such as search
games \cite{AG03}, locating submarines \cite{Shles06}, or human rescue
missions \cite{Stone07}. For biological organisms, the successful
location of food sources is crucial for their survival, as is
addressed within movement ecology \cite{Nath08,MCB14,Kla16}. On a
theoretical level, biological foraging processes are typically
modelled in terms of stochastic dynamics
\cite{BLMV11,MCB14,VLRS11}. An important paradigm for this modeling
was put forward by Karl Pearson, who suggested at the beginning of the
last century that organisms may migrate according to simple random
walks \cite{Pea06} characterised by Gaussian position distributions in
a suitable scaling limit. This paradigm was challenged two decades ago
by the experimental observation \cite{Vis96} and a corresponding
theory \cite{Vis99} that wandering albatrosses searching for food
performed flights according to non-Gaussian step length distributions
\cite{Edw07}. In this case, the mean square displacement (MSD) of an
ensemble of moving agents may not grow linearly in time like for
Gaussian spreading generated by random walks or Brownian
motion. Instead, it may grow nonlinearly, $\langle
\bm{x}^2(t)\rangle\sim t^{\alpha}$ with $\alpha\neq1$, where
$\bm{x}(t)$ is the position of an agent in space at time $t$. This
phenomenon is known as anomalous diffusion, where $\alpha>1$ denotes
superdiffusion and $\alpha<1$ subdiffusion while $\alpha=1$ refers to
normal diffusion \cite{MeKl00,CKW04,KRS08,sokol,MJJCB14}.

Motivated by these developments, much recent research was devoted to
explore the relevance of more non-trivial diffusion processes for
modeling foraging. Inspired by Refs.~\cite{Vis96,Vis99}, the focus was
on superdiffusive L\'evy walks (LWs) determined by power-law step
length distributions \cite{VLRS11}. Along similar lines normal
diffusive intermittent motion \cite{BLMV11} and correlated random
walks \cite{MCB14,BaCa16} have been analysed. Nevertheless, an
over-arching framework for stochastic search in movement ecology is
still missing. Especially for the wide variety of anomalous stochastic
processes beyond LWs \cite{MJJCB14}, efficiency of search has not very
much been investigated. This applies particularly to fractional
Brownian motion (fBm), a paradigmatic stochastic process that, in
contrast to L\'evy dynamics, exhibits Gaussian distributions and
power-law memory by generating the whole spectrum of anomalous
diffusion \cite{mandel}. The goal of our article is to explore in a
very systematic way, based on extensive computer simulations and
simple analytical arguments, the complexity of search exhibited by
fBm. We hope that our work will set the scene for further studies to
understand biological foraging on the basis of stochastic theory.

\subsection{Background}\label{sec:backg}

Already at the beginning of the 90's it was suggested that
superdiffusive motion obtained from L\'evy flights and walks may
optimise the search for targets by increasing the search efficiency
compared to Brownian motion \cite{SZF94}. L\'evy flights can be
modelled as a Markov process, where the instantaneous jumps over
certain distances are sampled randomly from a power law distribution
\cite{SZK93,KRS08}. In the special case of LWs, any jump length is
coupled with the time to perform the jump by assuming a constant speed
\cite{ZDK15}. In order to explain the experimental albatross data of
Ref.~\cite{Vis96}, Ref.~\cite{Vis99} proposed a simple two-dimensional
stochastic search model. It consists of a L\'evy walker searching for
targets, which are disks uniformly random distributed in the
plane. For a sparse field of replenishing, immobile targets a suitably
defined search efficiency yielded a maximum for power law jumps, while
Brownian motion was optimal when the density of the target
distribution increased \cite{Vis99,BHRSV01,RBLSSV03}. This result
became famous as the L\'evy flight foraging hypothesis (LFFH)
\cite{Vis99,VLRS11}. The LFFH initiated a long debate, particularly
when applied to finding many targets under biologically realistic
search conditions
\cite{Edw07,Benh07,Sims08,Sims10,BLMV11,VLRS11,dJWH11,HaBa12,LICCK12,PCM14,MCB14,ZDK15,Pyke15,Reyn15,Kla16,Reyn18,LTBV20,GuKo20,BRBHRLV21,LTBV21,VOCGTNA21}. A
special case of the LFFH is single target search in simplified
theoretical settings, which defines mathematically solvable first
passage (FP) and first arrival (FA) problems
\cite{Red01,MCB14,MOR14,PBL19,LBV21}. `Passage' corresponds to the
situation of a biological cruise forager, who can perceive a target
while moving. Thus, a target is found whenever a cruise searcher
`passes' it.  A saltatory forager, on the other hand, does not scan
for a target while moving but has to land, respectively to
  `arrive' on it, or within a suitable neighborhood of it, to
perceive it after performing a jump
\cite{JPP09,BLMV11,PCM14,PBL19}. Examples of cruise foragers are large
fish, such as tuna and sharks. Saltatory foraging was observed among
smaller fish, ground foraging birds, lizards and insects
\cite{OBBE90}.

The LFFH created awareness that apart from models related to classical
random walks and Brownian motion, which yield normal diffusion with
Gaussian distributions, more advanced stochastic processes are
available, and needed, in order to understand data of foraging
organisms (see, e.g.,
Refs.~\cite{Benh07,MCB14,Pyke15,ZDK15,Kla16,Reyn15,Reyn18} and
respective discussions). Most notably, it motivated the experimental
biological foraging community to look for power laws in data. Over the
past two decades many analyses of foraging data indeed suggested the
existence of dynamics governed by power laws
\cite{Sims08,Sims10,VLRS11,dJWH11,HaBa12,Reyn18}. At the same time,
however, evidence accumulated that in many cases questionable data
analyses were performed by checking for L\'evy dynamics, partially due
to a lack of full appreciation of the relevant theoretical background
\cite{Edw07,Benh07,JPP09,BLMV11,JPE11,MCB14,Pyke15,ZDK15,Kla16,LTBV20,LTBV21}. Another
fundamental problem was missing knowledge whether the observed
movements are intrinsic or extrinsic to a forager \cite{Nath08,MCB14},
being induced, say, by the food source distribution or other
environmental conditions
\cite{Vis96,BoRa06,Benh07,JPP09,Sims10,JPE11,Kla16}, which by
themselves could be governed by power laws. The complex interaction
between forager and environment during search defines a topic of more
recent research
\cite{RBLSSV03,Benh07,CBRM15,ZJLW15,BaCa16,KWPES16,VoVo17,ZWS17,GuKo20,Giu20,VOCGTNA21}.
A question related to this is to which extent the optimality of a
specific search strategy as predicted by a given model is robust with
respect to varying parameters of the process and the given
environment. Further, it remains unclear whether this optimality
depends on other details of the search scenario. Along these lines the
original theoretical results leading to the LFFH
\cite{Vis99,BHRSV01,RBLSSV03} were critically investigated by limiting
their range of application
\cite{Benh07,JPP09,JPE11,BLMV11,MCB14,PCM14,PCM14b,PCKM16,PMKMC17,GuKo20}.
In very recent work they were eventually largely refuted
\cite{LTBV20}; however, see Refs.~\cite{BRBHRLV21,LTBV21} for an
associated reply and comment, and also Ref.~\cite{GuKo20}. We will
come back to these important points at the end of our work in the
light of our new results.

Apart from using simple random walks, classical Brownian motion or
L\'evy dynamics for modelling foraging, other theoretical studies
considered intermittent motion \cite{LKMK08,BLMV11}, correlated random
walks \cite{bartu2,BaLe08,BaCa16}, multiscale random walks
\cite{CBRM15,PCKM16,PMKMC17}, more non-trivial one-dimensional motion
\cite{BRVL14} and generalisations of the original \cite{Vis99} LW
search model \cite{ZJLW15,VoVo17}. Yet beyond these models, there
exist numerous other types of, in particular, anomalous stochastic
processes \cite{MeKl00,CKW04,KRS08,KlSo11,MJJCB14} which, to our
knowledge, have not been assessed for optimising search.  One of the
most famous and important classes is fractional Brownian motion (fBm),
which was originally studied by Kolmogorov \cite{kolmogorov1940wiener}
and Yaglom \cite{Yag58} and has become more widely known through the
work of Mandelbrot and Van Ness \cite{mandel}. FBm defines a Gaussian
stochastic process generating anomalous diffusion
\cite{embrechts2009selfsimilar,jeon1,burov2011single,MOR14,fuli}.
However, while for non-Gaussian L\'evy dynamics anomalous diffusion
originates from sampling power law distributions, in the case of fBm
it is generated by a power law correlation function decay in time
yielding non-Markovian dynamics. This enables fBm to produce the whole
spectrum of anomalous diffusion under parameter variation, from sub-
to normal, to superdiffusion. LWs, on the other hand, are purely
superdiffusive. Hence these two stochastic processes represent
fundamentally different classes of anomalous dynamics. FBm has been
widely used to describe the experimentally observed anomalous
diffusion of a tracer particle in a visco-elastic system \cite{goy},
in artificially crowded environments
\cite{weis1,khadem,szymanski2009elucidating}, in complex intracellular
media \cite{jeonvivi,weisp} and in living cells
\cite{weiss2004anomalous}; for a wide range of further applications of
fBm see Ref.~\cite{Wiese19}. On the other hand, so far fBm has been
rarely used to understand the search by biological
organisms. Motivated by insect motion and other problems in ecology,
the efficiency of search of fBm and LWs has been compared in computer
simulations \cite{Rey09}. In Ref.~\cite{LICCK12}, non-trivial
autocorrelation function decay was observed in experimental data of
bumblebee flights, with the conclusion that correlation decay is
important to characterise biological search. While both exponential
and power law decay of correlations has been reported for biological
cell migration \cite{NHKD17}, extracting correlation functions from
experimental data has not been very prominent in the foraging
literature. By studying search in fBm we thus further advocate the use
of correlation functions, or memory, for understanding biological
foraging. We may also remark that very recently, subdiffusive search
related to fBm has even been studied on the human proteomic network,
as a possible explanation that via protein expression the COVID-19
virus only attacks a certain subset of organs \cite{Estr20}. FP
problems of fBm on both finite and infinite domains have been analysed
rigorously mathematically \cite{Molch99}, as well as in relation to
physical and biological applications
\cite{GuWe08,JCM11,SaAm12,MOR14,guerin,Wiese19}. However, apart from
Ref.~\cite{Rey09}, more complex search scenarios have not been
investigated for fBm.

\subsection{Scope of this work}

In this article we compute the efficiency of a searcher moving by fBm
in a two-dimensional array of uniformly random distributed
targets. The target distribution is determined by the number density
and the radius of the circular targets as two extrinsic parameters. In
addition, there are two parameters that are intrinsic to fBm, the
Hurst exponent determining the type of diffusion generated by fBm and
the mean jump length of the process characterising the strength of the
diffusive spreading.  Under these conditions, we consider both cruise
and saltaltory foragers (viz.\ FP and FA problems) for finding
targets.  We compare these two different settings by using two
generically different types of efficiencies. The first one employs an
ensemble average of moving fBm particles for finding one target only,
while the second one is given by the time-ensemble average of
searchers for finding many targets along their whole trajectories
\cite{Vis99,BHRSV01,RBLSSV03,BaLe08,JPP09,BRVL14,ZJLW15,LTBV20,VoVo17,PCM14,PCM14b,PCKM16,PMKMC17,PBL19}. These
efficiencies are numerically computed under variation of the two
intrinsic parameters of fBm.

In the former case of single target search we define the setting so
that boundaries play no major role. In the latter case of long
trajectories hitting many targets, however, we inevitably run into the
problem of boundary conditions. They are more than a technicality, as
in reality all media in which search may take place (such as a lake, a
forest or living cells) are surrounded by boundaries that the
searching element can (or will) not exceed. In our simulations we
model bounded regions by a box with reflecting boundaries. For
foraging we found this setup more realistic than using periodic
boundaries, apart from the problem that the long time memory in
superdiffusive fBm may lead to questionable effects in combination
with periodic boundaries.  Interestingly, recent experimental and
theoretical studies have demonstrated that so-called active particles,
which model self-propelled biological motion \cite{RBELS12,BeDiL16},
spend most of their time close to reflecting boundaries exhibiting
`stickiness' to the walls
\cite{duzgun2018active,das2015boundaries,elgeti2015run,kaiser2012capture,BeDiL16,VoVo17,ZWS17}. A
similar phenomenon has recently been reported for superdiffusive fBm
\cite{gugg,VHSJGM20} that by definition displays strong persistence,
similar to self-driven active Brownian particles. Therefore we explore
the interplay between non-Markovian persistence in fBm and reflecting
boundary conditions in comparison to using periodic ones. For
reflecting boundaries we find very intricate phenomena determining the
success or failure of search.

The main lesson to be learned from our research is that for fBm we do
not observe any universal search optimisation in terms of maximising a
search efficiency. On the contrary, exploring a range of different
search scenarios by both varying parameters and search settings we
identify different mechanisms determining search success. Boundary
effects play a crucial role for search, which to our knowledge has not
been sufficiently appreciated in previous work.  Our studies
demonstrate that search is a very flexible, complex process that
sensitively depends on the interplay between its different
ingredients.  We believe that these findings further open up the field
of biologically inspired search research \cite{Shles06}. In
particular, our work suggests to shift the focus from finding simple
universalities to developing a much broader picture of search. Within
this general framework the LFFH takes its place as a special case.

Our article is organised as follows: In Sec.~\ref{sec:model} we
briefly review the concept of fBm and define our basic search setting
by introducing all relevant model parameters. In Sec.~\ref{sec:es} we
study FA and FP problems, viz.\ saltatory and cruise foraging, of
finding only the first target in a field of resources under variation
of both intrinsic parameters. Numerical findings are explained by a
simple analytical argument. Section~\ref{sec:ss} reports results for
multi-target search along a trajectory, both in saltatory (arrival at
targets) and cruise (passage through targets) mode. This is done for
finding either replenishing or non-replenishing targets, which in this
setting becomes a non-trivial variation. Within this context, the
impact of boundary conditions turns out to be crucial, yielding a
wealth of different search mechanisms. In our concluding
Sec.~\ref{sec:concl}, we give a coherent overview of all different
search scenarios that we have investigated in our work and the
quantities they depend on by summarising our main results.

\section{The model}
\label{sec:model}

In this section, we first review the stochastic process of fBm by
explaining its main characteristics. We then define our specific
search problem by identifying all relevant model parameters.

\subsection{Fractional Brownian motion}

\begin{figure}[h!]
\includegraphics[width=1\linewidth]{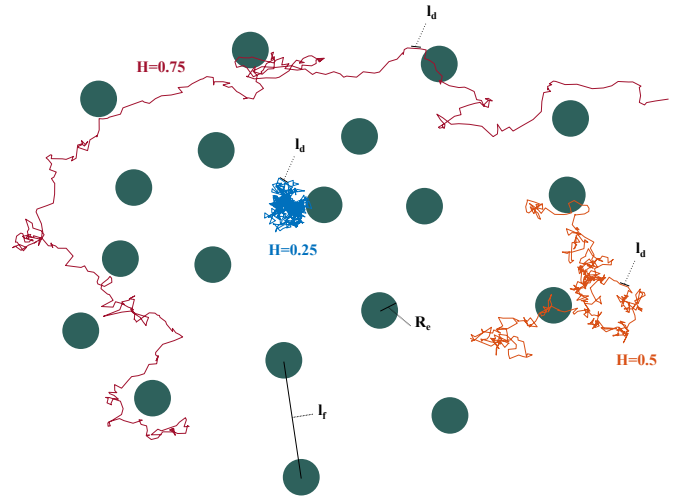}
\caption{The three generically different types of fractional Brownian
  motion (fBm) and basic parameters defining our search set-up. The
  trajectory for Hurst exponent $H=0.25$ displays anti-persistent
  subdiffusive dynamics, the one for $H=0.5$ normal diffusion, and the
  one for $H=0.75$ persistent superdiffusion. An fBm searcher moves
  with mean jump length $l_d$ in a field of uniformly random
  distributed targets of radius $R_e$, which are a mean distance $l_f$
  apart from each other.}
\label{fig:setup}
 \end{figure}

FBm in $d$ dimensions can be generated by the stochastic equation of
motion \cite{mandel,Yag58,kolmogorov1940wiener,jeon1,MOR14}
\be
\bm{x}^H (t)=\bm{B}^H (t)
\label{eq:fBm}
\ee
with position $\bm{x}^H\in\mathbb{R}^d$ at time $t$, where $H \in (0,1)$
is the Hurst exponent.  Here, $\bm{x}^H (t)$ holds for a Gaussian stochastic
process with zero mean,
\begin{equation}
 \langle\bm{x}^H(t)\rangle=0\:,
\label{fBmmean}
\end{equation}
and position autocorrelation function
\begin{equation}
\langle \bm{x}^H(t)\bm{x}^H(t')\rangle = d K_H\left( |t|^{2H}+ |t'|^{2H}-|t-t'|^{2H} \right)\:,
\label{fBmpp}
\end{equation}
where $K_H$ yields a generalized diffusion coefficient
\cite{burov2011single}. Here we represent $d$-dimensional fBm by $d$
independent one-dimensional fBms, i.e., one for each
component. Similar to recent work on two-dimensional L\'evy walks
\cite{ZFDB16}, other definitions of two-dimensional fBm may be
possible. The power law decay of the correlation function makes the
process non-Markovian in terms of a very slow decay of memory with
time. As we will argue below, this memory can be physically understood
in terms of non-trivial position correlation decay in time within a
heat bath in which a particle moving according to fBm is
immersed. Equation~(\ref{fBmpp}) results via the Taylor-Green-Kubo
formula in the MSD
\begin{equation}
\langle\left(\bm{x}^H(t)\right)^2\rangle = 2 d K_H t^{2H}
\label{fBmmsd}
\end{equation}
with $\alpha=2H$ as the exponent of anomalous diffusion.  Depending on
the value of the Hurst exponent, fBm thus leads to subdiffusion,
$H<1/2$, or to superdiffusion, $H>1/2$. This corresponds to
anti-persistent, respectively persistent motion of fBm particles, as
can be seen from calculating the velocity autocorrelation function of
the process \cite{jeon1}. That is, for $H<1/2$ it decays to zero from
negative values for long times reflecting
anti-correlations. Topologically this shows up as trajectories that
display a lot of turns, see the example for $H=0.25$ in
Fig.~\ref{fig:setup}. In contrast, for $H>1/2$ this function decays to
zero from positive values yielding positive correlations. Accordingly
particles move more in one direction displaying trajectories that are
more elongated, see the example for $H=0.75$ in
Fig.~\ref{fig:setup}. These two types of correlation function decay
have been experimentally observed for bumblebee flights
\cite{LICCK12}. In the limiting case of $H=1$ fBm generates
ballistic motion with $\alpha=2$. For $H=1/2 $ the correlation
function decay Eq.~(\ref{fBmpp}) boils down to a delta function, and
one recovers the Markovian Wiener process with normal diffusion,
$\alpha=1$; see again Fig.~\ref{fig:setup} for a third example. Hence
there are three generic cases of fBm dynamics, anti-persistent
subdiffusion, normal diffusion and persistent superdiffusion depending
on the value of $H$ as displayed in Fig.~\ref{fig:setup}.

This discussion suggests that fBm is a generalisation of Brownian
motion by correlating the position autocorrelation function. To see
this more clearly one rewrites Eq.~(\ref{eq:fBm}) in the form of a
stochastic differential equation \cite{sokol,burov2011single}, \be
\dot{\bm{x}}^H(t)=\bm{f}^H(t)\:, \label{fbm-fgn} \ee where $\bm{f}^H
(t)$ denotes $d$-dimensional fractional, i.e., power law correlated
Gaussian noise (fGn)
\cite{mandel,Yag58,kolmogorov1940wiener,rangarajan2003processes}; see
Refs.~\cite{jeon1,burov2011single,MOR14,fuli} for
details. Equation~(\ref{fbm-fgn}) can be understood as an overdamped
Langevin equation \cite{jeon1,MOR14}. Within this framework the left
hand side can be interpreted as a constant friction term without
memory and the right hand side as a correlated random force. The
latter models collisions of a tracer with heat bath particles that
perform dynamics with power law memory decay. We remark that related
equations have recently been used as models of active Brownian
particles \cite{FNCTVW16,USJ19}. These particles are self-propelled
due to the fact that by definition, as in Eq.~(\ref{fbm-fgn}), the
fluctuation-dissipation relation is broken \cite{ChKl12}. That is,
here we only have memory in the noise but not in the friction
\cite{jeon1}. This type of dynamics has also been applied to model
biological cell migration \cite{NHKD17}.

FBm can be generated numerically by different methods either
(theoretically) exactly \cite{hosking,asmus,davi} or approximately
\cite{willi,noros,diek}. In this work we obtain fBm from discrete time
fGn using Hosking's method \cite{hosking}.  According to
Eq.~(\ref{fbm-fgn}), the increments of fBm are then computed by the
integral $\bm{B}^H(t)= \int_0^t \bm{f}^H(t')dt'$. In practice, we
discretize the motion of an fBm particle to a series of successive
jumps at unit time steps, $t_0=1$. Then the integral breaks down to
the discrete sum $\bm{B}^H(t)\simeq \Sigma_{i=1}^n \bm{f}^H(t_i) t_0$,
where $t_i=i t_0$, $t=t_n$.  This means that such a time-discrete fBm
particle performs ballistic jumps during unit times $t_0$ according to
the increments generated from fGn at time $t$. The direction of the
subsequent ballistic step is then determined by the next increment of
the fGn. The underlying Hurst parameter controls whether these
consecutive steps are positively of negatively
correlated. Importantly, this time-discrete set-up allows us to study
both FP and FA search. In the latter case, a searcher may cross a
target without finding it. This is in contrast to rigorous
time-continuous fBm, where due to the strict self-similarity of the
trajectories FP is identical to FA. Evidently, this self-similarity is
absent for time-discrete trajectories generated at short times $t \sim
t_0$; see \cite{Rey09} for a related discussion. We note, however,
that in a suitable scaling limit the time-discrete fBm as defined
above converges to the mathematically exact fBm.

\subsection{Search in a random distribution of targets}
\label{sec:rdt}

In analogy to Ref.~\cite{Vis99}, we consider the basic setting where a
searcher moves in a plane, $d=2$, to find uniformly random distributed
targets, see again Fig.~\ref{fig:setup}.  The motion of the searcher
is obtained from the aforementioned fBm process. This means that
according to Eq.~(\ref{fBmpp}), the searcher has a memory of its past
positions. The memory could arise from extrinsic environmental
conditions, for instance when a biological cell diffuses in a
visco-elastic medium where the diffusing element is part of a larger
evolving system \cite{goy}. Or, it could be an intrinsic property of a
searcher, like internal memory during a foraging process
\cite{Nath08,MCB14,NHKD17}. Within our fBm framework we consider the
memory to be intrinsic. In terms of movement, one may think of memory
as generating persistence or anti-persistence between the different
steps in the process. We say the search process exhibits non-renewal
if the memory lasts during our whole measurement time. If the memory
has a certain duration, i.e., the memory kernel in an fBM process
possesses a cut-off, we say the process has a renewal. Accordingly,
below we may speak of resetting the process to some initial condition
if we truncate it, or of non-resetting. The cut-off time can be a
constant value or the process can be renewed after visiting a
target. The latter situation was considered in Ref.~\cite{Vis99} by
truncating a LW upon hitting a target. This aspect will become
important when we define two different types of search efficiencies
further in this article.

Within this setting we consider the two different search modes of
cruise and saltatory foraging
\cite{JPP09,BLMV11,PCM14,PCM14b,PCKM16,Kla16,PBL19,LBV21} briefly
mentioned in Sec.~\ref{sec:backg}. In detail they are defined as
follows:

\begin{enumerate}

\item A {\em cruise forager} perceives a target while moving.
  Accordingly, whenever a target is passed it is found. The special
  problem that a cruise forager detects only the first target yields
  mathematically a FP problem
  \cite{BHRSV01,RBLSSV03,JPP09,BLMV11,BRVL14,guerin,CBRM15,LTBV20,LBV21}.

\item A {\em saltatory forager} cannot find a target while moving. It
  has to land on the target after performing a jump, or sufficiently
  close to it, to find it. After a jump the searcher typically changes
  direction. In the case of first target search this can be formulated
  as a FA (or hitting) problem
  \cite{PCM14,PCM14b,PCKM16,PMKMC17,PBL19,GuKo20,LBV21}.

\end{enumerate}

However, our setting is not yet complete, as we have to specify how a
searcher `perceives' a target. This can be modelled in two ways
\cite{JPP09,BLMV11,Vis99,LTBV20}:

\begin{enumerate}

\item[A.] We may assign a radius of {\em perception} $R_p$ to a
  searcher. This means that within $R_p$ it perceives a target with
  certainty.  In this case even point targets are found. Here, the
  perception radius is a parameter intrinsic to the searcher.

\item[B.] The searcher is {\em blind} and only finds a target when
  hitting it. In this case the searcher is reduced to a point without
  any perception. It thus can only find targets that have an
  extension.

\end{enumerate}

Combining case 1 with cases A and B yields the following two rules for
cruise search \cite{JPP09,Benh07,PBL19}:

\begin{enumerate}

\item[1.A.] A {\em perceptive cruise forager:} If a target intersects
  with a tube of radius $R_p$ around the searcher's trajectory it is
  found.

\item[1.B.] A {\em blind cruise forager:} If the searcher's trajectory
  crosses a target it is found.

\end{enumerate}

For saltatory search we get accordingly \cite{JPP09,PBL19}:

\begin{enumerate}

\item[2.A.] A {\em perceptive saltatory forager:} If after a jump of
  duration $t_0$ the target intersects with a circle of perception
  radius $R_p$ around the searcher's position it is found.

\item[2.B.] A {\em blind saltatory forager:} If after a jump of
  duration $t_0$ the searcher's position is within a target it is
  found.

\end{enumerate}

Figure~\ref{fig:setup} depicts the special case of 1.B.\ where a blind
cruise forager searches for disks of radius $R_e$. This suggests that
$R_e$ defines an extrinsic parameter specifying the environment. But
for circular targets one can scale away either $R_p$ or $R_e$ by
combining both parameters. As we have already encountered in the case
of memory, speaking here of an extrinsic parameter is consequently
ambiguous. These interpretations thus depend on the specific situation
at hand. Consequently, for disks case 1.A.\ is mathematically
equivalent to 1.B.\ and 2.A.\ to 2.B. In the search scenario shown in
Fig.~\ref{fig:setup} we therefore only need to distinguish between
cruise and saltatory foraging, cases 1.\ and 2., which is what we
study in the following.

Our second extrinsic parameter quantifies the density of targets.  We
consider the case where $N$ point targets are uniformly random
distributed in a two-dimensional box of side length $L$. This results
in a target number density of $n=N/L^2$.  The mean (free flight)
distance $l_f$ between a point target and its nearest neighbor is thus
calculated to
\be
l_f = \sqrt{\frac1n}\:;
\label{eq:td}
\ee
see Fig.~\ref{fig:setup} for the pictorial meaning. If not said
otherwise, we choose reflecting boundaries for the simulation
box. That is, when a particle crosses a boundary during a displacement
drawn from fGn, it is specularly reflected back to the bulk. But this
does not affect its next step governed by the memory in the fGn. This
means we do not `truncate' the trajectory by hitting a wall but keep
its memory. An illustration of this process can be found in the inset
of Fig.~\ref{3}(a).

An important intrinsic length scale determining such a search process
is the mean jump length of a searcher. Note that this only comes into
play because of our choice of time-discrete fBm thus yielding a slight
variation of mathematically rigorous fBm. If we assume that the
trajectory of a searcher consists of a series of consecutive jumps
with time steps $t_0$, cf.\ Fig.~\ref{fig:setup}, using the MSD
suggests an average displacement of a searcher during $t_0$ as \be
l_d=\sqrt{\langle\left(\bm{x}^H(t_0)\right)^2\rangle}\:, \ee which is
the standard deviation of fGn, see Eq.~(\ref{fbm-fgn}).  By
Eq.~(\ref{fBmmsd}) one can express $l_d$ in terms of the generalized
diffusion coefficient $K_H$ and the time step $t_0$ as \be l_d=\sqrt{2
  d K_H t^{2H}_0}\:.
\label{eq:jl}
\ee
As mentioned before, in our simulations we set $t_0=1$.  This is
convenient, since that way in Eq.~(\ref{eq:jl}) we decouple $l_d$ from
$\alpha=2H$. We thus vary $l_d$ by varying $K_H$, which in turn
depends on the fGn strength $\bm{f}^H (t)$. While $l_d$ relates to the
strength of diffusion, the second intrinsic parameter $\alpha$, which
we introduced before, determines the type of diffusion.  For our
studies, we keep the two extrinsic parameters $R_e$ and $l_f$ fixed,
which defines a specific search environment. We then explore the
impact of varying the two intrinsic parameters $\alpha$ and $l_d$ on
two generic types of foraging efficiencies, which we define later in
the text. But first we select the basic environmental regime viz.\ the
properties of the targets that we focus on in this work in relation to
all three length scales $R_e$, $l_f$ and $l_d$ introduced above.

The situation of scarce targets is perhaps the most interesting one as
it poses the challenge to efficiently locate a target after many time
steps. This regime of target densities translates geometrically to the
condition that the effective target radius is much smaller than the
mean distance between two point targets, $R_e\ll l_f$. This means that
after finding a target, there exists no other target for a searcher
within radius $R_e$. Furthermore, for effectively modelling low target
densities, the mean jump length should be much smaller than the mean
target distance, $l_d \ll l_f$. We remark that it is typically assumed
for LWs that $l_d \ll R_e$ \cite{Vis99,RBLSSV03,VoVo17,LTBV20}. For
all our simulations, the box size, effective radius and mean target
distance were set to $L=10000$, $R_e=1$ and $l_f=40$. Instead we
varied the jump length $0.04\le l_d\le24$ and $0<\alpha<2$.

\section{Efficiencies for finding the first target}
\label{sec:es}

In this section, we study the problem of finding only the first
target. For this purpose, we consider an ensemble of searchers and
numerically compute two suitably defined search efficiencies. We do so
under variation of the exponent $\alpha=2H$ of anomalous diffusion for
different mean jump lengths $l_d$, in the case of both saltatory and
cruise foragers. We explain our numerical results heuristically and by
a simple analytical approximation.

\subsection{Efficiencies based on inverse mean search times}

For finding a single target, a search process consists of a starting
point represented by a respective initial condition of a searcher and
an end point, which is when a target has been found. For a saltatory
searcher, such a search process is characterised by the mean FA time
\cite{PCM14,PCM14b,PCKM16,PMKMC17,PBL19,GuKo20,LBV21}. For a cruise searcher
the corresponding quantity is the FP time
\cite{BHRSV01,RBLSSV03,JPP09,BLMV11,BRVL14,MCB14,guerin,CBRM15,LTBV20,LBV21}.
The situation of single target search applies, for instance, to
non-recurrent chemical processes, where as soon as a reactant finds a
target, the chemical reaction takes place and the search ends
\cite{Red01,MOR14,BLMV11,VLRS11}. To simulate this search scenario, we
let a searcher start from a randomly chosen initial position. Note
that the choice of initial condition is non-trivial
\cite{BRVL14,BaCa16,PCM14,PCM14b,PCKM16,PMKMC17,PBL19,LBV21}, as we will
discuss in Sec.~\ref{sec:ss} for replenishing targets. When the
searcher finds the first target, the process starts again from another
random initial position. This can be considered as a resetting
procedure, however, as eventually we average over all initial
conditions, in effect this yields an ensemble average of searchers
with respect to random initial conditions. In order to avoid here, as
far as possible, the impact of boundaries on the results, the initial
position is chosen from a small box in the center of the main
simulation box. The size of the box is such that for a smaller jump
length $l_d$ the probability that the searcher will find a target
before reaching the boundaries is very large. For larger $l_d$,
however, boundary effects do come into play, as we discuss below. They
will be investigated in full detail in Sec.~\ref{sec:ss}.

\begin{figure} [h!]
\includegraphics[width=1\linewidth]{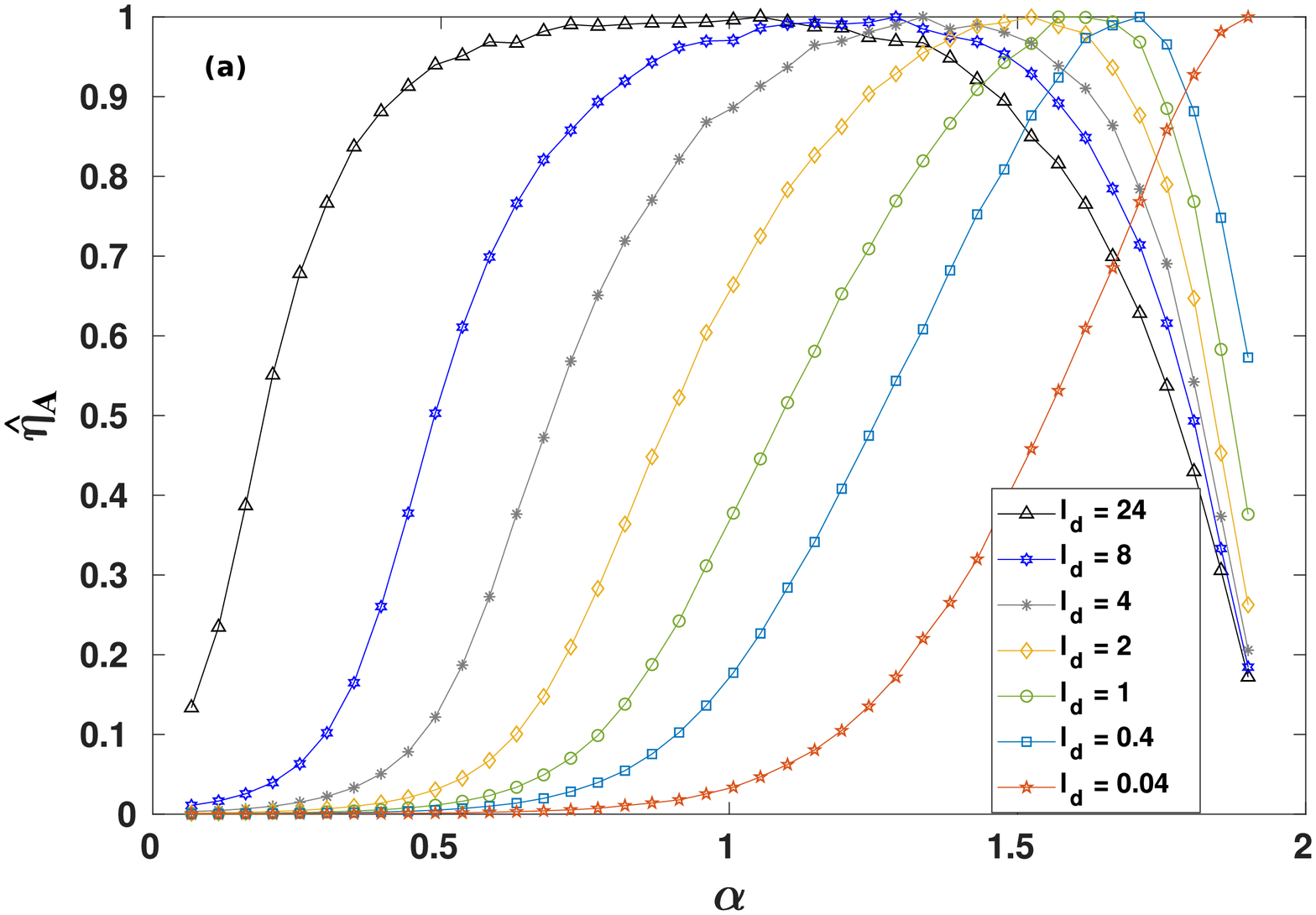}\quad\includegraphics[width=1\linewidth]{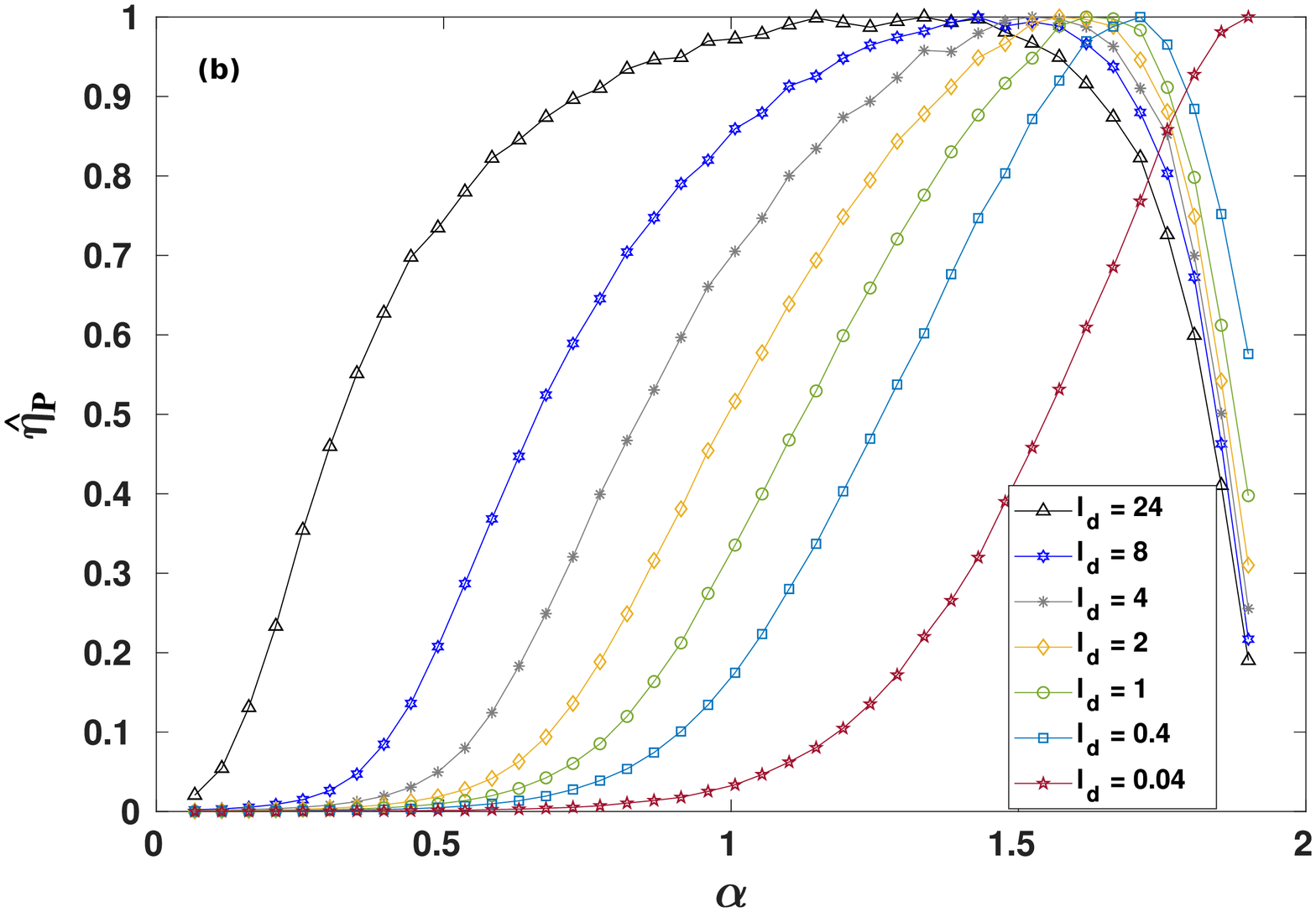}
\caption{Normalised efficiencies $\hat{\eta}_{A/P}$ defined by
  Eq.~(\ref{eq:etanorm}) for first arrival and first passage at a
  target under variation of the exponent $\alpha$ of the mean square
  displacement Eq.~(\ref{fBmmsd}) for different jump lengths
  $l_d$. Panel (a) shows the efficiencies $\hat{\eta}_A$ for first
  arrival, (b) $\hat{\eta}_P$ for first passage.}
\label{1}
\end{figure}

\begin{figure}[h!!]
\includegraphics[width=1 \linewidth]{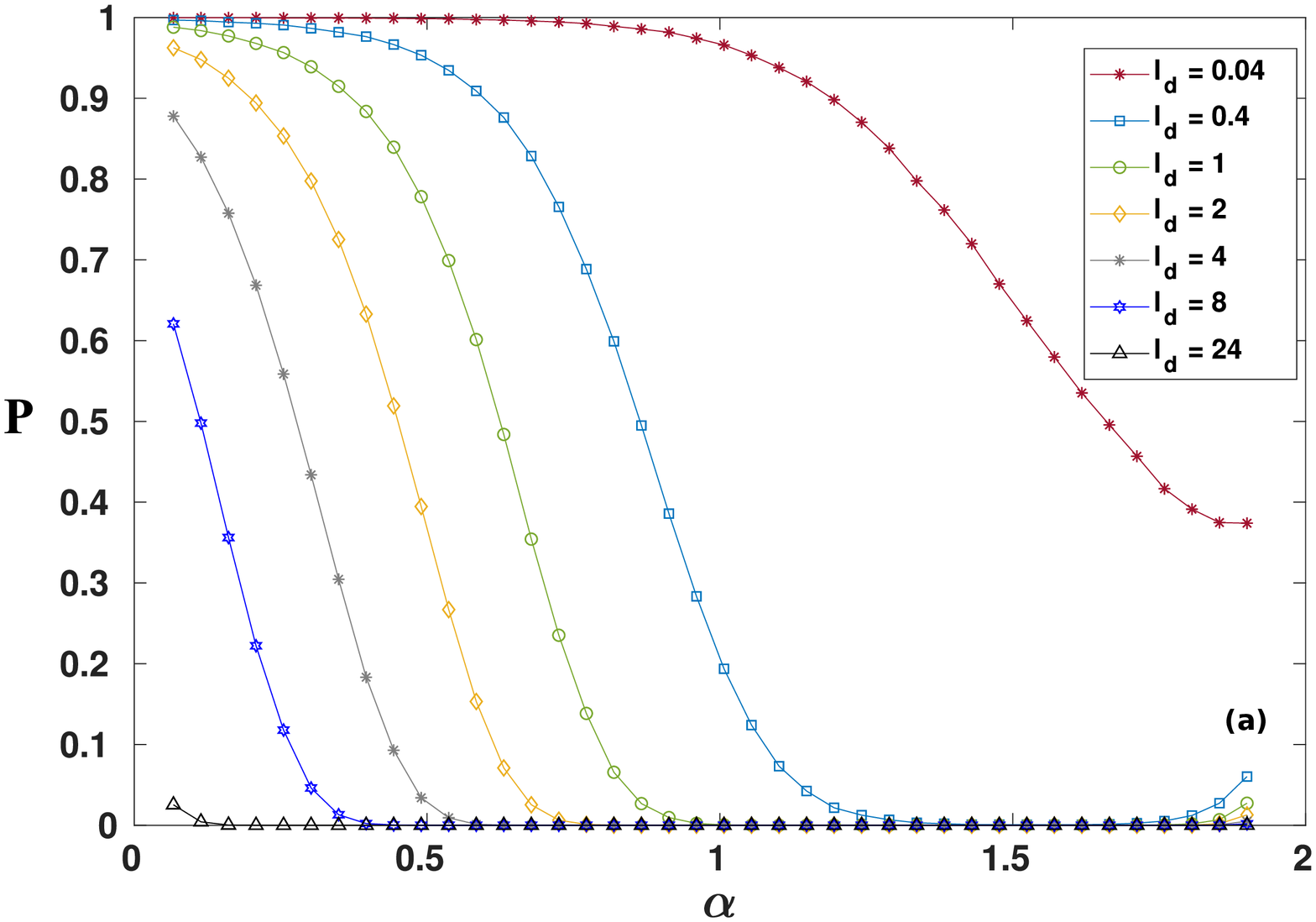}\quad\includegraphics[width=1 \linewidth]{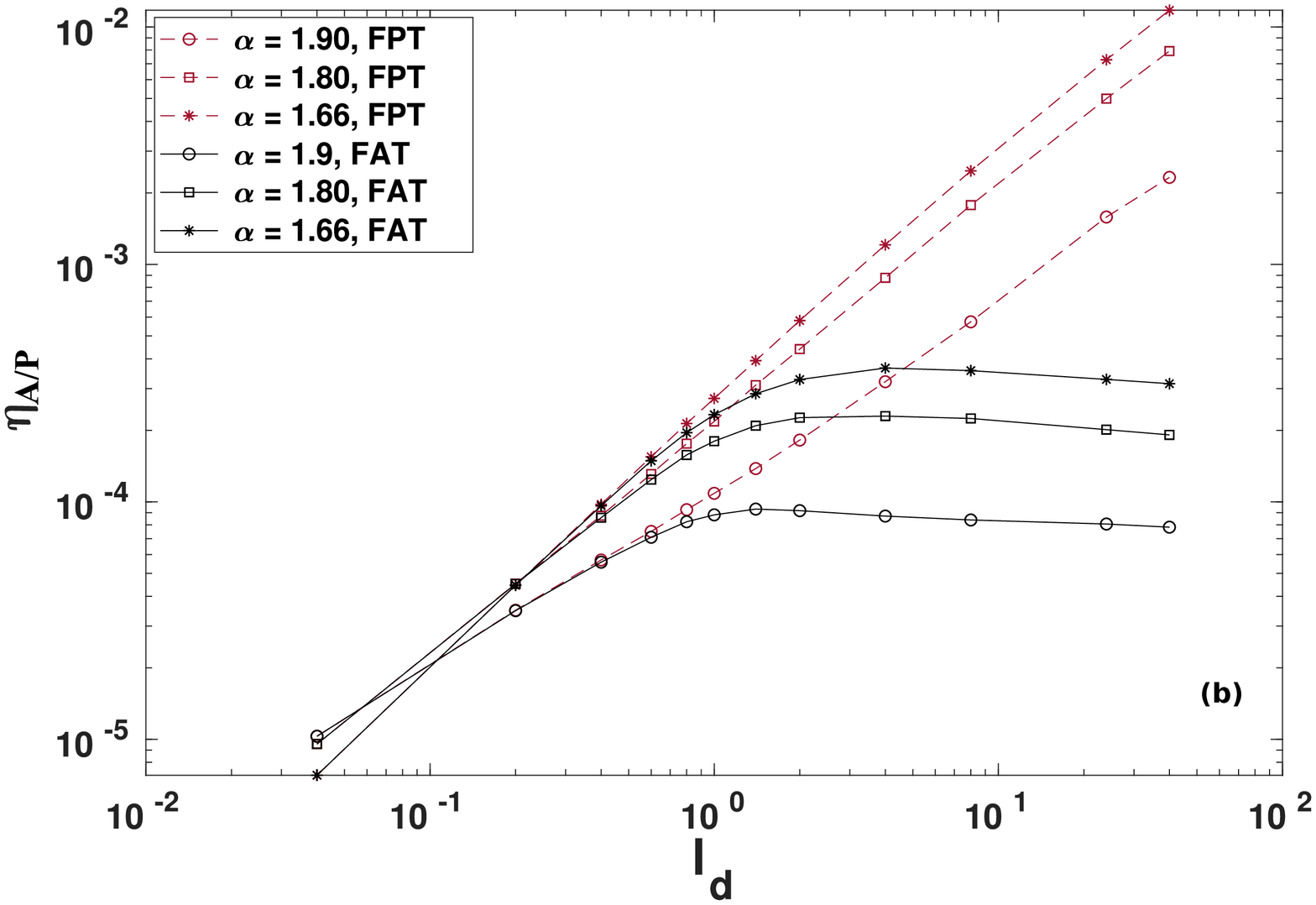}
\caption{(a) Fraction of failed tries to find a target within the
  numerically given time $T$ viz.\ the survival probability $P$ of
  first passage for the targets under variation of $\alpha$ for
  different $l_d$s. Results for first arrival are very similar. (b)
  Unnormalized efficiencies $\eta_{A/P}$, Eq.~(\ref{e1}), for both
  first arrival and first passage as functions of $l_d$. Shown are
  results for three different exponents $\alpha$ of anomalous
  diffusion as given by the legend.}
\label{1b}
\end{figure}

Based on the mean FA (FP) time written as $\left<t_{A/P}\right>$,
where the angular brackets denote an ensemble average over both
  random initial conditions of the searcher and different target
  positions, we define the corresponding efficiency $\eta_{A/P}$ as
\cite{Vis99,BHRSV01,JPP09,BRVL14,LTBV20,MCB14,VLRS11}
 \begin{equation}
 \eta_{A/P}=\frac{1}{\langle t_{A/P} \rangle}\:.
\label{e1}
 \end{equation}
For the ensemble average we considered $10^5$ simulation runs. After
$500$ runs, we regenerated the (uniform) distribution of point targets
thereby averaging over this distribution. Since we can follow
trajectories numerically for only finite times, we set the maximum
time for each search process to $T=10^5 t_0$. If until then no target
was found, we stop the search and use $T$ for the corresponding
trajectory to calculate the average in Eq.~(\ref{e1}). In that sense,
modulo small statistical errors we obtain an upper bound for the
efficiency $\eta_{A/P}$. The results of these simulations for
$\eta_{A/P}$ under variation of $\alpha$ for different jump lengths
$l_d$ are displayed in Fig.~\ref{1}. Panel (a) depicts the
efficiencies calculated using FA times, (b) is for FP
times. Each curve in the figure is normalised with respect to the
maximum value of the efficiency obtained at the corresponding jump
length, i.e.,
\be
\hat{\eta}_{A/P}=\frac{\eta_{A/P}}{\eta^{max}_{A/P}}\:.
\label{eq:etanorm}
\ee
The reason for the normalisation is that the unnormalised efficiencies
vary over several orders of magnitude with $l_d$, see Fig.~\ref{1b}
(b) that we discuss afterwards.
Figure~\ref{1} shows that for very small jump lengths $l_d$ compared
to the effective radius $R_e=1$ and the mean distance $l_f=40$, which
are both held constant, the exponent $\alpha$ of anomalous diffusion
that optimises both efficiencies $\hat{\eta}_A$ and $\hat{\eta}_P$
lies in the superdiffusive regime close to the ballistic limit
$\alpha=2$. The physical explanation is that a searcher needs to
compensate for small values of $l_d$ by performing quasi-ballistic
motion in order to move at all through in space. If $l_d$ increases for both
efficiencies, persistent motion close to $\alpha=2$ is not optimal
anymore for finding targets. This reflects undersampling, i.e.,
instead of efficiently exploring a given area, a persistent searcher
with large $\alpha$, which performs in addition jumps of large length $l_d$,
moves immediately to another area by starting there its search again
from scratch. But since it encounters a new random target
distribution in this area, the searcher loses time. This has two important
consequences for both $\hat{\eta}_A$ and $\hat{\eta}_P$. First, for
larger $l_d$ both efficiencies develop a minimum at larger $\alpha$
values. Second, accordingly a maximum emerges in both $\hat{\eta}_A$
and $\hat{\eta}_P$ for smaller $\alpha$. In this region there appears to be an
optimal interplay between a larger jump length $l_d$ and less
persistence in the motion of a searcher leading to more frequent
turns, which compensates for too large jumps by yielding an optimal
scanning of a given area. While this is observed for both FA and FP, a
crucial difference between both foraging modes is that for large
$l_d$, FA exhibits leapovers \cite{PCM14,PCM14b,PBL19,LBV21}, where a
searcher misses a target by jumping across the target without landing
on it. We will discuss this phenomenon in more detail in relation to
Fig.~\ref{1b} (b). Another observation is that, as $l_d$ increases
by approaching $R_e\simeq1$, the optimal exponent $\alpha$ for
$\hat{\eta}_P$ seems to accumulate around $\alpha=1.5$, while for
$\hat{\eta}_A$ it exhibits a shift towards the normal diffusive regime
around $\alpha=1$. This indicates that for FP, there is a special
regime of parameters $l_d$ and $\alpha$ that optimises search related
to the scanning procedure discussed above. This regime avoids both
undersampling for large $\alpha$, as well as oversampling
\cite{VoVo17,JPP09,PCM14b,PCM14,LKMK08,BaCa16,BLMV11,LTBV20} of a
small region by too many nearby trajectories for small $\alpha$. In
contrast, for $\hat{\eta}_A$ there is a less pronounced accumulation of
the maximum for larger $l_d$ with the peaks shifting and flattening
out. That is, $\hat{\eta}_A$ becomes to some extent independent of
$\alpha$ around $\alpha=1$ for larger $l_d$. This suggests that
oversampling for small $\alpha$ is less of a problem for FA processes.
which intuitively looks plausible, as this search process is
characterised by random points in space and not by the full
trajectories. We call this flattening out of $\hat{\eta}_A$ in
$\alpha$ the `paradise' regime, as for $l_d\ge l_f/2$ the jumps of the
searcher become large enough that it can always find a target
irrespective of its persistence if $\alpha$ is not too large or too
small. The factor of $1/2$ is due to choosing random initial positions
between two targets. Note that, to some extent, a paradise regime also
exists for FP search at the largest $l_d=24$, however, it is not as
pronounced as for FA processes.

For large $l_d$ and large $\alpha > 1.8$, boundary effects come into
play. However, here we argue that they do not play a role for the
results presented in Fig.~\ref{1} and their interpretation, as
follows: In Fig.~\ref{1b} (a) we plot the fraction of search
processes that failed for finding a target during the given maximum
time $T$. Accordingly, this quantity yields the survival probability
$P$, i.e., the probability of finding no target until time $T$, which
in a way is an inverse measure of search efficiency. Shown are FP
results for $P$ as a function of $\alpha$ for different $l_d$. But we
remark that $P$ is essentially the same for FA, as it is significantly
different from zero only for small $\alpha$ or small $l_d$, where
there are no leapovers. The most important result of
  Fig.~\ref{1b} (a) is the existence of two maxima at small and
large $\alpha$, which correspond to the respective minima of
$\hat{\eta}_{A/P}$ in Fig.~\ref{1}. The family of curves
  displaying small maxima around $\alpha=2$ for large $l_d$ values is
indeed due to boundary effects, which will be explicitly investigated
in Sec.~\ref{sec:ss}. In this case, if a searcher hits the reflecting
wall with a large velocity component perpendicular to it, due to the
strong persistence in the fBm motion for large $\alpha$, it will
bounce back and forth off the wall for a long time. Consequently it is
essentially stuck in one area by the wall, which means that no target
may be found anymore. This effect is worse for smaller $l_d$ values,
while for larger $l_d$ the searcher can still explore larger areas,
which explains why the small peak around $\alpha=2$ decreases for
larger $l_d$ values. Note that in any case, the fraction of searchers
affected by this boundary effect is very small (typically much below
$5\%$ for $l_d>0.4$, cf.\ Fig.~\ref{1b} (a)). Therefore, this does
not explain the drop-off in the efficiencies in Fig.~\ref{1} at larger
$\alpha$ values, which conversely becomes stronger for larger $l_d$
values. For small $\alpha$ values, in most cases of $l_d$ the
subdiffusive search times are so long that they go beyond the
numerically accessible regime. But as explained above, for $\alpha<1$
all efficiencies safely yield at least an upper bound for the exact
efficiency values.

Figure~\ref{1b} (b) supplements the analysis of normalised
efficiencies $\hat{\eta}_{A/P}$ presented in Fig.~\ref{1} by showing
the unnormalised counterparts $\eta_{A/P}$, here as functions of $l_d$
for three different $\alpha$. As we mentioned before, both quantities
vary over orders of magnitude. Most notably, $\eta_P$ increases
monotonically close to a power law in $l_d$ (for exponents see
Fig.~\ref{224} (a)) while $\eta_A$ saturates for larger $l_d$. The
saturation is a consequence of the leapover phenomenon discussed
above, which exists for FA but not for FP. Indeed, the saturation of
$\eta_A$ sets in around $l_d\simeq2=2R_e$, which matches exactly 
the condition where a searcher can jump over the full diameter of a
target without finding it. We also note that $\eta_A$ seems to
decrease slightly for larger $l_d$, which might be due to further
undersampling related to the leapovers. Furthermore, both efficiencies
decrease with larger $\alpha$. This is in line with Fig.~\ref{1},
where it was explained by undersampling.

\subsection{Comparison to efficiencies for L\'evy walks}

We now relate these results to previous works on LWs in which similar
search scenarios have been studied \cite{Vis99,VLRS11,LTBV20}. As
explained before, LWs and fBm define in principle two fundamentally
different stochastic processes. However, in both cases there are
important parameters that govern the type of diffusive spreading. For
fBm, this is the exponent $0<\alpha<2$ in the correlation function
decay Eq.~(\ref{fBmpp}), which determines the MSD Eq.~(\ref{fBmmsd}).
For LWs, in turn, the crucial quantity defining this process is the
distribution $P(r)$ from which the jump lengths $r$ are sampled
randomly at each time step, which is assumed to follow a power law,
$P(r)\propto |r|^{-\mu} $ with $1\le\mu\le3$. Via continuous time
random walk theory, the MSD can be calculated for this process, which
depends on $\mu$ in a more complicated way \cite{ZDK15}. Crucially,
for $\mu=3$, a LW reproduces normal diffusion, while $\mu=1$ yields the
limit of ballistic motion. As far as diffusion is concerned, one may
compare the LW parameter regime of $1\le\mu\le3$ with the
respective parameter regime of $0\le\alpha\le2$ for fBm. A basic
difference is that LWs do not generate subdiffusion, hence there is no
matching for $0\le\alpha<1$ in fBm to a corresponding parameter regime
for LWs. However, one may identify qualitatively $\alpha=2$ with
$\mu=1$ and $\alpha=1$ with $\mu=3$, since in the case of the former
parameter values both processes yield purely ballistic motion, while in
the latter case they reproduce normal (Brownian) diffusion.

It has now been claimed in the literature that for LWs, a universal
exponent around $\mu=2$ yields an optimal efficiency for finding
sparsely distributed targets
\cite{Vis99,BHRSV01,RBLSSV03,Sims08,Sims10,VLRS11,BaLe08,BRVL14}.
Intuitively, a LW for $\mu=2$ suitably combines properties of the two
extreme cases of ballistic and Brownian motion generating trajectories
that explore a given area by avoiding oversampling (too frequent
turns), as well as undersampling (too straight trajectories). However,
this points to our previous argument about the accumulation of optimal
efficiencies around $\alpha=1.5$ for fBm with FA for which we have
given an analogous microscopic explanation. One may thus conclude
that, under certain conditions, a search process that is intermediate
between two extreme cases may indeed optimise search efficiencies. We
emphasize, however, that we see no universality of such exponents for
optimising search, as their values depend very much on the precise
conditions of the search problem at hand. For example, our FA
efficiency depicted in Fig.~\ref{1} does not exhibit any clear
accumulation of maximal efficiencies around a particular $\alpha$, and
even for FP this effect is quite washed out. This is fully in line
with the conclusions in Ref.~\cite{CBV15} that the optimal exponent
$\mu$ for LWs largely depends on scaling, as has been substantiated
in the very recent Refs.~\cite{LTBV20,LTBV21}. For a LW search in more
complex settings, it has been found that while for similar
qualitative reasons an optimal exponent typically appears to exist,
its value depends on the variation of intrinsic and extrinsic
parameters, as well as the precise topography at hand
\cite{VoVo17,ZJLW15}.

\subsection{Efficiencies based on mean inverse search times}
  
\begin{figure}[h!!]
\includegraphics[width=1 \linewidth]{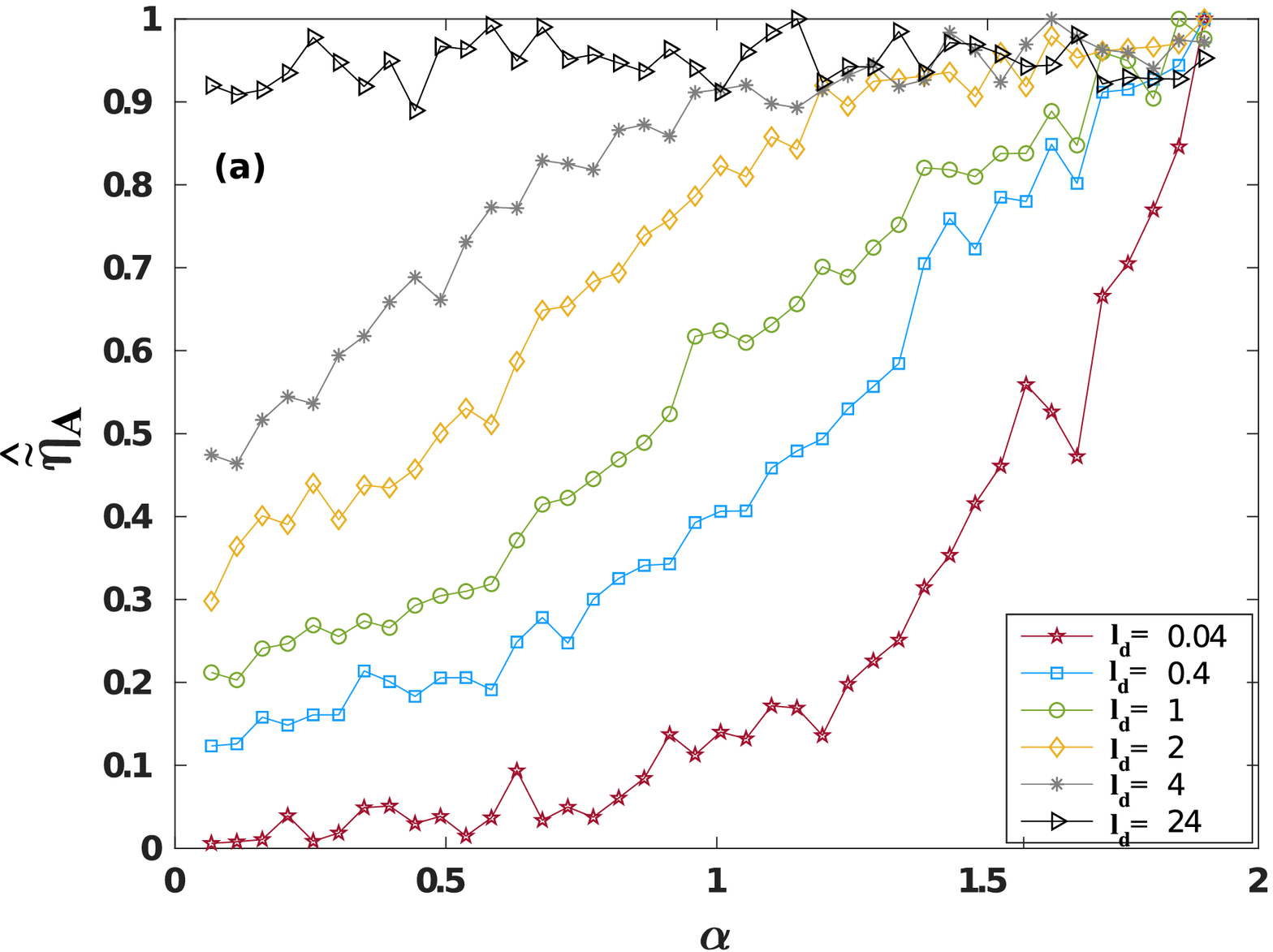}\quad\includegraphics[width=1 \linewidth]{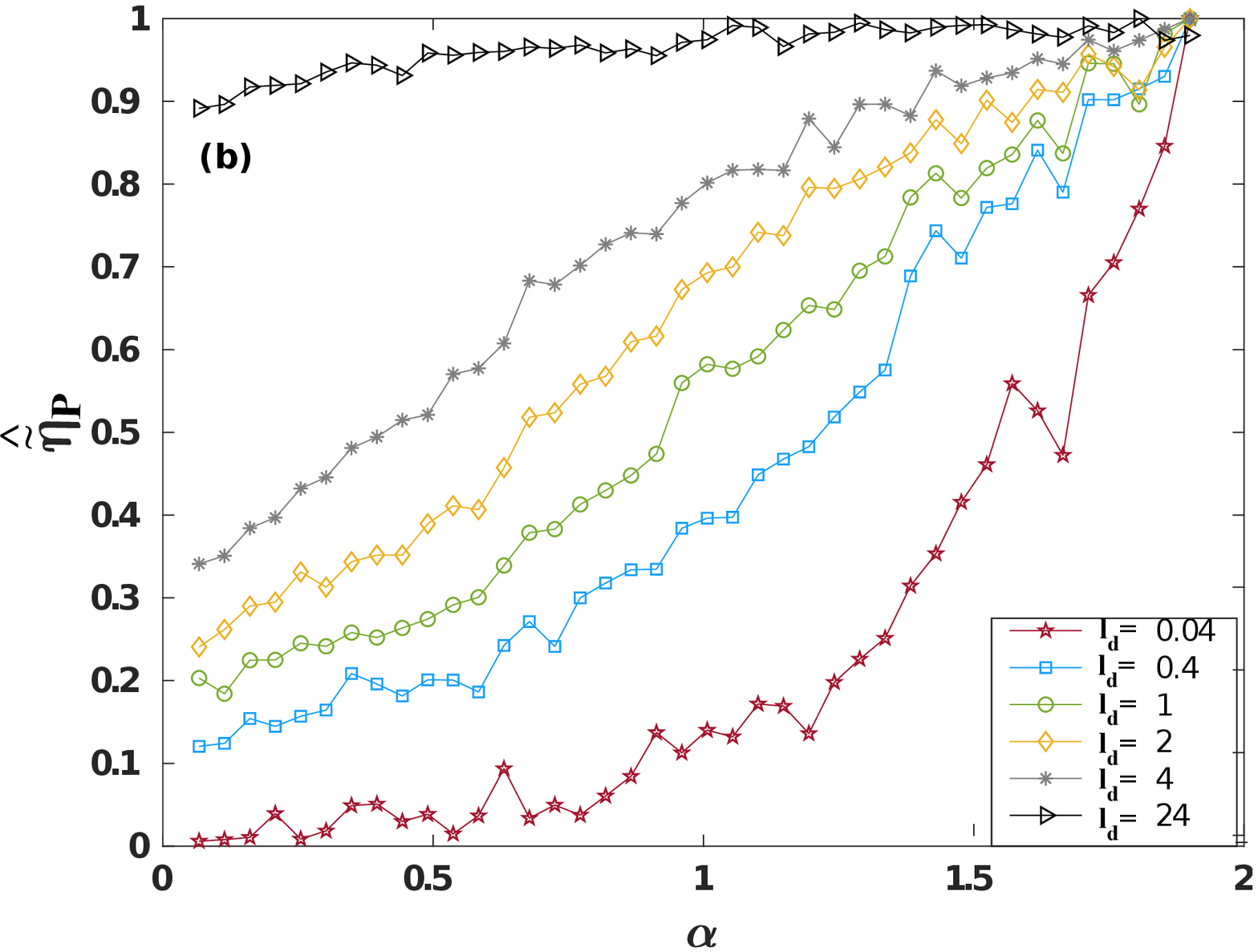}
\caption{Normalised efficiencies $\hat{\tilde{\eta}}_{A/P}$ based on
  the normalised Eq.~(\ref{e2}) for first arrival and first passage at
  a target under variation of the exponent $\alpha$ of the mean square
  displacement Eq.~(\ref{fBmmsd}) for different jump lengths
  $l_d$. Panel (a) shows $\hat{\eta}_A$ for first arrival, (b)
  $\hat{\eta}_P$ for first passage.}
\label{2}
\end{figure}

For calculating search efficiencies, Eq.~(\ref{e1}) does not always
yield a sensible definition, as in certain situations (like single
target search on the line by Brownian motion) the mean FA and/or FP
times may diverge \cite{PCM14b,Red01}. Hence, in this case one would
only obtain the trivial result $\eta_{A/P}=0$. It was thus proposed to
redefine efficiencies by using the different average
\cite{PCM14,PCM14b,PCKM16,PMKMC17}
 \begin{equation}
 \tilde{\eta}_{A/P}=\left< \frac{1}{t_{A/P}} \right> \:.
 \label{e2}
 \end{equation}
Note that this definition implies a completely different weighting of
search times compared to Eq.~(\ref{e1}): While long times now only
mildly suppress the value of $\tilde{\eta}_{A/P}$ yielding small
inverse values, the large inverse values for short search times
contribute to the efficiency more profoundly. This relates to the
specific properties of the tails of FA and FP time distributions
which, however, are numerically difficult to obtain because of their
extreme statistics. Figure~\ref{2} shows simulation results for the
renormalised efficiency $\hat{\tilde{\eta}}_{A/P}$ of both FA and FP
searches, where we set the time for failed processes to infinity. The
data is, as before, for $10^5$ runs with a length of $T=10^5 t_0$ time
steps. By comparing these curves with the ones in Fig.~\ref{1} we see,
first, that for the efficiency obtained from Eq.~(\ref{e2}),
regardless of the jump length, the most efficient way of finding a
target is to perform ballistic motion. We also observe that by
decreasing $\alpha$ the efficiency is decreasing, which for $\alpha<1$
reproduces roughly the same trend as in Fig.~\ref{1}. However, the new
efficiency wipes out any non-trivial dependencies due to long search
times. This means that the decay for large $\alpha$ is completely gone
and correspondingly there is no local maximum anymore. Note that
numerically there is a sharp drop of $\hat{\tilde{\eta}}_{A/P}$ for
$\alpha\to0$, which is very difficult to resolve computationally due
to difficulties with properly capturing very long search processes,
see Fig.~\ref{1b} (a). For practical purposes, one may thus need to
make a sensible choice as to which definition to use for calculating
efficiencies depending on the situation at hand. We also remark that
this example shows very clearly how profoundly optimality depends on
the definition of efficiency that one employs.

\subsection{Analytical approximation of efficiencies}
  
In view of the non-Markovian nature of fBm and the complexity of
two-dimensional multi-target search, providing a theory for respective
FA and FP problems is a non-trivial task
\cite{Vis99,RBLSSV03,VLRS11,guerin,LBGV18,GuKo20,LTBV20,BRBHRLV21,LTBV21}.
Here, we outline an extremely simple, handwaving argument that
analytically reproduces some of the parameter dependencies for the FP
efficiencies as seen before. We emphasize that this argument has to be
handled with much care, as we discuss in the following subsection,
where we relate it to more substantial, general theories published in
previous literature \cite{guerin,LBGV18}.

We start by simplifying our two-dimensional FP problem to a
one-dimensional setting as follows: We interpret the mean free
distance $l_f$ between two targets as the length of a line between two
absorbing boundaries. Let the searcher start the diffusion process
right in the middle of the line. Then it needs to travel a distance
$l_f/2$ towards the left or the right to hit a target. For estimating
the mean FP time, one may now simply use the MSD of the fBm process
given by Eq.~(\ref{fBmmsd}). Replacing $x^H(t)=l_f/2$ and solving for
time $t$ we obtain
\begin{equation}
  t = \left(\frac{l_f}{2\sqrt{2K_\alpha}}\right)^{2/\alpha}\:.
 \label{e21}
\end{equation}
The generalized diffusion constant $K_\alpha$ can be in turn
approximated to, see Eq.~(\ref{eq:jl}) with $d=1$,
\be
K_\alpha=\frac{l_d^2}{2 t_0^\alpha}\:,
\label{eq:gdk}
\ee
where in our case $t_0=1$. We can now calculate the FP efficiency
$\eta_P$ in terms of all relevant parameters by substituting the FP
time in Eq.~(\ref{e1}) by the one of Eq.~(\ref{e21}) supplemented by
Eq.~(\ref{eq:gdk}). This yields
\begin{equation}
  \eta_P\simeq {\left(\frac{\sqrt{2}l_d}{l_f}\right)}^{2/\alpha} \:.
 \label{e22}
\end{equation}
Our simple argument may be understood as a kind of mean field
approximation, in the sense that we use a single-target picture to
approximate multi-target search in a low density limit. It should also
apply to FA search as long as leapovers do not dominate, see
Fig.~\ref{1b} (b). We test the validity of our approach by
comparing the dependencies of $\eta_P$ on $l_d$ and $\alpha$ as
predicted by Eq.~(\ref{e22}) with numerical data.

\begin{figure}[h!!]
\includegraphics[width=1 \linewidth]{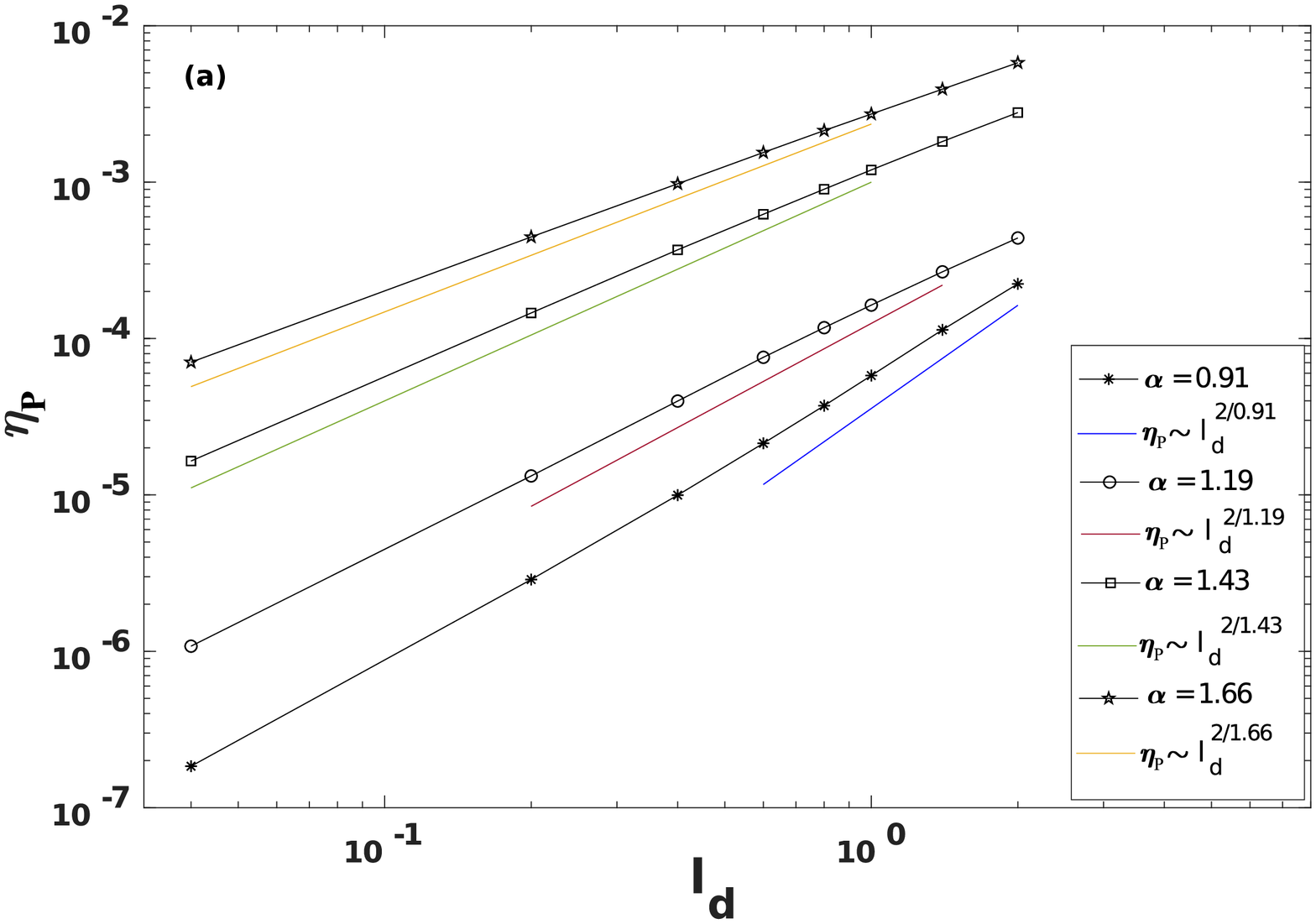}\quad\includegraphics[width=1 \linewidth]{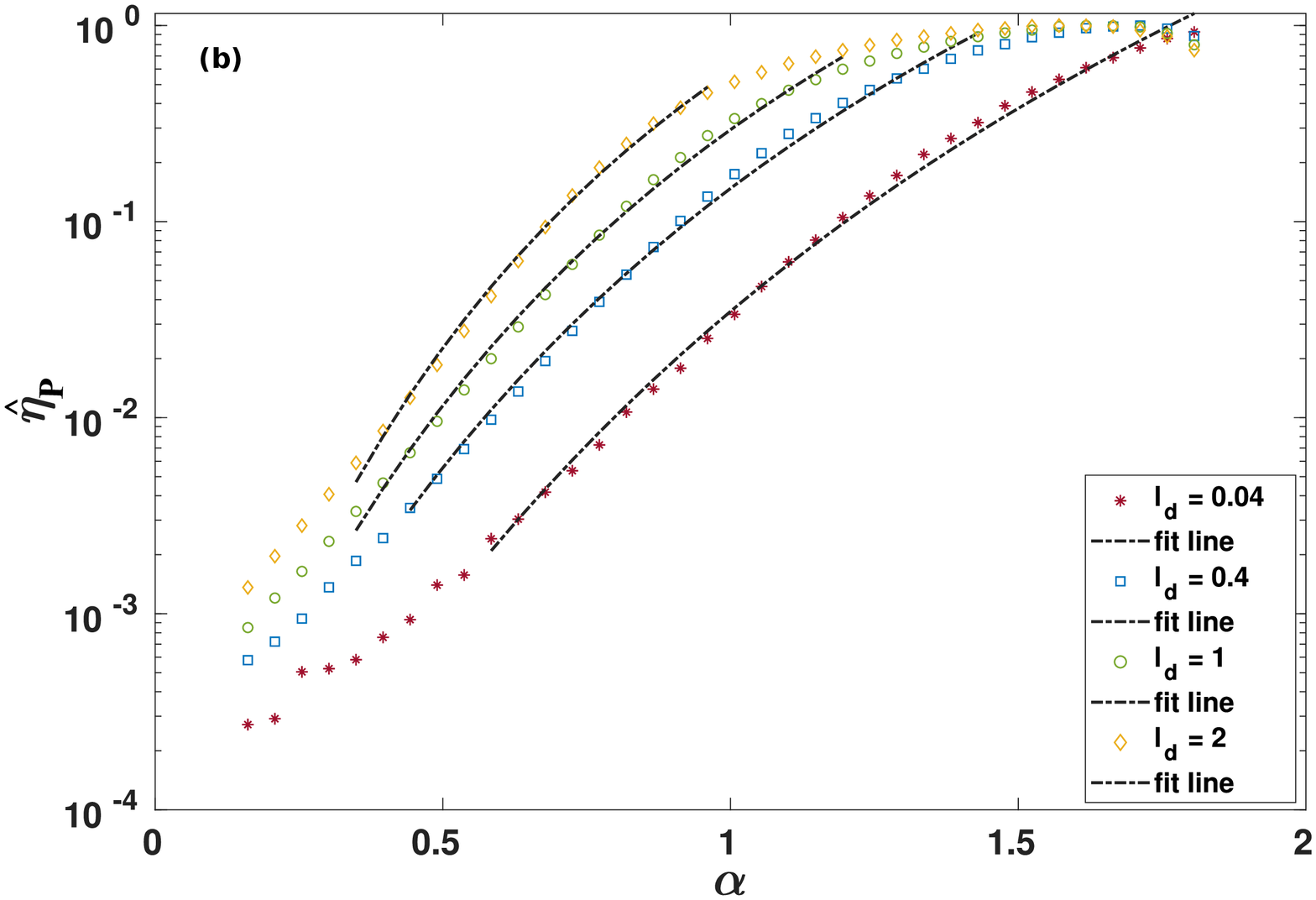}
\caption{(a) Unnormalized efficiency $\eta_P$, Eq.~(\ref{e1}), for
  first passage search as functions of the mean jump length $l_d$
  computed for different exponents $\alpha$ of anomalous diffusion.
  The data corresponding to $\alpha=1.43$ and $\alpha=1.66$ are
  multiplied with factors of $5$ and $10$ for sake of clarity.  Lines
  indicate power laws of the efficiencies with $\eta_P \sim
  l_d^{2/\alpha}$ in agreement with Eq.~(\ref{e22}). (b) Normalised
  efficiencies $\hat{\eta_P}$, Eq.~(\ref{eq:etanorm}), as functions of
  the exponent $\alpha$ of anomalous diffusion for different $l_d$
  fitted with Eq.~(\ref{e22}).}
\label{224}
\end{figure}

Figure~\ref{224} (a) shows the unnormalized efficiency $\eta_P$ as a
function of the mean jump length $l_d$ for different exponents
$\alpha$ of anomalous diffusion (cp.\ with Fig.~\ref{1b} (b)).
While due to the simplicity of the theoretical argument we may not
expect a full quantitative matching between data and approximation, at
least the power law dependence of $\eta_P$ with $l_d$ for different
$\alpha$ is reproduced surprisingly well. Note that for large $\alpha$
we have restricted our fit region to smaller $l_d$, as we may not
expect our theory to capture the effect of undersampling. For smaller
$\alpha$, we need to go to larger $l_d$ due to the otherwise high
percentage of failed searches, see Fig.~\ref{1b} (a). We did not
include results for $\alpha<0.91$, as here the efficiencies become so
small that they are difficult to compute numerically, cf.\ again
Fig.~\ref{1b} (a) and our respective diuscussion.  Figure~\ref{224}
(b) displays results for the normalised efficiency $\hat{\eta}_P$
as a function of $\alpha$ for different $l_d$ (selected range of data
from Fig.~\ref{1} (b)). The fits have been adjusted to an
intermediate region of $\alpha$ values, as for small $\alpha$ the data
is not reliable due to many failed searches, see Fig.~\ref{1b} (a),
which yields insufficient statistics to check for fine details. For
larger $\alpha$ we may not expect Eq.~(\ref{e22}) to reproduce the
non-monotonic dependencies of $\hat{\eta}_P$ due to undersampling, see
in Fig.~\ref{1} (b). Again, in this intermediate regime our
handwaving theory predicting an exponential dependence on $2/\alpha$
works surprisingly well.

\subsection{Relation between our approximation and other results
    for mean first passage times}

Here, we first show that our simple approximation Eq.~(\ref{e22})
forms a special case of a much more advanced theory for calculating
mean FP times of fBm \cite{guerin}. This theory, on the other hand, is
recovered within the framework of a more general theory \cite{LBGV18}
which, if applied to our particular setting, indicates important
limits of validity of our approach.

The first layer of embedding is provided by the theory of Gu\'erin et
al.\ \cite{guerin}, who derived analytical results for the mean FP
time $\langle t_P \rangle$ of compact (as explained below)
non-Markovian random walkers for finding a single target in a
$d$-dimensional volume $V$ confined by reflecting boundaries. By
applying a generalised form of a renewal equation, for fBm in one
dimension the mean FP time was calculated to
  \begin{equation}
\langle t_P \rangle= \frac{V \beta_H}{x_0}
\left(\frac{x_0}{\sqrt{K_H}}\right)^{1/H}\:,
 \label{eq:guerin}
 \end{equation}
where $x_0$ is the initial condition of the searcher with a target at
position $x=0$. The constant $\beta_H$ is a non-trivial quantity that
captures the non-Markovianity of the process and can only be computed
numerically \cite{guerin}. Comparing our trivial formula
Eq.~(\ref{e21}) with Eq.~(\ref{eq:guerin}), one can see that the
former yields the same dependence of the mean FP time on $K_H$ as the
latter. Furthermore, replacing $x_0=l_f/2$ in Eq.~(\ref{eq:guerin})
according to our simplified assumption, as well as using $V=l_f$ for
our one-dimensional setting, for the FP time in Eq.~(\ref{e21}) we
obtain exactly the same scaling with $l_f$ as in
Eq.~(\ref{eq:guerin}). To what extent the scaling of the FP time with
$\alpha$ for fixed $K_{\alpha}$ and $l_f$ is reproduced is not so
clear, as in detail this depends on $\beta_H$. The relation between
Eqs.~(\ref{e21}) and (\ref{eq:guerin}) also explains why our
analytical approximation cannot be applied to the efficiency
definition Eq.~(\ref{e2}). We thus conclude that our approximation
Eq.~(\ref{e21}) corresponds to a simplified version of
Eq.~(\ref{eq:guerin}) if strictly constrained to one dimension.

A more general theory for calculating mean FP times of non-Markovian
scale-invariant, aging diffusion processes, which includes the one of
Ref.~\cite{guerin} as a special case, was developed by Levernier et
al.\ \cite{LBGV18}. A crucial feature of this theory is to distinguish
between compact and non-compact stochastic processes, as these two
cases lead to different formulas for the corresponding mean FP
times. Compactness (or recurrence) means that a process comes back to
its starting point with probability one while non-compactness (or
transience) implies the opposite
\cite{Red01,MOR14,BCK10,BeVo14}. Consequently, this mathematical
property intimately relates to what before we called oversampling,
respectively undersampling. For non-ageing dynamics \cite{LBGV18}, a
simple criterion for compactness is derived from the walk dimension
$d_w$ of a stochastic process, defined by $\langle x^2(t)\rangle\sim
t^{2/d_w}\:(t\to\infty)$ \cite{LBGV18,MOR14,BeVo14}. If $d_w$ is
greater than the dimension $d$ of the embedding space of the walk,
$d_w>d$, a process is called compact. Conversely, non-compactness
means $d_w<d$; the case $d_w=d$ is called marginal
\cite{MOR14,BCK10,BeVo14}. For fBm with $d_w=1/H$,
cf.\ Eq.(\ref{fBmmsd}), we thus have that in dimension $d=1$ it is
compact for all $0<H<1$. Hence, strictly speaking Eqs.~(\ref{e21}) and
(\ref{eq:guerin}) only hold for compact one-dimensional fBm.

In higher dimensions, the theoretical results for mean FP times of
both Ref.~\cite{guerin,LBGV18} are in many respects much more
complicated. For the particular case of fBm in $d=2$ dimensions, the
above definition yields that fBm is compact for $H<1/2$ while for
$H>1/2$ it is non-compact. This is reflected in different formulas for
$\langle t_P \rangle$ in both regimes \cite{LBGV18}.  By using the
notation of Ref.~\cite{LBGV18} (see the results in Tables~I and II
therein), let $R$ denote the confining domain size, $r$ the distance
the walker starts from a target and $a$ the target size. Working in
the limits $R\gg 1$ and $a\ll r$, it was shown that for compact fBm in
$d$ dimensions \be \langle t_P \rangle \sim R^d r^{1/H-d}
\label{eq:tpc}
\ee
while in the non-compact case
\be
\langle t_P \rangle \sim R^d a^{1/H-d} \:.
\label{eq:tpnc}
\ee
Let us now assume that $R=l_f$ and $r=l_f/2$, as in our approximation,
and in our setting we have $a=2R_e$. For one-dimensional (compact) fBm
the scaling of $\langle t_P \rangle \sim l_f^{1/H}$ is then recovered
from Eq.~(\ref{eq:tpc}), in agreement with the one-dimensional fBm
results Eqs.~(\ref{e21}) and (\ref{eq:guerin}). The very same scaling
holds for $d=2$ but only in the compact case of $H<1/2$ while for
$H>1/2$ Eq.~(\ref{eq:tpnc}) leads to $\langle t_P \rangle \sim
(l_f/R_e)^d R_e^{1/H}$. For two-dimensional fBm the theory of
Ref.~\cite{LBGV18} thus predicts different laws for $\langle t_P
\rangle$ in the compact and non-compact regimes. In contrast, our
numerical results Fig.~\ref{224} (b) display a smooth transition
around $H=1/2$ viz.\ $\alpha=1$ under variation of $\alpha$, well
fitted by the single formula Eq.~(\ref{e21}) that according to the
above theory should only hold for $H<1/2$ in one dimension.

These seemingly different facts can be reconciled as follows:
References \cite{guerin,LBGV18} consider the case of a walker starting
at a given initial condition that searches for a single target at
another given position. In contrast, we reported results for finding
one out of many randomly distributed targets by averaging over random
initial conditions of the searcher. In our approximation, this is
roughly captured by assuming $R=l_f$ and $r=l_f/2$ (see above). But
these two particular assumptions cancel out the dimensionality
dependence $d$ in the compact case Eq.~(\ref{eq:tpc}). Hence, the
respective one-dimensional result carries over to two dimensions,
which in turn is what we assumed for our approximation. We also note
that both Eqs.~(\ref{eq:tpc}) and (\ref{eq:tpnc}) are
proportionalities (up to a prefactor independent of geometric
parameters) yielding no information about the generalised diffusion
coefficient $K_H$ and the associated jump length $l_d$ that we
varied. Finally, for our choice of fixed parameters we have $a=2R_e=2$
while $r=l_f/2=20$, which does not quite match to the condition $a\ll
r$ underlying the derivation of both Eqs.~(\ref{eq:tpc}),
(\ref{eq:tpnc}), nor do we necessarily work at low densities of
targets. In that respect the theory developed in Ref.~\cite{LBGV18}
describes a different asymptotic parameter regime compared to our
setting. We thus argue that it cannot be applied directly to
understand the transition in the efficiency around $\alpha=1$
displayed in Fig.~\ref{224} (b). To explain in further detail why
for our data the transition in $\alpha$ looks smooth and is reproduced
by the formula that according to Eqs.~(\ref{eq:tpc}), (\ref{eq:tpnc})
should only hold for the compact regime thus remains an interesting
open question.

In view of the above theoretical results it becomes clear, however,
that in more general situations our simple approximation
Eq.~(\ref{e21}) can only be of rather limited validity. Typically, the
dimensionality dependence of $\langle t_P \rangle$ will not cancel
out, and one may thus not trivially extrapolate a result for one to
higher dimensions; see Refs.~\cite{LTBV20,BRBHRLV21,LTBV21} for an
analogous discussion in the case of LWs. Exactly for that reason we
expect Eq.~(\ref{e21}) to be wrong under variation of $l_f$. To study
the dependence of $\langle t_P \rangle$ on the density of targets
hence yields another interesting problem for further study. While thus
there remain challenging open questions, we see no contradiction
between neither our numerical results nor our handwaving approximation
and the theories developed in Refs.~\cite{guerin,LBGV18}.

\section{Efficiencies for subsequently finding many targets} \label{sec:ss}

We now study the problem of a searcher that, without resetting, finds
many targets during one run. First we introduce the basic framework of
this search problem by defining a suitably adapted search
efficiency. This type of search very much depends on details of the
environment, in particular properties of the target and boundary
conditions. We first explore the search for targets in the bulk, here
both replenishing and non-replenishing resources, before we
investigate the impact of boundary conditions on finding replenishing
targets. The latter case establishes a cross-link to the very recent
field of active particles.

\subsection{Multi-target search along a trajectory}
  
In numerous realistic situations, such as recurrent chemical reactions
or animals looking for food \cite{MOR14,BLMV11,VLRS11,MCB14} many
targets need to be found by a single searcher. Compared to the search
problem of Sec.~\ref{sec:es}, one may call this scenario
non-resetting, as here a searcher consumes targets along a single path
generated by its continuous movements in time. In Sec.~\ref{sec:es},
efficiencies were calculated by the inverse of FA and FP times, which
in turn were defined as ensemble averages over many searchers starting
at different initial conditions, see Eq.~(\ref{e1}). In order to
adequately describe multi-target search along a trajectory, one
replaces this average by a combined time-ensemble average as follows:
First one counts the number $N$ of targets that have been visited
along a trajectory during a given time $T$
\cite{Vis99,BaLe08,JPP09,VLRS11,BRVL14,LTBV20,ZJLW15,MCB14}. The
fraction $T/N$ then yields the time average for finding $N$ targets
along the trajectory of a searcher. This is easily seen by
  $<t>=1/N\sum_{k=0}^{N-1}(T_{k+1}-T_k)$, where $T_0=0$ is the initial
  time and $T_k$ the time to find the $k$-th target. For large $N$ we
  can neglect $(T-T_N)/(N+1)$ and have $<t>\simeq T/N$. If not said
otherwise, for the following results we fix $T=10^6$. Since for this
finite search time the outcome may still depend on the specific path
of the searcher, in addition we average over an ensemble of searchers
starting at random initial positions. And we typically choose $10^4$
simulation runs. In analogy to Eq.~(\ref{e1}), an adequate definition
of efficiency is then obtained by
\cite{BaLe08,BRVL14,LTBV20,ZJLW15,MCB14}
\begin{equation}
 \eta_{A/P}^*=\frac{\langle N_{A/P} \rangle}{T}\:.
 \label{e3}
\end{equation}
As before, the angular brackets denote the average over an ensemble of
searchers starting at random initial positions, here with respect to
the number $N_{A/P}$ of targets found during the total time $T$ while
arriving (A) at or passing (P) through targets. It is trivially clear
that the above equation gives nothing else than the inverse of the
combined time-ensemble average for the average time $\langle
T_{A/P}\rangle=T / \langle N_{A/P}\rangle$ along a path.

In analogy to Sec.~\ref{sec:es}, in what follows we numerically
investigate this search situation for both saltatory and cruise
searchers, viz.\ arrival at and passage through targets, by varying
the exponent $\alpha$ of anomalous diffusion. We do so for the same
fixed extrinsic parameters $L,R_e,l_f$ as before,
cf.\ Sec.~\ref{sec:model}, by choosing different jump lengths
$l_d$. The key quantity to compute is again the efficiency, in this
case defined by Eq.~(\ref{e3}). However, in contrast to first target
search, for consecutively finding many targets one needs to
distinguish between two different types of resources
\cite{Vis99,bartu2,LTBV20,MCB14,VLRS11,ZJLW15}.

\begin{enumerate}

\item {\em Non-destructive, or replenishing targets}: After a target
  is visited by a searcher, it remains intact and can be revisited.
  Typically, such a target is modelled as a replenishing resource,
  i.e., it reappears either when a certain delay time has passed after
  its consumption \cite{RBLSSV03}, or after the searcher has passed a
  certain cut-off distance away from it \cite{Benh07,LTBV20}. This
  prevents the searcher from artifically consuming the same target
  over and over again.  For our subsequent studies, we choose a
  cut-off distance of $d_c=2 R_e= 2$.

\item {\em Destructive, or non-replenishing targets}
  \cite{BaLe08,VoVo17}: After being visited by a searcher the target
  disappears forever.

  \end{enumerate}

These two extrinsic environmental conditions for the targets define
completely different search scenarios
\cite{Vis99,bartu2,LTBV20,VLRS11}. An additional crucial complication
that we will explore in detail is the interplay between these two
different target types and the boundary conditions.

\subsection{Replenishing and non-replenishing targets in the bulk}

\begin{figure} [h!]
\includegraphics[width=1\linewidth]{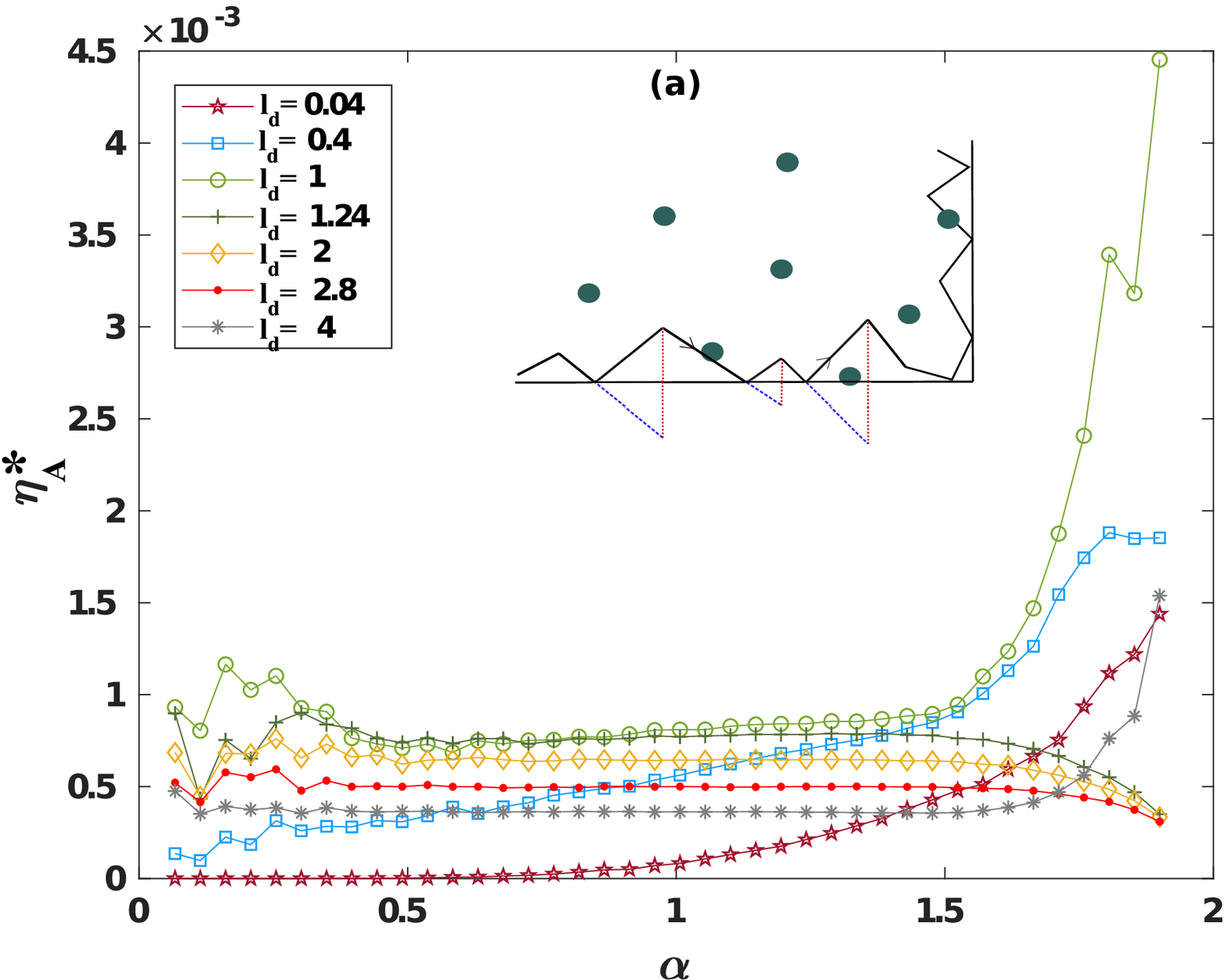}\quad\includegraphics[width=1\linewidth]{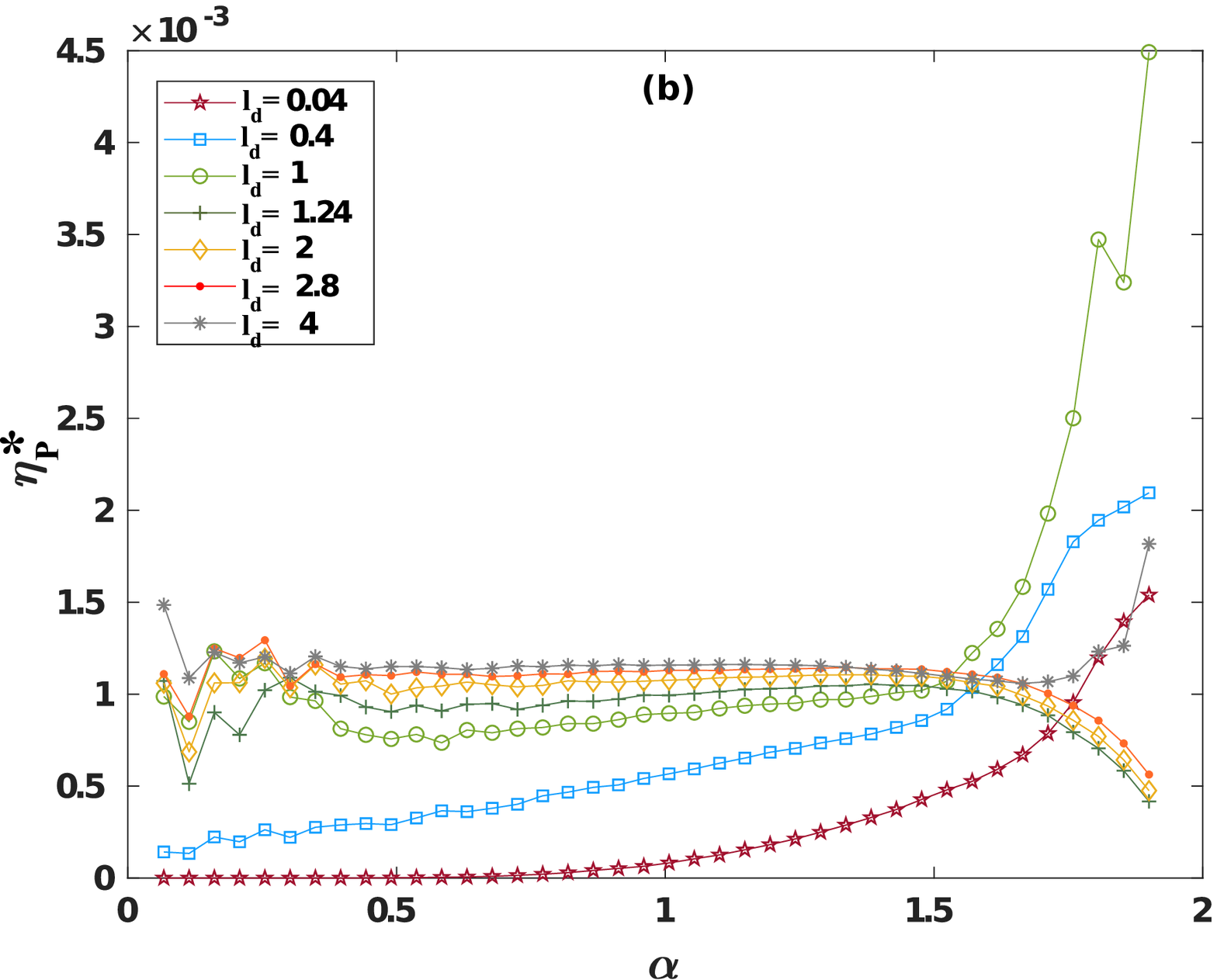}
\caption{Efficiencies $\eta_{A/P}^*$ defined by Eq.~(\ref{e3}) for
  arrival (A) at and passage (P) through replenishing targets along
  the trajectory of a searcher during a given time $T$ under variation
  of the exponent $\alpha$ of the mean square displacement
  Eq.~(\ref{fBmmsd}) for different jump lengths $l_d$. Panel (a) shows
  the efficiencies $\eta_A^*$ for arrival (saltatory), (b) $\eta_P^*$
  for passage (cruise) search. The inset in (a) illustrates our
  definition of reflecting boundaries. It shows a corner of the wall
  confining the simulation box.  The black line with the arrows
  represents the trajectory of an fBm particle that has hit the wall
  at least once already. The blue dotted lines depict how it would
  continue without the presence of the wall. The red dotted line
  indicates how the particle is specularly reflected at the wall.  For
  a particle with persistence, this leads to the phenomenon of
  `stickiness' at the wall. The grey disks are scatterers, as in
  Fig.~\ref{fig:setup}.}
\label{3}
\end{figure}

We start by investigating the situation of replenishing target search
in the bulk, i.e., without elaborating on the impact of
boundaries. Figure~\ref{3} shows simulation results for the
efficiencies $\eta_{A/P}^*$, Eq.~(\ref{e3}), of both arrival (a) at
and passage (b) through replenishing targets under variation of
the exponent $\alpha$ of anomalous diffusion for different jump
lengths $l_d$. We see that for the smallest $l_d=0.04$, in both cases
ballistic motion outperforms any other type of motion, in analogy to
Fig.~\ref{1}. The physical explanation is the same as for
Fig.~\ref{1}: A large $\alpha$ needs to compensate for a small $l_d$
for the searcher to move anywhere. Furthermore, if a target is found,
for small jump lengths the cut-off distance $d_c$ translates into long
delay times before the visited target reappears. Hence, the return
time to revisit the same target is very long, which explains why
efficiencies are close to zero in the subdiffusive regime of
$\alpha<1$, where returns dominate the search due to anti-persistence
in fBm. This particular type of return dynamics, which we call the
revisiting target mechanism, will subsequently become very
important. Note that not any subdiffusive dynamics is
  characterised by returns. A counterexample is a continuous time
  random walk, where subdiffusion originates from power law waiting
  times at a given position \cite{MeKl00}. This dynamic should thus
  yield very different search efficiencies compared to fBm.

For the slightly larger jump length $l_d=0.4$, the region of maximal
values of the two efficiency curves widens a bit by including slightly
smaller $\alpha$ values. One may speculate that, as in Fig.~\ref{1},
the maximal efficiency now starts to shift to smaller $\alpha$ values
due to undersampling around $\alpha=2$. Overall, both efficiencies are
getting larger for all $\alpha$ values. That this happens in the
region of $\alpha<1$ can again be explained by the revisiting target
mechanism described above: Notably, for larger $l_d$ it now switches
from slowing down search to enhancing it, because the searcher leaves
the target more quickly after finding it, leading it to replenish
quickly. However, due to the anti-persistence of fBm, for $\alpha<1$
the searcher more frequently returns to the same target. Hence, by
increasing $l_d$ subdiffusive dynamics yields a new important search
strategy for exploiting replenishing targets. That subdiffusion can
enhance search success has also been reported for a very different
search setting in Ref.~\cite{GuWe08}. Very recently, by analysing
  experimental data it has been found that subdiffusion adequately
  describes the area-restricted search of avian predators
  \cite{VOCGTNA21}.

For the next largest jump length $l_d=1$, instead of the maximum
around $\alpha=2$ further shifting to smaller $\alpha$ values as in
Fig.~\ref{1}, there is a dramatic increase of both efficiencies
towards $\alpha=2$ again, which is in sharp contrast to the first
target search. We will argue in the next Sec.~\ref{sec:bc} that this
is due to an interplay between persistence in fBm and the reflecting
boundaries that we have chosen. But as this is a highly non-trivial
problem by itself, we now primarily focus on $\alpha<1.5$ where
searchers typically do not hit the boundaries and bulk dynamics
dominates the search. Here, both efficiency curves further increase by
increasing $l_d$ compared to the two previous smaller jump lengths by
flattening out up to $\alpha<1.5$. This increase may again be
explained by smaller delay times for a visited target to reappear when
$l_d$ is getting larger, which enables normal and sub-diffusive search
to more strongly contribute to search success due to the revisiting
target mechanism. Note that at the smallest $\alpha$ values the
  search times become very large and have to be numerically truncated,
  cf.\ Fig.~\ref{1b} (a) and our respective discussion. Hence,
  beyond the numerically supported general trend to small
  efficiencies, we cannot resolve whether the shown fluctuations are
  due to statistical errors or hint at more subtle local
  non-monotonicities.

If we further increase the jump length to $l_d>1$, we obtain two
different families of efficiencies for arrival compared to passage
search for all $\alpha$ values. This is because of the onset of
leapovers for arrival search as discussed in Sec.~\ref{sec:es}, see
particularly Fig.~\ref{1b} (b). That is, while $\eta_A^*$
generally starts to slightly decrease for all $\alpha$ by increasing
$l_d$, in line with our results for first target search in
Fig.~\ref{1b} (b), $\eta_P^*$ generally keeps slightly increasing
with $l_d$ until a quite perfect plateau region has been reached for
$\alpha<1.5$ up to the largest $l_d$ considered here. It is
intuitively clear that for passage search, a larger $l_d$ should
generally increase the search efficiency unless there are other
effects mitigating this mechanism.

Remarkably, we observe two rather spectacular transitions at the
largest $\alpha$ values, from maximal efficiencies for $l_d<1$ to
minimal values for $l_d>1$, and then the reverse between $l_d=2.8$ and
$l_d=4$. This happens for both arrival and passage search and thus
cannot be attributed to the onset of leapovers as discussed above. In
the following section, we will argue that these two transitions are
again subtle boundary effects. Correspondingly, for $1<l_d < 2.8$ now
non-ballistic search with $\alpha<1.5$ yields the largest efficiencies
for both passage and arrival search. This can be understood again by
the revisting target mechanism introduced above. We argue that,
surprisingly, subdiffusive search maximises our search efficiencies
within this $l_d$ parameter regime, cf.\ again Ref.~\cite{GuWe08} for
related results.

\begin{figure}  [h!]
\includegraphics[width=1\linewidth]{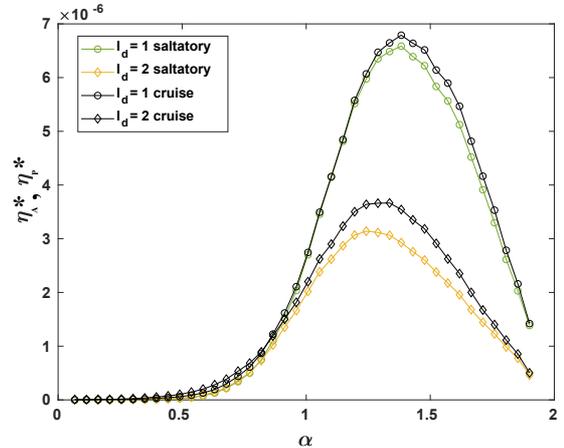}
\caption{Efficiencies $\eta_{A/P}^*$ defined by Eq.~(\ref{e3}) for
  arrival (A) at and passage (P) through non-replenishing targets
  along the trajectory of a searcher during a given time $T$. Shown
  are results under variation of the exponent $\alpha$ of the mean
  square displacement Eq.~(\ref{fBmmsd}) for two different jump
  lengths $l_d$.}
\label{6}
\end{figure}

Figure~\ref{6} demonstrates that this mechanism provides indeed the
correct explanation. Similar to Fig.~\ref{3}, it displays results for
both efficiencies $\eta_A^*$ and $\eta_P^*$ as functions of $\alpha$,
here for the two particular jump lengths $l_d=1,2$. But in this case,
we have non-replenishing targets that are destroyed after a visit,
hence the revisiting target mechanism cannot contribute to search
success anymore. Note that the y-axis for the values of the
efficiencies is scaled by a factor of $10^{-6}$, in sharp contrast to
Fig.~\ref{3} where the scale is $10^{-3}$. Overall, both search
efficiencies are diminuished dramatically in the case of
non-replenishing targets. In the subdiffusive regime of $\alpha<1$,
the efficiencies are indeed close to zero, which confirms the
importance of the revisiting target mechanism. We furthermore observe,
at least with respect to two values of $l_d$, that both efficiencies
decrease for all $\alpha$ by increasing $l_d>1$. While for arrival
this is in line with Figs.~\ref{3} (a) and \ref{1b} (b) due to
the onset of leapovers, for passage this is exactly the opposite to
Fig.~\ref{3} (b). One may speculate that this reflects some
interplay between the periodic boundaries and FP search. However,
based on only two curves more detailed conclusions cannot be drawn and
remain open for further research. The existence of local maxima in
$1<\alpha<1.5$ viz.\ the suppression of efficiencies for $\alpha\to2$
reflects again boundary effects, as will be explained in the next
section.

\subsection{Replenishing targets and boundary conditions}
\label{sec:bc}

\begin{figure} [h!]
\includegraphics[width=1\linewidth]{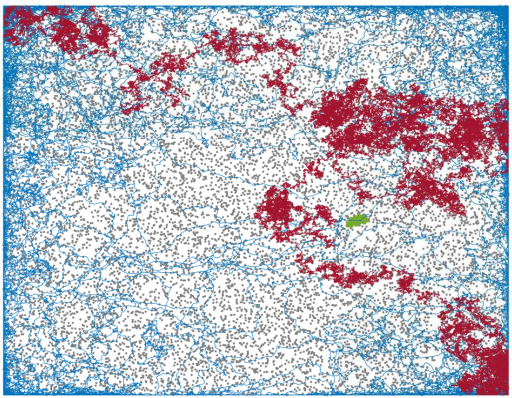}
\caption{Three examples of trajectories of a searcher moving according
  to fBm with exponents of anomalous diffusion $\alpha=0.49$ (green),
  $1.2$ (red) and $1.8$ (blue) in a medium containing uniformly
  distributed point targets. The jump length is $l_d=4$, the total
  iteration time $T=10^6$. All the other parameters are as explained
  in Sec.~\ref{sec:model}.}
\label{4}
\end{figure}

We now explain the two transitions between maxima and minima around
$\alpha=2$ in both $\eta_A^*$ and $\eta_P^*$ shown in Fig.~\ref{3} to
which we referred in passing above. In the strongly superdiffusive,
quasi-ballistic regime of $\alpha>1.5$, a searcher quickly approaches
the boundaries of the system. But since we have chosen reflective
walls, the searcher is thrown back into the bulk after hitting a
boundary. However, since for $\alpha>1.5$ fBm displays strong
persistence in the motion, the searcher will immediately move back to
the wall by getting reflected again, and so on. This microscopic
mechanism is illustrated in the inset of Fig.~\ref{3} (a). It leads
to the phenomenon of stickiness to the wall, see the blue trajectory
in Fig.~\ref{4}, recently investigated for fBm in
Refs.~\cite{gugg,VHSJGM20}. This is more widely known for active
Brownian particles that by definition of activity exhibit persistence
in their motion
\cite{duzgun2018active,das2015boundaries,elgeti2015run,kaiser2012capture,VoVo17,ZWS17,BeDiL16}. In
more detail, when hitting the wall one can decompose the velocity of a
searcher into a component parallel and one perpendicular to it. Since
the probability that a searcher hits the wall strictly perpendicular
to it is zero (with respect to Lebesgue measure in angular space),
there will always be a component of the velocity parallel to the
wall. Typically, an fBm searcher will thus for a long time move along
the wall by displaying zig-zag quasi one-dimensional quasi-ballistic
creeps. These are constrained to a boundary layer of an approximate
width $l_d$, which defines a crucial boundary length scale. But on top
of that, two further length scales come into play, governing the
search process in this boundary layer. The second one is the effective
(perception) radius $R_e=1$, which defines the relevant parameter for
first finding a target. If a searcher is moving in a boundary layer of
width $l_d$ in which there are randomly distributed (point) targets,
the efficiency for finding targets will be best if approximately
$l_d\le R_e$, where $R_e$ can alternatively be interpreted as
determining the average extension of a circular target,
cf.\ Sec.~\ref{sec:model}. We call this maximisation of efficiency the
pac-man effect, in analogy to an old computer game where a searcher
subsequently eats targets by moving along channels in a maze, as this
channeling helps to locate targets by increasing the search
efficiency; see again the blue colored regions at the boundaries in
Fig.~\ref{4}. However, once $l_d>R_e$, the $l_d$ boundary layer is
getting too wide compared with $R_e$, and the searcher starts to miss
targets. This explains the breakdown of the maximum at $\alpha=2$ from
$l_d=1$ to $l_d=1.24$ by a breakdown of the pac-man effect. Note that
for $l_d\le R_e$ the pac-man effect equally applies to arrival and
passage search, which elucidates why here the maxima are very similar
for both efficiencies (for $l_d=1$ the maxima look even
identical). But this mechanism does not clarify why the maxima when
$l_d\le1$ become minima for $1<l_d<4$.

However, there is a third length scale that plays a crucial role at
the boundary, which is the cut-off distance $d_c=2$ specific to
replenishing targets. If we now consider the motion approximately
perpendicular to the wall, a searcher must leave a target region of
approximately $2d_c=4$ before the target can replenish. But this
defines yet another boundary layer of respective width that a searcher
should leave as quickly as possible to benefit from the revisiting
target mechanism, here induced by anti-persistence due to the
reflecting boundaries. Indeed, a target will not replenish at all for a
searcher bouncing multiple times perpendicularly to the wall with jump
length $l_d<4$. We call this the bouncing fly effect, as this is
similar to a fly hitting a window many times, sometimes at almost the
same spot, by trying to escape. While this effect is always present
when $l_d<4$, for $l_d<1$, it seems to be dominated by the above pac-man
effect. However, as pac-man breaks down for $l_d>1$, our explanation of
the minimum at $\alpha=2$ for $1<l_d<4$ is that here a searcher moves
in a boundary layer that is deficient for revisiting target search
induced by anti-persistence due to the reflecting boundaries.

This effect minimising efficiencies breaks down again
when $l_d>4$, as then for the first time a searcher can jump over a
distance larger than the cut-off replenishing target region. This now
reactivates the revisiting target mechanism as a beneficient search
strategy since during a jump, a target replenishes and is available
again to be found. This explains why for $l_d=4$, we have again maximal
efficiencies at $\alpha=2$. In the transition from smaller to larger
$\alpha$ values at $l_d=4$, there even appears to be a slight minimum
around $\alpha\simeq1.7$ in $\eta_P^*$ while, conversely, the
corresponding efficiency for FP search in Fig.~\ref{1} (b) is
rather maximal in the same parameter region. This demonstrates again
the sensitivity of optimal search on the variation of both internal
and external parameters of the whole process.

Similarly, the minima in Fig.~\ref{6} for the $l_d=1$ curves around
$\alpha=2$ can be explained. Since in this case the targets are
non-replenishing, the boundary layer of width $l_d$ will become
depleted of targets, which is detrimental to repeated pac-man search
success by generating very small efficiencies. Minima for $l_d=2$
around $\alpha=2$ existed before already both in Fig.~\ref{3} and in
Fig.~\ref{6}, and there is no other mechanism in place that could
yield any larger efficiency here.

\begin{figure}  [h!]
\includegraphics[width=1\linewidth]{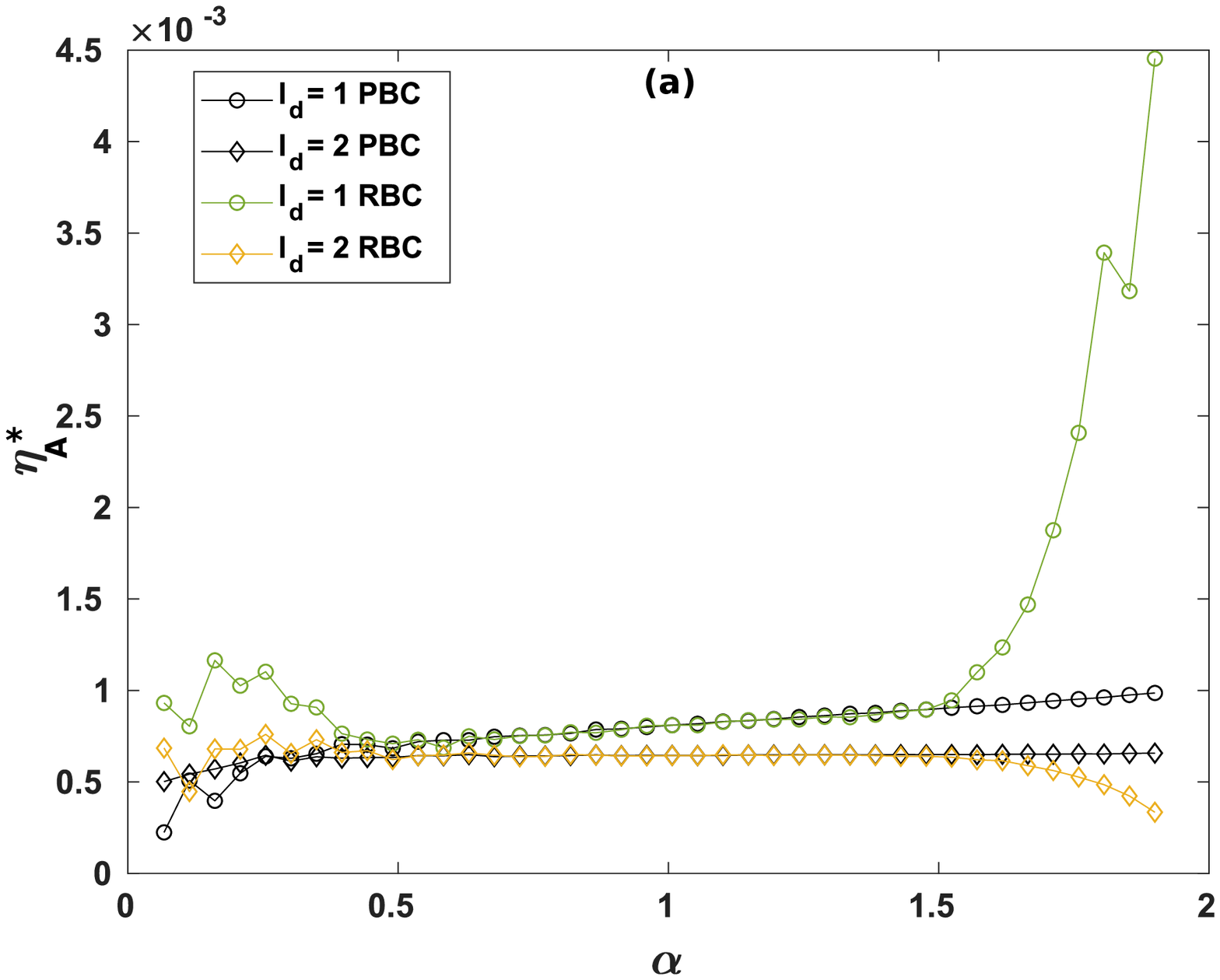}\quad\includegraphics[width=1\linewidth]{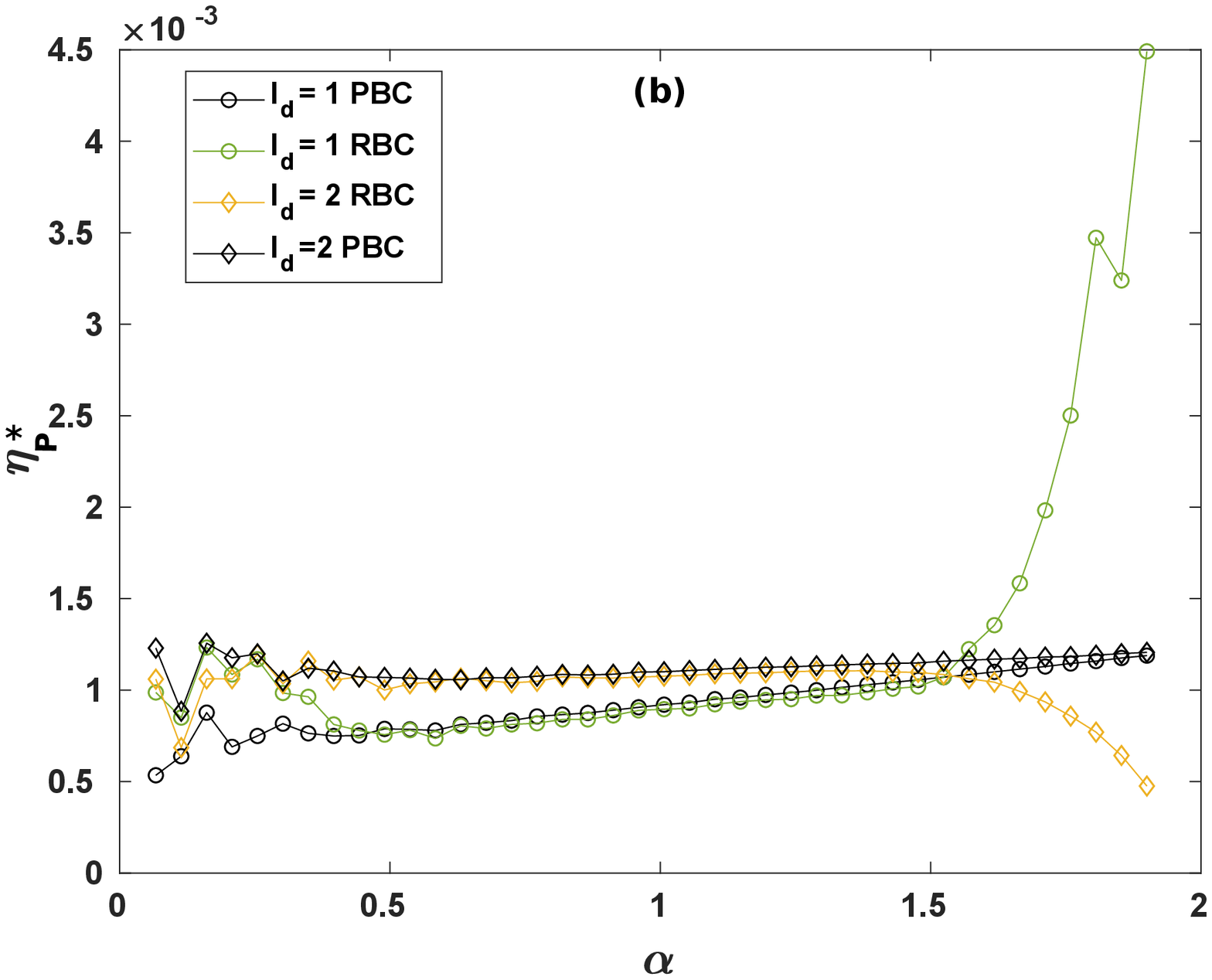}
\caption{Efficiencies $\eta_{A/P}^*$ defined by Eq.~(\ref{e3}) for
  saltatory (arrival at targets) (panel (a)) and cruise (passage
  through targets) (panel (b)) search of replenishing resources along
  the trajectory of a searcher during a given time $T$. Shown are
  results under variation of the exponent $\alpha$ of the mean square
  displacement Eq.~(\ref{fBmmsd}) for two different jump lengths $l_d$
  for both periodic (PBC; black symbols) and reflecting boundary
  conditions (RBC; colored symbols).}
\label{5}
 \end{figure}

Finally, to confirm the impact of the boundary conditions on the
efficiencies, Fig.~\ref{5} shows again $\eta_{A/P}^*$ for reflecting
boundaries, cf.\ $\eta_{A/P}^*$ at $l_d=1,2$ in Fig.~\ref{3}, in
comparison to the ones for periodic boundaries. One can see that the
periodic boundary conditions eliminate the maxima and minima at
$\alpha=2$ in all cases. This unambiguously demonstrates that all
these extrema are indeed due to boundary effects, as argued above. We
furthermore remark that while for passage search ballistic motion with
$\alpha=2$ now yields optimal efficiencies for replenishing targets in
Fig.~\ref{5} (b) (if we neglect the fluctuations at small
$\alpha$ values), this is not the case for arrival processes with
$l_d=2$. This might be due to leapovers that become stronger for
larger $\alpha$ values.

 \section{Summary and conclusions}
 \label{sec:concl}

\begin{figure*}
\includegraphics[width=1\linewidth]{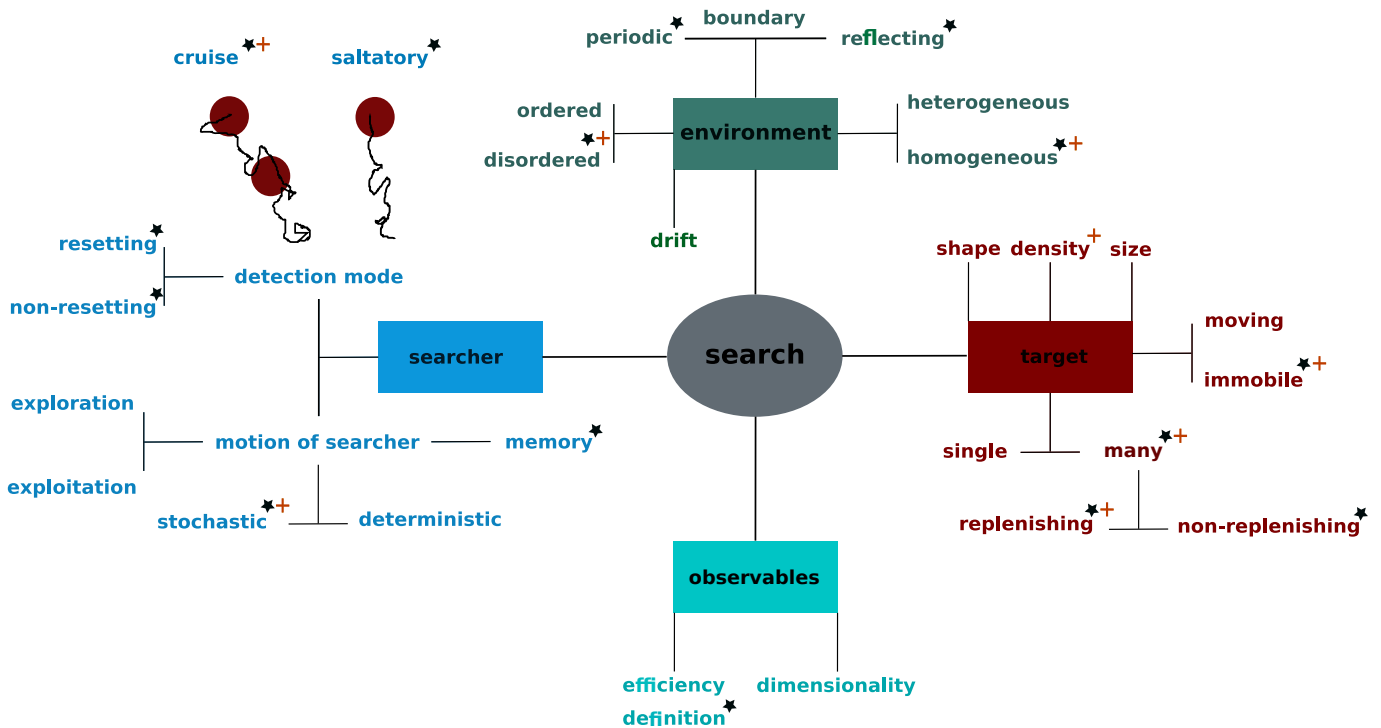}
\caption{Search is a complex process: This diagram categorises
  conditions on which search success depends. Some of them are
  intrinsic to the searcher by determining its modes of motion and
  target detection. Some of them are extrinsic to the searcher by
  characterising the search environment and the type of target.  Yet
  others are specific to the statistical analysis of the problem in
  terms of how to assess chosen observables. Black stars mark
  properties investigated in this article for fractional Brownian
  motion. Red crosses indicate the search scenario on which the
  L\'evy Flight Foraging Hypothesis is based.}
\label{fig:search}
 \end{figure*}

In this work, we studied the efficiency of search generated by fBm in
a random field of targets. In more general terms we explored the
sensitivity of search succes on the specific setting and the
parameters defining the search process, the environment and the
interaction of both with each other. That way our essentially
computational study suggests a conceptual framework that should apply
to {\em any} theoretical description of a respective search problem,
irrespective of whether one considers fBm, LWs or other types of
motion. Figure~\ref{fig:search} identifies important ingredients on
which such a generic search setting depends. They can be broadly
classified as intrinsic to a searcher by characterising its dynamics,
or extrinsic to it by defining the search environment
\cite{Nath08,MCB14}. The results also depend on the quantity by which
search success is assessed in terms of statistical analysis
\cite{PCM14}. For fBm we have investigated all of the conditions
marked by (black) stars, as we will briefly summarise below by going
through this figure. We emphasize that the picture put forward in
Fig.~\ref{fig:search} provides only a first sketch, which invites to
be amended in future research.

The given properties of the searcher (light blue box to the left and
associated tree structure in Fig.~\ref{fig:search}) define a key
aspect of search. For search dynamics one needs to distinguish between
the motion of the searcher and its associated modes of target
detection.  In our case, the motion was stochastic and governed by the
two intrinsic parameters defining fBm, the jump length $l_d$
Eq.~(\ref{eq:jl}) and the exponent $\alpha$ of the mean square
displacement Eq.~(\ref{fBmmsd}). The latter determines in turn the
memory of the process via the position autocorrelation function decay
Eq.~(\ref{fBmpp}). In case of intermittent search, one has a
separation of the search dynamics into local exploitation and
long-range exploration \cite{BaCa16,CBRM15}. Concerning target
detection, one distinguishes between cruise searchers that perceive a
target while moving and saltatory foragers that only find targets
after landing within the perception radius $R_e$ next to it
\cite{JPP09}. The former relate mathematically to (first) passage
problems of finding a target while the latter are (first) arrival, or
(first) hitting problems \cite{PBL19}. Yet another distinction is
whether one looks at the problem of finding only one target, which we
modelled by a resetting procedure of the searcher to a random initial
position, or many of them along a single trajectory, which we denoted
as non-resetting after visiting a target.

These properties in turn determine how a search process is assessed by
evaluating observables (turquoise box at the bottom in
Fig.~\ref{fig:search}). Here we defined two generically different
efficiencies depending on the statistical averaging applied, see
Eqs.~(\ref{e1}) \cite{Vis99}, respectively (\ref{e3})
\cite{BaLe08}. The averaging was even obtained in a third way leading
to yet another type of efficiency, see Eq.~(\ref{e2}) \cite{PCM14}.
We found that for non-Markovian processes with memory these three
definitions yield very different values. Recent results showed that
for LWs the dimensionality of the search process plays a crucial role
for determining the values of efficiencies \cite{LTBV20}, which
however we did not explore in this work.

The search environment (green-grey box at the top) we modelled by a
disordered, i.e., uniformly random distribution of homogeneous targets
in the plane. Computer simulations are constrained to a finite area or
volume determined by the length of the simulation box $L$ as an
extrinsic parameter. This quantity becomes a non-trivial parameter
when an fBm searcher interacts with the walls of the system depending
on the boundary conditions \cite{gugg}. Here we considered primarily
reflecting boundaries but compared some of our results to the
situation of periodic ones. In some works, the impact of drift on
(L\'evy flight) search has been investigated \cite{PCM14,PCM14b}.

As targets (red-brown box to the right) we chose disks of radius
$R_e$, see Sec.~\ref{sec:rdt}, with a density measured by the mean
target distance $l_f$ Eq.~(\ref{eq:td}), which yields another two
extrinsic parameters. As explained before, $R_e$ can be reinterpreted
as the radius of perception of a searcher for finding point targets,
hence it is ambiguous to categorise it as an intrinsic or extrinsic
parameter. The disks were immobile, and we had many of them. One also
needs to choose the type of resource \cite{Vis99}: A target may be
non-destructive in the sense that it replenishes after having been
found, or it is destroyed upon finding it and does not replenish. In
the former case, one needs to introduce another extrinsic parameter,
which in our case was the cut-off distance $d_c$ the searcher has to
be away from the target after having found it, in order for it to
replenish \cite{LTBV20}, see Sec.~\ref{sec:ss}.

We performed computer simulations to study the dependence of our three
different search efficiencies on the above two intrinsic parameters by
keeping the three extrinsic ones fixed. We also tested the impact of
the boundary conditions. Our main results are summarised as follows:

\begin{enumerate}

  \item For FA and FP search in a replenishing field of targets, we
    observe the existence of maxima in both search efficiencies for
    intermediate $\alpha$ values, $0.5<\alpha<1.7$, see
    Fig.~\ref{1}. The maxima are especially sharp in this regime if
    $R_e\simeq l_d\ll l_f$, and for FP with $R_e<l_d<l_f$ they
    accumulate around $\alpha\simeq1.5$. The latter seems to reflect
    an optimal sampling of the target space in-between over- and
    undersampling.

    Similar results have been reported for FP search by LWs leading to
    the LFFH, i.e., that search strategies right between ballistic
    dynamics and normal diffusion are optimal to find sparse,
    replenishing targets in bounded domains \cite{Vis99}. The specific
    search scenario yielding the LFFH is marked by red stars in
    Fig.~\ref{fig:search}. It is thus recovered as a special case in
    our over-arching conceptual framework.
    
  \item For FA search, we encounter a `paradise regime' when the jump
    length starts to exceed the mean target distance, $R_e\ll
    l_d\simeq l_f$. In this case, $l_d$ is large enough for the
    searcher to always find a target. This is essentially independent
    of the persistence in fBm if $\alpha$ is not too large or too
    small. In Fig.~\ref{1} (a), this is represented by the efficiency
    curve flattening out, which implies that targets are found with
    maximal efficiency over a wide range of $\alpha$ values,
    $0.5<\alpha<1.5$.

  \item In contrast to FP problems, FA processes exhibit leapovers
    when $l_d\ge R_e$, that is, a searcher can jump over a target
    without finding it \cite{PCM14}. This mechanism diminuishes search
    efficiencies, as is clearly seen in Fig.~\ref{2} (b) for FP
    compared to FA, and also in Fig.~\ref{6} by comparing the
    efficiencies for arrival (saltatory) search to the ones for
    passage (cruise) search under variation of $l_d$.

  \item Optimising search by maximising efficiencies depends very much
    on the definition of efficiency that one chooses. This is
    demonstrated in Figs.~\ref{1}, \ref{3} and \ref{6}, which display
    results for the three different efficiency definitions
    Eqs.~(\ref{e1}), (\ref{e2}) and (\ref{e3}), respectively. One can
    see that these three different efficiencies yield totally
    different results.
        
  \item We put forward a very simple analytical argument, which may be
    considered as a boiled-down version of the theory in
    Ref.~\cite{guerin}, that analytically reproduced the functional
    forms of the FP efficiencies under variation of $l_d$ and
    $\alpha$, respectively, see Fig.~\ref{224}.

  \item Subdiffusion can optimise search efficiencies for multi-target
    search along a trajectory of both arrival and passage processes in
    an area with reflecting boundaries when $R_e< l_d \ll l_f$, see
    Fig.~\ref{3}.  This is due to a revisiting target mechanism, which
    in turn is generated by the anti-persistence in fBm for $\alpha<1$
    \cite{GuWe08}. The mechanism also holds for arrival processes
    under periodic boundary conditions in this regime of $l_d$ as
    shown in Fig.~\ref{5} (a).

  \item Reflecting boundaries can generate very intricate memory
    effects in search governed by fBm \cite{gugg}. This is represented
    by multiple transitions between maxima and minima in the
    efficiencies of multi-target search along a path for both arrival
    and passage processes, cf.\ again Fig.~\ref{3} for
    $\alpha>1.5$. We explained these variations microscopically by
    what we called pac-man and bouncing fly effects. There is a
    cross-link between these effects and the well-known stickiness of
    active particles to walls due to self-driven persistent motion
    \cite{BeDiL16}.
  
\end{enumerate}

We conclude that a seemingly simple problem of stochastic search in a
random distribution of targets delivered highly non-trivial numerical
results. Search turned out to be an extremely sensitive process,
exhibiting all signatures of a complex system, where the whole is more
than the sum of its single parts: Testing different search scenarios
by differently combining search process, environment and their
interaction with each other yielded entirely different results for
respective search efficiencies, as is reflected in a high sensitivity
of all results on variation of model parameters. A somewhat related
sensitivity was already observed in foraging experiments
\cite{LICCK12,VOCGTNA21}. Our simulation results are thus in sharp
contrast to claims of robust, universal optimal search strategies
suggested by the LFFH and reported verifications of it by experiments
conducted in the wild \cite{VLRS11}. They are, however, fully in line
with recent work limiting the range of validity of the LFFH
\cite{LTBV20,LTBV21} by re-evaluating the theoretical model
underpinning it. While here we did not investigate LWs, for the LW
search that led to the LFFH the general framework is the same as the
one summarised in Fig.~\ref{fig:search}. The specific search situation
that applies to the LFFH does have its place in this figure, but it
only defines a small subset in it, as marked by the (red) crosses. We
believe that this gives proper credit to the LFFH by adequately
embedding it into a more general framework. For a similar search
setting, we have found that maximisations of search efficiencies for
fBm are not robust under variation of the search conditions and are
thus not universally valid. In Refs.~\cite{LTBV20,LTBV21} it was shown
that the same holds to quite some extent for LWs.

To our knowledge, this is the first work where it has been explored
how search by fBm depends on the variation of a large number of
important conditions defining a specific search scenario. Our main
results summarised above, in combination with Fig.~\ref{fig:search},
reveal the need to investigate further search scenarios on a case by
case basis. The most straightforward extension would be to study
search under variation of the three external parameters of our model,
which deliberately we kept fixed, i.e., the target size $R_e$, the
density of targets governed by $l_f$, and the cut-off distance for the
replenishing of targets $d_c$. While we believe that we have
identified generic boundary effects in our work, checking for their
robustness under parameter variation would be interesting. More
complicated problems would be search for moving targets, as studied to
some extent already in Ref.~\cite{AG03}. Biologically, one might be
interested in search restricted to the home range of a forager
\cite{SHG19}. And investigating swarms of searchers is important for
robotic applications \cite{DOT16}. We furthermore remark that
stochastic dynamics with persistence became very prominent as models
for active Brownian particles reproducing self-propelled biological
motion \cite{FNCTVW16,USJ19}. Within the latter context, it might be
interesting to study biologically relevant search problems in more
detail \cite{KWPES16,ZWS17}. Finally, a more comprehensive analytical
description of fBm search \cite{guerin,LBGV18} going beyond our simple
handwaving argument would be highly desirable. This theory should
explain the intricate dependence of search efficiencies on the
variation of model parameters and other settings as observed in our
simulations.\\

{\bf Acknowledgements:} This work was funded by the Deutsche
Forschungsgemeinschaft (DFG, German Research Foundation) —
Projektnummer~163436311 — SFB~910. RK thanks the SFB for a Mercator
Visiting Professorship, which initiated this project and during which
most parts of it have been performed. He is also grateful for funding
from the Office of Naval Research Global and acknowledges an External
Fellowship from the London Mathematical Laboratory. All three authors
thank two anonymous referees for their very helpful comments that led
to a profoundly improved version of their manuscript.


\begin{thebibliography}{110}%
\makeatletter
\providecommand \@ifxundefined [1]{%
 \@ifx{#1\undefined}
}%
\providecommand \@ifnum [1]{%
 \ifnum #1\expandafter \@firstoftwo
 \else \expandafter \@secondoftwo
 \fi
}%
\providecommand \@ifx [1]{%
 \ifx #1\expandafter \@firstoftwo
 \else \expandafter \@secondoftwo
 \fi
}%
\providecommand \natexlab [1]{#1}%
\providecommand \enquote  [1]{``#1''}%
\providecommand \bibnamefont  [1]{#1}%
\providecommand \bibfnamefont [1]{#1}%
\providecommand \citenamefont [1]{#1}%
\providecommand \href@noop [0]{\@secondoftwo}%
\providecommand \href [0]{\begingroup \@sanitize@url \@href}%
\providecommand \@href[1]{\@@startlink{#1}\@@href}%
\providecommand \@@href[1]{\endgroup#1\@@endlink}%
\providecommand \@sanitize@url [0]{\catcode `\\12\catcode `\$12\catcode
  `\&12\catcode `\#12\catcode `\^12\catcode `\_12\catcode `\%12\relax}%
\providecommand \@@startlink[1]{}%
\providecommand \@@endlink[0]{}%
\providecommand \url  [0]{\begingroup\@sanitize@url \@url }%
\providecommand \@url [1]{\endgroup\@href {#1}{\urlprefix }}%
\providecommand \urlprefix  [0]{URL }%
\providecommand \Eprint [0]{\href }%
\providecommand \doibase [0]{http://dx.doi.org/}%
\providecommand \selectlanguage [0]{\@gobble}%
\providecommand \bibinfo  [0]{\@secondoftwo}%
\providecommand \bibfield  [0]{\@secondoftwo}%
\providecommand \translation [1]{[#1]}%
\providecommand \BibitemOpen [0]{}%
\providecommand \bibitemStop [0]{}%
\providecommand \bibitemNoStop [0]{.\EOS\space}%
\providecommand \EOS [0]{\spacefactor3000\relax}%
\providecommand \BibitemShut  [1]{\csname bibitem#1\endcsname}%
\let\auto@bib@innerbib\@empty
\bibitem [{\citenamefont {Chupeau}\ \emph {et~al.}(2015)\citenamefont
  {Chupeau}, \citenamefont {B{\'e}nichou},\ and\ \citenamefont
  {Voituriez}}]{CBV15}%
  \BibitemOpen
  \bibfield  {author} {\bibinfo {author} {\bibfnamefont {M.}~\bibnamefont
  {Chupeau}}, \bibinfo {author} {\bibfnamefont {O.}~\bibnamefont
  {B{\'e}nichou}}, \ and\ \bibinfo {author} {\bibfnamefont {R.}~\bibnamefont
  {Voituriez}},\ }\bibfield  {title} {\enquote {\bibinfo {title} {Cover times
  of random searches},}\ }\href@noop {} {\bibfield  {journal} {\bibinfo
  {journal} {Nat. Phys.}\ }\textbf {\bibinfo {volume} {11}},\ \bibinfo {pages}
  {844} (\bibinfo {year} {2015})}\BibitemShut {NoStop}%
\bibitem [{\citenamefont {B{\'e}nichou}\ \emph {et~al.}(2011)\citenamefont
  {B{\'e}nichou}, \citenamefont {Loverdo}, \citenamefont {Moreau},\ and\
  \citenamefont {Voituriez}}]{BLMV11}%
  \BibitemOpen
  \bibfield  {author} {\bibinfo {author} {\bibfnamefont {O.}~\bibnamefont
  {B{\'e}nichou}}, \bibinfo {author} {\bibfnamefont {C.}~\bibnamefont
  {Loverdo}}, \bibinfo {author} {\bibfnamefont {M.}~\bibnamefont {Moreau}}, \
  and\ \bibinfo {author} {\bibfnamefont {R.}~\bibnamefont {Voituriez}},\
  }\bibfield  {title} {\enquote {\bibinfo {title} {Intermittent search
  strategies},}\ }\href@noop {} {\bibfield  {journal} {\bibinfo  {journal}
  {Rev. Mod. Phys.}\ }\textbf {\bibinfo {volume} {83}},\ \bibinfo {pages} {81}
  (\bibinfo {year} {2011})}\BibitemShut {NoStop}%
\bibitem [{\citenamefont {Raichlen}\ \emph {et~al.}(2014)\citenamefont
  {Raichlen}, \citenamefont {Wood}, \citenamefont {Gordon}, \citenamefont
  {Mabulla}, \citenamefont {Marlowe},\ and\ \citenamefont
  {Pontzer}}]{RWGMMP14}%
  \BibitemOpen
  \bibfield  {author} {\bibinfo {author} {\bibfnamefont {D.~A.}\ \bibnamefont
  {Raichlen}}, \bibinfo {author} {\bibfnamefont {B.~M.}\ \bibnamefont {Wood}},
  \bibinfo {author} {\bibfnamefont {A.~D.}\ \bibnamefont {Gordon}}, \bibinfo
  {author} {\bibfnamefont {A.~Z.~P.}\ \bibnamefont {Mabulla}}, \bibinfo
  {author} {\bibfnamefont {F.~W.}\ \bibnamefont {Marlowe}}, \ and\ \bibinfo
  {author} {\bibfnamefont {H.}~\bibnamefont {Pontzer}},\ }\bibfield  {title}
  {\enquote {\bibinfo {title} {Evidence of {L}{\'e}vy walk foraging patterns in
  human hunter{\textendash}gatherers},}\ }\href@noop {} {\bibfield  {journal}
  {\bibinfo  {journal} {Proc. Natl. Acad. Sci.}\ }\textbf {\bibinfo {volume}
  {111}},\ \bibinfo {pages} {728} (\bibinfo {year} {2014})}\BibitemShut
  {NoStop}%
\bibitem [{\citenamefont {Alpern}\ and\ \citenamefont {Gal}(2003)}]{AG03}%
  \BibitemOpen
  \bibfield  {author} {\bibinfo {author} {\bibfnamefont {S.}~\bibnamefont
  {Alpern}}\ and\ \bibinfo {author} {\bibfnamefont {S.}~\bibnamefont {Gal}},\
  }\href@noop {} {\emph {\bibinfo {title} {The theory of search games and
  rendezvous}}},\ International Series in Operations Research and Managment
  Science\ (\bibinfo  {publisher} {Kluwer Academic Publishers},\ \bibinfo
  {address} {Boston},\ \bibinfo {year} {2003})\BibitemShut {NoStop}%
\bibitem [{\citenamefont {Shlesinger}(2006)}]{Shles06}%
  \BibitemOpen
  \bibfield  {author} {\bibinfo {author} {\bibfnamefont {M.~F.}\ \bibnamefont
  {Shlesinger}},\ }\bibfield  {title} {\enquote {\bibinfo {title} {Search
  research},}\ }\href@noop {} {\bibfield  {journal} {\bibinfo  {journal}
  {Nature}\ }\textbf {\bibinfo {volume} {443}},\ \bibinfo {pages} {281}
  (\bibinfo {year} {2006})}\BibitemShut {NoStop}%
\bibitem [{\citenamefont {Stone}(2007)}]{Stone07}%
  \BibitemOpen
  \bibfield  {author} {\bibinfo {author} {\bibfnamefont {L.~A.}\ \bibnamefont
  {Stone}},\ }\href@noop {} {\emph {\bibinfo {title} {Theory of Optimal
  Search}}},\ \bibinfo {edition} {2nd}\ ed.\ (\bibinfo  {publisher} {Informs},\
  \bibinfo {address} {Hanover, MD},\ \bibinfo {year} {2007})\BibitemShut
  {NoStop}%
\bibitem [{\citenamefont {Nathan}\ \emph {et~al.}(2008)\citenamefont {Nathan},
  \citenamefont {Getz}, \citenamefont {Revilla}, \citenamefont {Holyoak},
  \citenamefont {Kadmon}, \citenamefont {Saltz},\ and\ \citenamefont
  {Smouse}}]{Nath08}%
  \BibitemOpen
  \bibfield  {author} {\bibinfo {author} {\bibfnamefont {R.}~\bibnamefont
  {Nathan}}, \bibinfo {author} {\bibfnamefont {W.~M.}\ \bibnamefont {Getz}},
  \bibinfo {author} {\bibfnamefont {E.}~\bibnamefont {Revilla}}, \bibinfo
  {author} {\bibfnamefont {M.}~\bibnamefont {Holyoak}}, \bibinfo {author}
  {\bibfnamefont {R.}~\bibnamefont {Kadmon}}, \bibinfo {author} {\bibfnamefont
  {D.}~\bibnamefont {Saltz}}, \ and\ \bibinfo {author} {\bibfnamefont {P.~E.}\
  \bibnamefont {Smouse}},\ }\bibfield  {title} {\enquote {\bibinfo {title} {A
  movement ecology paradigm for unifying organismal movement research},}\
  }\href@noop {} {\bibfield  {journal} {\bibinfo  {journal} {Proc. Natl. Acad.
  Sci.}\ }\textbf {\bibinfo {volume} {105}},\ \bibinfo {pages} {19052}
  (\bibinfo {year} {2008})}\BibitemShut {NoStop}%
\bibitem [{\citenamefont {M{\'e}ndez}\ \emph {et~al.}(2014)\citenamefont
  {M{\'e}ndez}, \citenamefont {Campos},\ and\ \citenamefont
  {Bartumeus}}]{MCB14}%
  \BibitemOpen
  \bibfield  {author} {\bibinfo {author} {\bibfnamefont {V.}~\bibnamefont
  {M{\'e}ndez}}, \bibinfo {author} {\bibfnamefont {D.}~\bibnamefont {Campos}},
  \ and\ \bibinfo {author} {\bibfnamefont {F.}~\bibnamefont {Bartumeus}},\
  }\href@noop {} {\emph {\bibinfo {title} {Stochastic Foundations in Movement
  Ecology}}}\ (\bibinfo  {publisher} {Springer},\ \bibinfo {address} {Berlin},\
  \bibinfo {year} {2014})\BibitemShut {NoStop}%
\bibitem [{\citenamefont {Klages}(2016)}]{Kla16}%
  \BibitemOpen
  \bibfield  {author} {\bibinfo {author} {\bibfnamefont {R.}~\bibnamefont
  {Klages}},\ }\href@noop {} {\enquote {\bibinfo {title} {Search for food of
  birds, fish and insects},}\ } (\bibinfo {year} {2016}),\ \bibinfo {note}
  {book chapter in: A. Bunde, J. Caro, J.Kaerger, G. Vogl (Eds.), Diffusive
  Spreading in Nature, Technology and Society (Springer, Berlin)}\BibitemShut
  {NoStop}%
\bibitem [{\citenamefont {Viswanathan}\ \emph {et~al.}(2011)\citenamefont
  {Viswanathan}, \citenamefont {{Da Luz}}, \citenamefont {Raposo},\ and\
  \citenamefont {Stanley}}]{VLRS11}%
  \BibitemOpen
  \bibfield  {author} {\bibinfo {author} {\bibfnamefont {G.~M.}\ \bibnamefont
  {Viswanathan}}, \bibinfo {author} {\bibfnamefont {M.~G.~E.}\ \bibnamefont
  {{Da Luz}}}, \bibinfo {author} {\bibfnamefont {E.~P.}\ \bibnamefont
  {Raposo}}, \ and\ \bibinfo {author} {\bibfnamefont {H.~E.}\ \bibnamefont
  {Stanley}},\ }\href@noop {} {\emph {\bibinfo {title} {The Physics of
  Foraging}}}\ (\bibinfo  {publisher} {Cambridge University Press},\ \bibinfo
  {address} {Cambridge},\ \bibinfo {year} {2011})\BibitemShut {NoStop}%
\bibitem [{\citenamefont {Pearson}(1906)}]{Pea06}%
  \BibitemOpen
  \bibfield  {author} {\bibinfo {author} {\bibfnamefont {K.}~\bibnamefont
  {Pearson}},\ }\bibfield  {title} {\enquote {\bibinfo {title} {Mathematical
  contributions to the theory of evolution - a mathematical theory of random
  migration},}\ }\href@noop {} {\bibfield  {journal} {\bibinfo  {journal}
  {Biometric ser.}\ }\textbf {\bibinfo {volume} {3}},\ \bibinfo {pages} {54}
  (\bibinfo {year} {1906})}\BibitemShut {NoStop}%
\bibitem [{\citenamefont {Viswanathan}\ \emph {et~al.}(1996)\citenamefont
  {Viswanathan}, \citenamefont {Afanasyev}, \citenamefont {Buldyrev},
  \citenamefont {Murphy}, \citenamefont {Prince},\ and\ \citenamefont
  {Stanley}}]{Vis96}%
  \BibitemOpen
  \bibfield  {author} {\bibinfo {author} {\bibfnamefont {G.~M.}\ \bibnamefont
  {Viswanathan}}, \bibinfo {author} {\bibfnamefont {V.}~\bibnamefont
  {Afanasyev}}, \bibinfo {author} {\bibfnamefont {S.~V.}\ \bibnamefont
  {Buldyrev}}, \bibinfo {author} {\bibfnamefont {E.~J.}\ \bibnamefont
  {Murphy}}, \bibinfo {author} {\bibfnamefont {P.~A.}\ \bibnamefont {Prince}},
  \ and\ \bibinfo {author} {\bibfnamefont {H.~E.}\ \bibnamefont {Stanley}},\
  }\bibfield  {title} {\enquote {\bibinfo {title} {{L}{\'e}vy flight search
  patterns of wandering albatrosses},}\ }\href@noop {} {\bibfield  {journal}
  {\bibinfo  {journal} {Nature}\ }\textbf {\bibinfo {volume} {381}},\ \bibinfo
  {pages} {413} (\bibinfo {year} {1996})}\BibitemShut {NoStop}%
\bibitem [{\citenamefont {Viswanathan}\ \emph {et~al.}(1999)\citenamefont
  {Viswanathan}, \citenamefont {Buldyrev}, \citenamefont {Havlin},
  \citenamefont {Luz}, \citenamefont {Raposo},\ and\ \citenamefont
  {Stanley}}]{Vis99}%
  \BibitemOpen
  \bibfield  {author} {\bibinfo {author} {\bibfnamefont {G.~M.}\ \bibnamefont
  {Viswanathan}}, \bibinfo {author} {\bibfnamefont {S.~V.}\ \bibnamefont
  {Buldyrev}}, \bibinfo {author} {\bibfnamefont {S.}~\bibnamefont {Havlin}},
  \bibinfo {author} {\bibfnamefont {M.~G. E.~Da}\ \bibnamefont {Luz}}, \bibinfo
  {author} {\bibfnamefont {E.~P.}\ \bibnamefont {Raposo}}, \ and\ \bibinfo
  {author} {\bibfnamefont {H.~E.}\ \bibnamefont {Stanley}},\ }\bibfield
  {title} {\enquote {\bibinfo {title} {Optimizing the success of random
  searches},}\ }\href@noop {} {\bibfield  {journal} {\bibinfo  {journal}
  {Nature}\ }\textbf {\bibinfo {volume} {401}},\ \bibinfo {pages} {911}
  (\bibinfo {year} {1999})}\BibitemShut {NoStop}%
\bibitem [{\citenamefont {Edwards}\ \emph {et~al.}(2007)\citenamefont
  {Edwards}, \citenamefont {Phillips}, \citenamefont {Watkins}, \citenamefont
  {Freeman}, \citenamefont {Murphy}, \citenamefont {Afanasyev}, \citenamefont
  {Buldyrev}, \citenamefont {Luz}, \citenamefont {Raposo}, \citenamefont
  {Stanley},\ and\ \citenamefont {Viswanathan}}]{Edw07}%
  \BibitemOpen
  \bibfield  {author} {\bibinfo {author} {\bibfnamefont {A.~M.}\ \bibnamefont
  {Edwards}}, \bibinfo {author} {\bibfnamefont {R.~A.}\ \bibnamefont
  {Phillips}}, \bibinfo {author} {\bibfnamefont {N.~W.}\ \bibnamefont
  {Watkins}}, \bibinfo {author} {\bibfnamefont {M.~P.}\ \bibnamefont
  {Freeman}}, \bibinfo {author} {\bibfnamefont {E.~J.}\ \bibnamefont {Murphy}},
  \bibinfo {author} {\bibfnamefont {V.}~\bibnamefont {Afanasyev}}, \bibinfo
  {author} {\bibfnamefont {S.~V.}\ \bibnamefont {Buldyrev}}, \bibinfo {author}
  {\bibfnamefont {M.~G. E.~Da}\ \bibnamefont {Luz}}, \bibinfo {author}
  {\bibfnamefont {E.~P.}\ \bibnamefont {Raposo}}, \bibinfo {author}
  {\bibfnamefont {H.~E.}\ \bibnamefont {Stanley}}, \ and\ \bibinfo {author}
  {\bibfnamefont {G.~M.}\ \bibnamefont {Viswanathan}},\ }\bibfield  {title}
  {\enquote {\bibinfo {title} {Revisiting {L}{\'e}vy flight search patterns of
  wandering albatrosses, bumblebees and deer},}\ }\href@noop {} {\bibfield
  {journal} {\bibinfo  {journal} {Nature}\ }\textbf {\bibinfo {volume} {449}},\
  \bibinfo {pages} {1044} (\bibinfo {year} {2007})}\BibitemShut {NoStop}%
\bibitem [{\citenamefont {Metzler}\ and\ \citenamefont
  {Klafter}(2000)}]{MeKl00}%
  \BibitemOpen
  \bibfield  {author} {\bibinfo {author} {\bibfnamefont {R.}~\bibnamefont
  {Metzler}}\ and\ \bibinfo {author} {\bibfnamefont {J.}~\bibnamefont
  {Klafter}},\ }\bibfield  {title} {\enquote {\bibinfo {title} {The random
  walk's guide to anomalous diffusion: A fractional dynamics approach},}\
  }\href@noop {} {\bibfield  {journal} {\bibinfo  {journal} {Phys. Rep.}\
  }\textbf {\bibinfo {volume} {339}},\ \bibinfo {pages} {1} (\bibinfo {year}
  {2000})}\BibitemShut {NoStop}%
\bibitem [{\citenamefont {Coffey}\ \emph {et~al.}(2004)\citenamefont {Coffey},
  \citenamefont {Kalmykov},\ and\ \citenamefont {Waldron}}]{CKW04}%
  \BibitemOpen
  \bibfield  {author} {\bibinfo {author} {\bibfnamefont {W.}~\bibnamefont
  {Coffey}}, \bibinfo {author} {\bibfnamefont {Y.~P.}\ \bibnamefont
  {Kalmykov}}, \ and\ \bibinfo {author} {\bibfnamefont {J.~T.}\ \bibnamefont
  {Waldron}},\ }\href@noop {} {\emph {\bibinfo {title} {The Langevin
  Equation}}}\ (\bibinfo  {publisher} {World Scientific},\ \bibinfo {address}
  {Singapore},\ \bibinfo {year} {2004})\BibitemShut {NoStop}%
\bibitem [{\citenamefont {Klages}\ \emph {et~al.}(2008)\citenamefont {Klages},
  \citenamefont {Radons},\ and\ \citenamefont {Sokolov}}]{KRS08}%
  \BibitemOpen
  \bibinfo {editor} {\bibfnamefont {R.}~\bibnamefont {Klages}}, \bibinfo
  {editor} {\bibfnamefont {G.}~\bibnamefont {Radons}}, \ and\ \bibinfo {editor}
  {\bibfnamefont {I.~M.}\ \bibnamefont {Sokolov}},\ eds.,\ \href@noop {} {\emph
  {\bibinfo {title} {Anomalous transport: Foundations and Applications}}}\
  (\bibinfo  {publisher} {Wiley-VCH},\ \bibinfo {address} {Berlin},\ \bibinfo
  {year} {2008})\BibitemShut {NoStop}%
\bibitem [{\citenamefont {Sokolov}(2012)}]{sokol}%
  \BibitemOpen
  \bibfield  {author} {\bibinfo {author} {\bibfnamefont {I.~M.}\ \bibnamefont
  {Sokolov}},\ }\bibfield  {title} {\enquote {\bibinfo {title} {Models of
  anomalous diffusion in crowded environments},}\ }\href@noop {} {\bibfield
  {journal} {\bibinfo  {journal} {Soft Matter}\ }\textbf {\bibinfo {volume}
  {8}},\ \bibinfo {pages} {9043} (\bibinfo {year} {2012})}\BibitemShut
  {NoStop}%
\bibitem [{\citenamefont {Metzler}\ \emph
  {et~al.}(2014{\natexlab{a}})\citenamefont {Metzler}, \citenamefont {Jeon},
  \citenamefont {Cherstvy},\ and\ \citenamefont {Barkai}}]{MJJCB14}%
  \BibitemOpen
  \bibfield  {author} {\bibinfo {author} {\bibfnamefont {R.}~\bibnamefont
  {Metzler}}, \bibinfo {author} {\bibfnamefont {J.-H.}\ \bibnamefont {Jeon}},
  \bibinfo {author} {\bibfnamefont {A.~G.}\ \bibnamefont {Cherstvy}}, \ and\
  \bibinfo {author} {\bibfnamefont {E.}~\bibnamefont {Barkai}},\ }\bibfield
  {title} {\enquote {\bibinfo {title} {Anomalous diffusion models and their
  properties: non-stationarity{,} non-ergodicity{,} and ageing at the centenary
  of single particle tracking},}\ }\href@noop {} {\bibfield  {journal}
  {\bibinfo  {journal} {Phys. Chem. Chem. Phys.}\ }\textbf {\bibinfo {volume}
  {16}},\ \bibinfo {pages} {24128} (\bibinfo {year}
  {2014}{\natexlab{a}})}\BibitemShut {NoStop}%
\bibitem [{\citenamefont {Bartumeus}\ \emph {et~al.}(2016)\citenamefont
  {Bartumeus}, \citenamefont {Campos}, \citenamefont {Ryu}, \citenamefont
  {Lloret-Cabot}, \citenamefont {Méndez},\ and\ \citenamefont
  {Catalan}}]{BaCa16}%
  \BibitemOpen
  \bibfield  {author} {\bibinfo {author} {\bibfnamefont {F.}~\bibnamefont
  {Bartumeus}}, \bibinfo {author} {\bibfnamefont {D.}~\bibnamefont {Campos}},
  \bibinfo {author} {\bibfnamefont {W.~S.}\ \bibnamefont {Ryu}}, \bibinfo
  {author} {\bibfnamefont {R.}~\bibnamefont {Lloret-Cabot}}, \bibinfo {author}
  {\bibfnamefont {V.}~\bibnamefont {Méndez}}, \ and\ \bibinfo {author}
  {\bibfnamefont {J.}~\bibnamefont {Catalan}},\ }\bibfield  {title} {\enquote
  {\bibinfo {title} {Foraging success under uncertainty: search tradeoffs and
  optimal space use},}\ }\href@noop {} {\bibfield  {journal} {\bibinfo
  {journal} {Ecol. Lett.}\ }\textbf {\bibinfo {volume} {19}},\ \bibinfo {pages}
  {1299.} (\bibinfo {year} {2016})}\BibitemShut {NoStop}%
\bibitem [{\citenamefont {Mandelbrot}\ and\ \citenamefont
  {Van~Ness}(1968)}]{mandel}%
  \BibitemOpen
  \bibfield  {author} {\bibinfo {author} {\bibfnamefont {B.~B.}\ \bibnamefont
  {Mandelbrot}}\ and\ \bibinfo {author} {\bibfnamefont {J.~W.}\ \bibnamefont
  {Van~Ness}},\ }\bibfield  {title} {\enquote {\bibinfo {title} {Fractional
  {B}rownian motions, fractional noises and applications},}\ }\href@noop {}
  {\bibfield  {journal} {\bibinfo  {journal} {SIAM Rev.}\ }\textbf {\bibinfo
  {volume} {10}},\ \bibinfo {pages} {422} (\bibinfo {year} {1968})}\BibitemShut
  {NoStop}%
\bibitem [{\citenamefont {Shlesinger}\ \emph {et~al.}(1994)\citenamefont
  {Shlesinger}, \citenamefont {Zaslavsky},\ and\ \citenamefont
  {Frisch}}]{SZF94}%
  \BibitemOpen
  \bibinfo {editor} {\bibfnamefont {M.~F.}\ \bibnamefont {Shlesinger}},
  \bibinfo {editor} {\bibfnamefont {G.~M.}\ \bibnamefont {Zaslavsky}}, \ and\
  \bibinfo {editor} {\bibfnamefont {U.}~\bibnamefont {Frisch}},\ eds.,\
  \href@noop {} {\emph {\bibinfo {title} {L{\'e}vy flights and related
  topics}}}\ (\bibinfo  {publisher} {Springer},\ \bibinfo {address} {Berlin},\
  \bibinfo {year} {1994})\BibitemShut {NoStop}%
\bibitem [{\citenamefont {Shlesinger}\ \emph {et~al.}(1993)\citenamefont
  {Shlesinger}, \citenamefont {Zaslavsky},\ and\ \citenamefont
  {Klafter}}]{SZK93}%
  \BibitemOpen
  \bibfield  {author} {\bibinfo {author} {\bibfnamefont {M.~F.}\ \bibnamefont
  {Shlesinger}}, \bibinfo {author} {\bibfnamefont {G.~M.}\ \bibnamefont
  {Zaslavsky}}, \ and\ \bibinfo {author} {\bibfnamefont {J.}~\bibnamefont
  {Klafter}},\ }\bibfield  {title} {\enquote {\bibinfo {title} {Strange
  kinetics},}\ }\href@noop {} {\bibfield  {journal} {\bibinfo  {journal}
  {Nature}\ }\textbf {\bibinfo {volume} {363}},\ \bibinfo {pages} {31}
  (\bibinfo {year} {1993})}\BibitemShut {NoStop}%
\bibitem [{\citenamefont {Zaburdaev}\ \emph {et~al.}(2015)\citenamefont
  {Zaburdaev}, \citenamefont {Denisov},\ and\ \citenamefont {Klafter}}]{ZDK15}%
  \BibitemOpen
  \bibfield  {author} {\bibinfo {author} {\bibfnamefont {V.}~\bibnamefont
  {Zaburdaev}}, \bibinfo {author} {\bibfnamefont {S.}~\bibnamefont {Denisov}},
  \ and\ \bibinfo {author} {\bibfnamefont {J.}~\bibnamefont {Klafter}},\
  }\bibfield  {title} {\enquote {\bibinfo {title} {L\'{e}vy walks},}\
  }\href@noop {} {\bibfield  {journal} {\bibinfo  {journal} {Rev. Mod. Phys.}\
  }\textbf {\bibinfo {volume} {87}},\ \bibinfo {pages} {483} (\bibinfo {year}
  {2015})}\BibitemShut {NoStop}%
\bibitem [{\citenamefont {Buldyrev}\ \emph {et~al.}(2001)\citenamefont
  {Buldyrev}, \citenamefont {Havlin}, \citenamefont {Raposo}, \citenamefont
  {Stanley},\ and\ \citenamefont {Viswanathan}}]{BHRSV01}%
  \BibitemOpen
  \bibfield  {author} {\bibinfo {author} {\bibfnamefont {S.~V.}\ \bibnamefont
  {Buldyrev}}, \bibinfo {author} {\bibfnamefont {S.}~\bibnamefont {Havlin}},
  \bibinfo {author} {\bibfnamefont {E.~P.}\ \bibnamefont {Raposo}}, \bibinfo
  {author} {\bibfnamefont {H.~E.}\ \bibnamefont {Stanley}}, \ and\ \bibinfo
  {author} {\bibfnamefont {G.~M.}\ \bibnamefont {Viswanathan}},\ }\bibfield
  {title} {\enquote {\bibinfo {title} {Average time spent by {L}\'evy flights
  and walks on an interval with absorbing boundaries},}\ }\href@noop {}
  {\bibfield  {journal} {\bibinfo  {journal} {Phys. Rev. E}\ }\textbf {\bibinfo
  {volume} {64}},\ \bibinfo {pages} {18} (\bibinfo {year} {2001})}\BibitemShut
  {NoStop}%
\bibitem [{\citenamefont {Raposo}\ \emph {et~al.}(2003)\citenamefont {Raposo},
  \citenamefont {Buldyrev}, \citenamefont {Da~Luz}, \citenamefont {Santos},
  \citenamefont {Stanley},\ and\ \citenamefont {Viswanathan}}]{RBLSSV03}%
  \BibitemOpen
  \bibfield  {author} {\bibinfo {author} {\bibfnamefont {E.~P.}\ \bibnamefont
  {Raposo}}, \bibinfo {author} {\bibfnamefont {Sergey~V.}\ \bibnamefont
  {Buldyrev}}, \bibinfo {author} {\bibfnamefont {M.~G.~E.}\ \bibnamefont
  {Da~Luz}}, \bibinfo {author} {\bibfnamefont {M.~C.}\ \bibnamefont {Santos}},
  \bibinfo {author} {\bibfnamefont {H.~E.}\ \bibnamefont {Stanley}}, \ and\
  \bibinfo {author} {\bibfnamefont {G.~M.}\ \bibnamefont {Viswanathan}},\
  }\bibfield  {title} {\enquote {\bibinfo {title} {Dynamical robustness of
  {L}\'evy search strategies},}\ }\href@noop {} {\bibfield  {journal} {\bibinfo
   {journal} {Phys. Rev. Lett.}\ }\textbf {\bibinfo {volume} {91}},\ \bibinfo
  {pages} {240601} (\bibinfo {year} {2003})}\BibitemShut {NoStop}%
\bibitem [{\citenamefont {Benhamou}(2007{\natexlab{a}})}]{Benh07}%
  \BibitemOpen
  \bibfield  {author} {\bibinfo {author} {\bibfnamefont {S.}~\bibnamefont
  {Benhamou}},\ }\bibfield  {title} {\enquote {\bibinfo {title} {How many
  animals really do the {L}\'evy walk?}}\ }\href@noop {} {\bibfield  {journal}
  {\bibinfo  {journal} {Ecology}\ }\textbf {\bibinfo {volume} {88}},\ \bibinfo
  {pages} {1962} (\bibinfo {year} {2007}{\natexlab{a}})}\BibitemShut {NoStop}%
\bibitem [{\citenamefont {Sims}\ \emph {et~al.}(2008)\citenamefont {Sims},
  \citenamefont {Southall}, \citenamefont {Humphries}, \citenamefont {Hays},
  \citenamefont {Bradshaw}, \citenamefont {Pitchford}, \citenamefont {James},
  \citenamefont {Ahmed}, \citenamefont {Brierley}, \citenamefont {Hindell},
  \citenamefont {Morritt}, \citenamefont {Musyl}, \citenamefont {Righton},
  \citenamefont {Shepard}, \citenamefont {Wearmouth}, \citenamefont {Wilson},
  \citenamefont {Witt},\ and\ \citenamefont {Metcalfe}}]{Sims08}%
  \BibitemOpen
  \bibfield  {author} {\bibinfo {author} {\bibfnamefont {D.~W.}\ \bibnamefont
  {Sims}}, \bibinfo {author} {\bibfnamefont {E.~J.}\ \bibnamefont {Southall}},
  \bibinfo {author} {\bibfnamefont {N.~E.}\ \bibnamefont {Humphries}}, \bibinfo
  {author} {\bibfnamefont {G.~C.}\ \bibnamefont {Hays}}, \bibinfo {author}
  {\bibfnamefont {C.~J.~A.}\ \bibnamefont {Bradshaw}}, \bibinfo {author}
  {\bibfnamefont {J.~W.}\ \bibnamefont {Pitchford}}, \bibinfo {author}
  {\bibfnamefont {A.}~\bibnamefont {James}}, \bibinfo {author} {\bibfnamefont
  {M.~Z.}\ \bibnamefont {Ahmed}}, \bibinfo {author} {\bibfnamefont {A.~S.}\
  \bibnamefont {Brierley}}, \bibinfo {author} {\bibfnamefont {M.~A.}\
  \bibnamefont {Hindell}}, \bibinfo {author} {\bibfnamefont {D.}~\bibnamefont
  {Morritt}}, \bibinfo {author} {\bibfnamefont {M.~K.}\ \bibnamefont {Musyl}},
  \bibinfo {author} {\bibfnamefont {D.}~\bibnamefont {Righton}}, \bibinfo
  {author} {\bibfnamefont {E.~L.~C.}\ \bibnamefont {Shepard}}, \bibinfo
  {author} {\bibfnamefont {V.~J.}\ \bibnamefont {Wearmouth}}, \bibinfo {author}
  {\bibfnamefont {R.~P.}\ \bibnamefont {Wilson}}, \bibinfo {author}
  {\bibfnamefont {M.~J.}\ \bibnamefont {Witt}}, \ and\ \bibinfo {author}
  {\bibfnamefont {J.~D.}\ \bibnamefont {Metcalfe}},\ }\bibfield  {title}
  {\enquote {\bibinfo {title} {Scaling laws of marine predator search
  behaviour},}\ }\href@noop {} {\bibfield  {journal} {\bibinfo  {journal}
  {Nature}\ }\textbf {\bibinfo {volume} {451}},\ \bibinfo {pages} {1098}
  (\bibinfo {year} {2008})}\BibitemShut {NoStop}%
\bibitem [{\citenamefont {Humphries}\ \emph {et~al.}(2010)\citenamefont
  {Humphries}, \citenamefont {Queiroz}, \citenamefont {Dyer}, \citenamefont
  {Pade}, \citenamefont {Musy}, \citenamefont {Schaefer}, \citenamefont
  {Fuller}, \citenamefont {Brunnschweiler}, \citenamefont {Doyle},
  \citenamefont {Houghton}, \citenamefont {Hays}, \citenamefont {Jones},
  \citenamefont {Noble}, \citenamefont {Wearmouth}, \citenamefont {Southall},\
  and\ \citenamefont {Sims}}]{Sims10}%
  \BibitemOpen
  \bibfield  {author} {\bibinfo {author} {\bibfnamefont {N.~E.}\ \bibnamefont
  {Humphries}}, \bibinfo {author} {\bibfnamefont {N.}~\bibnamefont {Queiroz}},
  \bibinfo {author} {\bibfnamefont {J.~R~.M.}\ \bibnamefont {Dyer}}, \bibinfo
  {author} {\bibfnamefont {N.~G.}\ \bibnamefont {Pade}}, \bibinfo {author}
  {\bibfnamefont {M.~K.}\ \bibnamefont {Musy}}, \bibinfo {author}
  {\bibfnamefont {K.~M.}\ \bibnamefont {Schaefer}}, \bibinfo {author}
  {\bibfnamefont {D.~W.}\ \bibnamefont {Fuller}}, \bibinfo {author}
  {\bibfnamefont {J.~M.}\ \bibnamefont {Brunnschweiler}}, \bibinfo {author}
  {\bibfnamefont {T.~K.}\ \bibnamefont {Doyle}}, \bibinfo {author}
  {\bibfnamefont {J.~D.~R.}\ \bibnamefont {Houghton}}, \bibinfo {author}
  {\bibfnamefont {G.~C.}\ \bibnamefont {Hays}}, \bibinfo {author}
  {\bibfnamefont {C.~S.}\ \bibnamefont {Jones}}, \bibinfo {author}
  {\bibfnamefont {L.~R.}\ \bibnamefont {Noble}}, \bibinfo {author}
  {\bibfnamefont {V.~J.}\ \bibnamefont {Wearmouth}}, \bibinfo {author}
  {\bibfnamefont {E.~J.}\ \bibnamefont {Southall}}, \ and\ \bibinfo {author}
  {\bibfnamefont {D.~W.}\ \bibnamefont {Sims}},\ }\bibfield  {title} {\enquote
  {\bibinfo {title} {Environmental context explains {L}\'evy and {B}rownian
  movement patterns of marine predators},}\ }\href@noop {} {\bibfield
  {journal} {\bibinfo  {journal} {Nature}\ }\textbf {\bibinfo {volume} {465}},\
  \bibinfo {pages} {1066} (\bibinfo {year} {2010})}\BibitemShut {NoStop}%
\bibitem [{\citenamefont {de~Jager}\ \emph {et~al.}(2011)\citenamefont
  {de~Jager}, \citenamefont {Weissing}, \citenamefont {Herman}, \citenamefont
  {Nolet},\ and\ \citenamefont {van~de Koppel}}]{dJWH11}%
  \BibitemOpen
  \bibfield  {author} {\bibinfo {author} {\bibfnamefont {M.}~\bibnamefont
  {de~Jager}}, \bibinfo {author} {\bibfnamefont {F.~J.}\ \bibnamefont
  {Weissing}}, \bibinfo {author} {\bibfnamefont {P.~M.~J.}\ \bibnamefont
  {Herman}}, \bibinfo {author} {\bibfnamefont {B.~A.}\ \bibnamefont {Nolet}}, \
  and\ \bibinfo {author} {\bibfnamefont {J.}~\bibnamefont {van~de Koppel}},\
  }\bibfield  {title} {\enquote {\bibinfo {title} {L{\'e}vy walks evolve
  through interaction between movement and environmental complexity},}\
  }\href@noop {} {\bibfield  {journal} {\bibinfo  {journal} {Science}\ }\textbf
  {\bibinfo {volume} {332}},\ \bibinfo {pages} {1551} (\bibinfo {year}
  {2011})}\BibitemShut {NoStop}%
\bibitem [{\citenamefont {Harris}\ \emph {et~al.}(2012)\citenamefont {Harris},
  \citenamefont {Banigan}, \citenamefont {Christian}, \citenamefont {Konradt},
  \citenamefont {Wojno}, \citenamefont {Norose}, \citenamefont {Wilson},
  \citenamefont {John}, \citenamefont {Weninger},\ and\ \citenamefont
  {Luster}}]{HaBa12}%
  \BibitemOpen
  \bibfield  {author} {\bibinfo {author} {\bibfnamefont {T.H.}\ \bibnamefont
  {Harris}}, \bibinfo {author} {\bibfnamefont {E.J.}\ \bibnamefont {Banigan}},
  \bibinfo {author} {\bibfnamefont {D.A.}\ \bibnamefont {Christian}}, \bibinfo
  {author} {\bibfnamefont {C.}~\bibnamefont {Konradt}}, \bibinfo {author}
  {\bibfnamefont {E.D.~Tait}\ \bibnamefont {Wojno}}, \bibinfo {author}
  {\bibfnamefont {K.}~\bibnamefont {Norose}}, \bibinfo {author} {\bibfnamefont
  {E.H.}\ \bibnamefont {Wilson}}, \bibinfo {author} {\bibfnamefont
  {B.}~\bibnamefont {John}}, \bibinfo {author} {\bibfnamefont {W.}~\bibnamefont
  {Weninger}}, \ and\ \bibinfo {author} {\bibfnamefont {A.D.}\ \bibnamefont
  {Luster}},\ }\bibfield  {title} {\enquote {\bibinfo {title} {Generalized
  {L}\'{e}vy walks and the role of chemokines in migration of effector cd8{+} t
  cells},}\ }\href@noop {} {\bibfield  {journal} {\bibinfo  {journal} {Nature}\
  }\textbf {\bibinfo {volume} {486}},\ \bibinfo {pages} {545} (\bibinfo {year}
  {2012})}\BibitemShut {NoStop}%
\bibitem [{\citenamefont {Lenz}\ \emph {et~al.}(2012)\citenamefont {Lenz},
  \citenamefont {Ings}, \citenamefont {Chittka}, \citenamefont {Chechkin},\
  and\ \citenamefont {Klages}}]{LICCK12}%
  \BibitemOpen
  \bibfield  {author} {\bibinfo {author} {\bibfnamefont {F.}~\bibnamefont
  {Lenz}}, \bibinfo {author} {\bibfnamefont {T.~C.}\ \bibnamefont {Ings}},
  \bibinfo {author} {\bibfnamefont {L.}~\bibnamefont {Chittka}}, \bibinfo
  {author} {\bibfnamefont {A.~V.}\ \bibnamefont {Chechkin}}, \ and\ \bibinfo
  {author} {\bibfnamefont {R.}~\bibnamefont {Klages}},\ }\bibfield  {title}
  {\enquote {\bibinfo {title} {Spatiotemporal dynamics of bumblebees foraging
  under predation risk},}\ }\href@noop {} {\bibfield  {journal} {\bibinfo
  {journal} {Phys. Rev. Lett.}\ }\textbf {\bibinfo {volume} {108}},\ \bibinfo
  {pages} {098103} (\bibinfo {year} {2012})}\BibitemShut {NoStop}%
\bibitem [{\citenamefont {Palyulin}\ \emph
  {et~al.}(2014{\natexlab{a}})\citenamefont {Palyulin}, \citenamefont
  {Chechkin},\ and\ \citenamefont {Metzler}}]{PCM14}%
  \BibitemOpen
  \bibfield  {author} {\bibinfo {author} {\bibfnamefont {V.~V.}\ \bibnamefont
  {Palyulin}}, \bibinfo {author} {\bibfnamefont {A.~V.}\ \bibnamefont
  {Chechkin}}, \ and\ \bibinfo {author} {\bibfnamefont {R.}~\bibnamefont
  {Metzler}},\ }\bibfield  {title} {\enquote {\bibinfo {title} {{L\'{e}vy
  flights do not always optimize random blind search for sparse targets}},}\
  }\href@noop {} {\bibfield  {journal} {\bibinfo  {journal} {Proc. Nat. Acad.
  Sci.}\ }\textbf {\bibinfo {volume} {111}},\ \bibinfo {pages} {2931} (\bibinfo
  {year} {2014}{\natexlab{a}})}\BibitemShut {NoStop}%
\bibitem [{\citenamefont {Pyke}(2015)}]{Pyke15}%
  \BibitemOpen
  \bibfield  {author} {\bibinfo {author} {\bibfnamefont {G.~H.}\ \bibnamefont
  {Pyke}},\ }\bibfield  {title} {\enquote {\bibinfo {title} {Understanding
  movements of organisms: it's time to abandon the {L}\'evy foraging
  hypothesis},}\ }\href@noop {} {\bibfield  {journal} {\bibinfo  {journal}
  {Meth. Ecol. Evol.}\ }\textbf {\bibinfo {volume} {6}},\ \bibinfo {pages} {1}
  (\bibinfo {year} {2015})}\BibitemShut {NoStop}%
\bibitem [{\citenamefont {Reynolds}(2015)}]{Reyn15}%
  \BibitemOpen
  \bibfield  {author} {\bibinfo {author} {\bibfnamefont {A.~M.}\ \bibnamefont
  {Reynolds}},\ }\bibfield  {title} {\enquote {\bibinfo {title} {Liberating
  {L}{\'e}vy walk research from the shackles of optimal foraging},}\
  }\href@noop {} {\bibfield  {journal} {\bibinfo  {journal} {Phys. Life Rev.}\
  }\textbf {\bibinfo {volume} {14}},\ \bibinfo {pages} {59} (\bibinfo {year}
  {2015})}\BibitemShut {NoStop}%
\bibitem [{\citenamefont {Reynolds}(2018)}]{Reyn18}%
  \BibitemOpen
  \bibfield  {author} {\bibinfo {author} {\bibfnamefont {A.~M.}\ \bibnamefont
  {Reynolds}},\ }\bibfield  {title} {\enquote {\bibinfo {title} {Current status
  and future directions of {L}{\'e}vy walk research},}\ }\href@noop {}
  {\bibfield  {journal} {\bibinfo  {journal} {Biol. Open}\ }\textbf {\bibinfo
  {volume} {7}},\ \bibinfo {pages} {03016} (\bibinfo {year}
  {2018})}\BibitemShut {NoStop}%
\bibitem [{\citenamefont {Levernier}\ \emph {et~al.}(2020)\citenamefont
  {Levernier}, \citenamefont {Textor}, \citenamefont {B\'enichou},\ and\
  \citenamefont {Voituriez}}]{LTBV20}%
  \BibitemOpen
  \bibfield  {author} {\bibinfo {author} {\bibfnamefont {N.}~\bibnamefont
  {Levernier}}, \bibinfo {author} {\bibfnamefont {J.}~\bibnamefont {Textor}},
  \bibinfo {author} {\bibfnamefont {O.}~\bibnamefont {B\'enichou}}, \ and\
  \bibinfo {author} {\bibfnamefont {R.}~\bibnamefont {Voituriez}},\ }\bibfield
  {title} {\enquote {\bibinfo {title} {Inverse square {L}\'evy walks are not
  optimal search strategies for $d\ensuremath{\ge}2$},}\ }\href@noop {}
  {\bibfield  {journal} {\bibinfo  {journal} {Phys. Rev. Lett.}\ }\textbf
  {\bibinfo {volume} {124}},\ \bibinfo {pages} {080601} (\bibinfo {year}
  {2020})}\BibitemShut {NoStop}%
\bibitem [{\citenamefont {Guinard}\ and\ \citenamefont
  {Korman}(2021)}]{GuKo20}%
  \BibitemOpen
  \bibfield  {author} {\bibinfo {author} {\bibfnamefont {B.}~\bibnamefont
  {Guinard}}\ and\ \bibinfo {author} {\bibfnamefont {A.}~\bibnamefont
  {Korman}},\ }\bibfield  {title} {\enquote {\bibinfo {title} {Intermittent
  inverse-square {L}{\'e}vy walks are optimal for finding targets of all
  sizes},}\ }\href@noop {} {\bibfield  {journal} {\bibinfo  {journal} {Sci.
  Adv.}\ }\textbf {\bibinfo {volume} {7}},\ \bibinfo {pages} {eabe8211}
  (\bibinfo {year} {2021})}\BibitemShut {NoStop}%
\bibitem [{\citenamefont {Buldyrev}\ \emph {et~al.}(2021)\citenamefont
  {Buldyrev}, \citenamefont {Raposo}, \citenamefont {Bartumeus}, \citenamefont
  {Havlin}, \citenamefont {Rusch}, \citenamefont {Da~Luz},\ and\ \citenamefont
  {Viswanathan}}]{BRBHRLV21}%
  \BibitemOpen
  \bibfield  {author} {\bibinfo {author} {\bibfnamefont {S.~V.}\ \bibnamefont
  {Buldyrev}}, \bibinfo {author} {\bibfnamefont {E.~P.}\ \bibnamefont
  {Raposo}}, \bibinfo {author} {\bibfnamefont {F.}~\bibnamefont {Bartumeus}},
  \bibinfo {author} {\bibfnamefont {S.}~\bibnamefont {Havlin}}, \bibinfo
  {author} {\bibfnamefont {F.~R.}\ \bibnamefont {Rusch}}, \bibinfo {author}
  {\bibfnamefont {M.~G.~E.}\ \bibnamefont {Da~Luz}}, \ and\ \bibinfo {author}
  {\bibfnamefont {G.~M.}\ \bibnamefont {Viswanathan}},\ }\bibfield  {title}
  {\enquote {\bibinfo {title} {Comment on ``inverse square {L}\'evy walks are
  not optimal search strategies for $d\ensuremath{\ge}2$''},}\ }\href@noop {}
  {\bibfield  {journal} {\bibinfo  {journal} {Phys. Rev. Lett.}\ }\textbf
  {\bibinfo {volume} {126}},\ \bibinfo {pages} {048901} (\bibinfo {year}
  {2021})}\BibitemShut {NoStop}%
\bibitem [{\citenamefont {Levernier}\ \emph
  {et~al.}(2021{\natexlab{a}})\citenamefont {Levernier}, \citenamefont
  {Textor}, \citenamefont {B\'enichou},\ and\ \citenamefont
  {Voituriez}}]{LTBV21}%
  \BibitemOpen
  \bibfield  {author} {\bibinfo {author} {\bibfnamefont {N.}~\bibnamefont
  {Levernier}}, \bibinfo {author} {\bibfnamefont {J.}~\bibnamefont {Textor}},
  \bibinfo {author} {\bibfnamefont {O.}~\bibnamefont {B\'enichou}}, \ and\
  \bibinfo {author} {\bibfnamefont {R.}~\bibnamefont {Voituriez}},\ }\bibfield
  {title} {\enquote {\bibinfo {title} {Reply to ``comment on `inverse square
  {L}\'evy walks are not optimal search strategies for
  $d\ensuremath{\ge}2$'''},}\ }\href@noop {} {\bibfield  {journal} {\bibinfo
  {journal} {Phys. Rev. Lett.}\ }\textbf {\bibinfo {volume} {126}},\ \bibinfo
  {pages} {048902} (\bibinfo {year} {2021}{\natexlab{a}})}\BibitemShut
  {NoStop}%
\bibitem [{\citenamefont {Vilka}\ \emph {et~al.}(2021)\citenamefont {Vilka},
  \citenamefont {Orchan}, \citenamefont {Charter}, \citenamefont {Ganot},
  \citenamefont {Toledo}, \citenamefont {Nathan},\ and\ \citenamefont
  {Assaf}}]{VOCGTNA21}%
  \BibitemOpen
  \bibfield  {author} {\bibinfo {author} {\bibfnamefont {O.}~\bibnamefont
  {Vilka}}, \bibinfo {author} {\bibfnamefont {Y.}~\bibnamefont {Orchan}},
  \bibinfo {author} {\bibfnamefont {M.}~\bibnamefont {Charter}}, \bibinfo
  {author} {\bibfnamefont {N.}~\bibnamefont {Ganot}}, \bibinfo {author}
  {\bibfnamefont {S.}~\bibnamefont {Toledo}}, \bibinfo {author} {\bibfnamefont
  {R.}~\bibnamefont {Nathan}}, \ and\ \bibinfo {author} {\bibfnamefont
  {M.}~\bibnamefont {Assaf}},\ }\bibfield  {title} {\enquote {\bibinfo {title}
  {Ergodicity breaking and lack of a typical waiting time in area-restricted
  search of avian predators},}\ }\href@noop {} {\bibfield  {journal} {\bibinfo
  {journal} {preprint arXiv:2101.11527}\ } (\bibinfo {year}
  {2021})}\BibitemShut {NoStop}%
\bibitem [{\citenamefont {Redner}(2001)}]{Red01}%
  \BibitemOpen
  \bibfield  {author} {\bibinfo {author} {\bibfnamefont {S.}~\bibnamefont
  {Redner}},\ }\href@noop {} {\emph {\bibinfo {title} {A guide to first-passage
  processes}}}\ (\bibinfo  {publisher} {Cambridge University Press},\ \bibinfo
  {address} {Cambridge},\ \bibinfo {year} {2001})\BibitemShut {NoStop}%
\bibitem [{\citenamefont {Metzler}\ \emph
  {et~al.}(2014{\natexlab{b}})\citenamefont {Metzler}, \citenamefont
  {Oshanin},\ and\ \citenamefont {Redner}}]{MOR14}%
  \BibitemOpen
  \bibfield  {author} {\bibinfo {author} {\bibfnamefont {R.}~\bibnamefont
  {Metzler}}, \bibinfo {author} {\bibfnamefont {G.}~\bibnamefont {Oshanin}}, \
  and\ \bibinfo {author} {\bibfnamefont {S.}~\bibnamefont {Redner}},\
  }\href@noop {} {\emph {\bibinfo {title} {First-Passage Phenomena and Their
  Applications}}}\ (\bibinfo  {publisher} {World Scientific},\ \bibinfo
  {address} {Singapore},\ \bibinfo {year} {2014})\BibitemShut {NoStop}%
\bibitem [{\citenamefont {Palyulin}\ \emph {et~al.}(2019)\citenamefont
  {Palyulin}, \citenamefont {Blackburn}, \citenamefont {Lomholt}, \citenamefont
  {Watkins}, \citenamefont {Metzler}, \citenamefont {Klages},\ and\
  \citenamefont {Chechkin}}]{PBL19}%
  \BibitemOpen
  \bibfield  {author} {\bibinfo {author} {\bibfnamefont {V.~V.}\ \bibnamefont
  {Palyulin}}, \bibinfo {author} {\bibfnamefont {G.}~\bibnamefont {Blackburn}},
  \bibinfo {author} {\bibfnamefont {M.~A.}\ \bibnamefont {Lomholt}}, \bibinfo
  {author} {\bibfnamefont {Ni.~W.}\ \bibnamefont {Watkins}}, \bibinfo {author}
  {\bibfnamefont {R.}~\bibnamefont {Metzler}}, \bibinfo {author} {\bibfnamefont
  {R.}~\bibnamefont {Klages}}, \ and\ \bibinfo {author} {\bibfnamefont {A.~V.}\
  \bibnamefont {Chechkin}},\ }\bibfield  {title} {\enquote {\bibinfo {title}
  {First passage and first hitting times of {L}{\'{e}}vy flights and
  {L}{\'{e}}vy walks},}\ }\href@noop {} {\bibfield  {journal} {\bibinfo
  {journal} {New J. Phys.}\ }\textbf {\bibinfo {volume} {21}},\ \bibinfo
  {pages} {103028} (\bibinfo {year} {2019})}\BibitemShut {NoStop}%
\bibitem [{\citenamefont {Levernier}\ \emph
  {et~al.}(2021{\natexlab{b}})\citenamefont {Levernier}, \citenamefont
  {B\'enichou},\ and\ \citenamefont {Voituriez}}]{LBV21}%
  \BibitemOpen
  \bibfield  {author} {\bibinfo {author} {\bibfnamefont {N.}~\bibnamefont
  {Levernier}}, \bibinfo {author} {\bibfnamefont {O.}~\bibnamefont
  {B\'enichou}}, \ and\ \bibinfo {author} {\bibfnamefont {R.}~\bibnamefont
  {Voituriez}},\ }\bibfield  {title} {\enquote {\bibinfo {title} {Universality
  classes of hitting probabilities of jump processes},}\ }\href@noop {}
  {\bibfield  {journal} {\bibinfo  {journal} {Phys. Rev. Lett.}\ }\textbf
  {\bibinfo {volume} {126}},\ \bibinfo {pages} {100602} (\bibinfo {year}
  {2021}{\natexlab{b}})}\BibitemShut {NoStop}%
\bibitem [{\citenamefont {James}\ \emph {et~al.}(2009)\citenamefont {James},
  \citenamefont {Pitchford},\ and\ \citenamefont {Plank}}]{JPP09}%
  \BibitemOpen
  \bibfield  {author} {\bibinfo {author} {\bibfnamefont {A.}~\bibnamefont
  {James}}, \bibinfo {author} {\bibfnamefont {J.~W.}\ \bibnamefont
  {Pitchford}}, \ and\ \bibinfo {author} {\bibfnamefont {M.~J.}\ \bibnamefont
  {Plank}},\ }\bibfield  {title} {\enquote {\bibinfo {title} {Efficient or
  inaccurate? {A}nalytical and numerical modelling of random search
  strategies},}\ }\href@noop {} {\bibfield  {journal} {\bibinfo  {journal}
  {Bull. Math. Biol.}\ }\textbf {\bibinfo {volume} {72}},\ \bibinfo {pages}
  {896} (\bibinfo {year} {2009})}\BibitemShut {NoStop}%
\bibitem [{\citenamefont {O'Brien}\ \emph {et~al.}(1990)\citenamefont
  {O'Brien}, \citenamefont {Browman},\ and\ \citenamefont {Evans}}]{OBBE90}%
  \BibitemOpen
  \bibfield  {author} {\bibinfo {author} {\bibfnamefont {W.J.}\ \bibnamefont
  {O'Brien}}, \bibinfo {author} {\bibfnamefont {H.I.}\ \bibnamefont {Browman}},
  \ and\ \bibinfo {author} {\bibfnamefont {B.I.}\ \bibnamefont {Evans}},\
  }\bibfield  {title} {\enquote {\bibinfo {title} {Search strategies of
  foraging animals},}\ }\href@noop {} {\bibfield  {journal} {\bibinfo
  {journal} {Am. Sci.}\ }\textbf {\bibinfo {volume} {78}},\ \bibinfo {pages}
  {152} (\bibinfo {year} {1990})}\BibitemShut {NoStop}%
\bibitem [{\citenamefont {James}\ \emph {et~al.}(2011)\citenamefont {James},
  \citenamefont {Plank},\ and\ \citenamefont {Edwards}}]{JPE11}%
  \BibitemOpen
  \bibfield  {author} {\bibinfo {author} {\bibfnamefont {A.}~\bibnamefont
  {James}}, \bibinfo {author} {\bibfnamefont {M.~J.}\ \bibnamefont {Plank}}, \
  and\ \bibinfo {author} {\bibfnamefont {A.~M.}\ \bibnamefont {Edwards}},\
  }\bibfield  {title} {\enquote {\bibinfo {title} {Assessing {L}\'{e}vy walks
  as models of animal foraging},}\ }\href@noop {} {\bibfield  {journal}
  {\bibinfo  {journal} {Roy. Soc. Interf.}\ }\textbf {\bibinfo {volume} {8}},\
  \bibinfo {pages} {1233} (\bibinfo {year} {2011})}\BibitemShut {NoStop}%
\bibitem [{\citenamefont {Boyer}\ \emph {et~al.}(2006)\citenamefont {Boyer},
  \citenamefont {Ramos-Fernández}, \citenamefont {Miramontes}, \citenamefont
  {Mateos}, \citenamefont {Cocho}, \citenamefont {Larralde}, \citenamefont
  {Ramos},\ and\ \citenamefont {Rojas}}]{BoRa06}%
  \BibitemOpen
  \bibfield  {author} {\bibinfo {author} {\bibfnamefont {D.}~\bibnamefont
  {Boyer}}, \bibinfo {author} {\bibfnamefont {G.}~\bibnamefont
  {Ramos-Fernández}}, \bibinfo {author} {\bibfnamefont {O.}~\bibnamefont
  {Miramontes}}, \bibinfo {author} {\bibfnamefont {J.~L.}\ \bibnamefont
  {Mateos}}, \bibinfo {author} {\bibfnamefont {G.}~\bibnamefont {Cocho}},
  \bibinfo {author} {\bibfnamefont {H.}~\bibnamefont {Larralde}}, \bibinfo
  {author} {\bibfnamefont {H.}~\bibnamefont {Ramos}}, \ and\ \bibinfo {author}
  {\bibfnamefont {F.}~\bibnamefont {Rojas}},\ }\bibfield  {title} {\enquote
  {\bibinfo {title} {Scale-free foraging by primates emerges from their
  interaction with a complex environment},}\ }\href@noop {} {\bibfield
  {journal} {\bibinfo  {journal} {Proc. R. Soc. B.}\ }\textbf {\bibinfo
  {volume} {273}},\ \bibinfo {pages} {1743} (\bibinfo {year}
  {2006})}\BibitemShut {NoStop}%
\bibitem [{\citenamefont {Campos}\ \emph {et~al.}(2015)\citenamefont {Campos},
  \citenamefont {Bartumeus}, \citenamefont {Raposo},\ and\ \citenamefont
  {M{\'{e}}ndez}}]{CBRM15}%
  \BibitemOpen
  \bibfield  {author} {\bibinfo {author} {\bibfnamefont {D.}~\bibnamefont
  {Campos}}, \bibinfo {author} {\bibfnamefont {F.}~\bibnamefont {Bartumeus}},
  \bibinfo {author} {\bibfnamefont {E.~P.}\ \bibnamefont {Raposo}}, \ and\
  \bibinfo {author} {\bibfnamefont {V.}~\bibnamefont {M{\'{e}}ndez}},\
  }\bibfield  {title} {\enquote {\bibinfo {title} {{First-passage times in
  multiscale random walks: The impact of movement scales on search
  efficiency}},}\ }\href@noop {} {\bibfield  {journal} {\bibinfo  {journal}
  {Phys. Rev. E}\ }\textbf {\bibinfo {volume} {92}},\ \bibinfo {pages} {1}
  (\bibinfo {year} {2015})}\BibitemShut {NoStop}%
\bibitem [{\citenamefont {Zhao}\ \emph {et~al.}(2015)\citenamefont {Zhao},
  \citenamefont {Jurdak}, \citenamefont {Liu}, \citenamefont {Westcott},
  \citenamefont {Kusy}, \citenamefont {Parry}, \citenamefont {Sommer},\ and\
  \citenamefont {McKeown}}]{ZJLW15}%
  \BibitemOpen
  \bibfield  {author} {\bibinfo {author} {\bibfnamefont {K.}~\bibnamefont
  {Zhao}}, \bibinfo {author} {\bibfnamefont {R.}~\bibnamefont {Jurdak}},
  \bibinfo {author} {\bibfnamefont {J.}~\bibnamefont {Liu}}, \bibinfo {author}
  {\bibfnamefont {D.}~\bibnamefont {Westcott}}, \bibinfo {author}
  {\bibfnamefont {B.}~\bibnamefont {Kusy}}, \bibinfo {author} {\bibfnamefont
  {H.}~\bibnamefont {Parry}}, \bibinfo {author} {\bibfnamefont
  {P.}~\bibnamefont {Sommer}}, \ and\ \bibinfo {author} {\bibfnamefont
  {A.}~\bibnamefont {McKeown}},\ }\bibfield  {title} {\enquote {\bibinfo
  {title} {{Optimal L{\'{e}}vy-flight foraging in a finite landscape}},}\
  }\href@noop {} {\bibfield  {journal} {\bibinfo  {journal} {J. R. Soc.
  Interface}\ }\textbf {\bibinfo {volume} {12}} (\bibinfo {year}
  {2015})}\BibitemShut {NoStop}%
\bibitem [{\citenamefont {Khatami}\ \emph {et~al.}(2016)\citenamefont
  {Khatami}, \citenamefont {Wolff}, \citenamefont {Pohl}, \citenamefont
  {Ejtehadi},\ and\ \citenamefont {Stark}}]{KWPES16}%
  \BibitemOpen
  \bibfield  {author} {\bibinfo {author} {\bibfnamefont {M.}~\bibnamefont
  {Khatami}}, \bibinfo {author} {\bibfnamefont {K.}~\bibnamefont {Wolff}},
  \bibinfo {author} {\bibfnamefont {O.}~\bibnamefont {Pohl}}, \bibinfo {author}
  {\bibfnamefont {M.~R.}\ \bibnamefont {Ejtehadi}}, \ and\ \bibinfo {author}
  {\bibfnamefont {H.}~\bibnamefont {Stark}},\ }\bibfield  {title} {\enquote
  {\bibinfo {title} {Active {B}rownian particles and run-and-tumble particles
  separate inside a maze},}\ }\href@noop {} {\bibfield  {journal} {\bibinfo
  {journal} {Sci. Rep.}\ }\textbf {\bibinfo {volume} {6}},\ \bibinfo {pages}
  {1} (\bibinfo {year} {2016})}\BibitemShut {NoStop}%
\bibitem [{\citenamefont {Volpe}\ and\ \citenamefont {Volpe}(2017)}]{VoVo17}%
  \BibitemOpen
  \bibfield  {author} {\bibinfo {author} {\bibfnamefont {G.}~\bibnamefont
  {Volpe}}\ and\ \bibinfo {author} {\bibfnamefont {G.}~\bibnamefont {Volpe}},\
  }\bibfield  {title} {\enquote {\bibinfo {title} {The topography of the
  environmkent alters the optimal search strategy for active particles},}\
  }\href@noop {} {\bibfield  {journal} {\bibinfo  {journal} {Proc. Natl. Acad.
  Sci.}\ }\textbf {\bibinfo {volume} {114}},\ \bibinfo {pages} {11350}
  (\bibinfo {year} {2017})}\BibitemShut {NoStop}%
\bibitem [{\citenamefont {Zeitz}\ \emph {et~al.}(2017)\citenamefont {Zeitz},
  \citenamefont {Wolff},\ and\ \citenamefont {Stark}}]{ZWS17}%
  \BibitemOpen
  \bibfield  {author} {\bibinfo {author} {\bibfnamefont {M.}~\bibnamefont
  {Zeitz}}, \bibinfo {author} {\bibfnamefont {K.}~\bibnamefont {Wolff}}, \ and\
  \bibinfo {author} {\bibfnamefont {H.}~\bibnamefont {Stark}},\ }\bibfield
  {title} {\enquote {\bibinfo {title} {Active {B}rownian particles moving in a
  random {L}orentz gas},}\ }\href@noop {} {\bibfield  {journal} {\bibinfo
  {journal} {Eur. Phys. J. E}\ }\textbf {\bibinfo {volume} {40}},\ \bibinfo
  {pages} {1} (\bibinfo {year} {2017})}\BibitemShut {NoStop}%
\bibitem [{\citenamefont {Giuggioli}(2020)}]{Giu20}%
  \BibitemOpen
  \bibfield  {author} {\bibinfo {author} {\bibfnamefont {L.}~\bibnamefont
  {Giuggioli}},\ }\bibfield  {title} {\enquote {\bibinfo {title} {Exact
  spatiotemporal dynamics of confined lattice random walks in arbitrary
  dimensions: A century after {S}moluchowski and {P}\'olya},}\ }\href@noop {}
  {\bibfield  {journal} {\bibinfo  {journal} {Phys. Rev. X}\ }\textbf {\bibinfo
  {volume} {10}},\ \bibinfo {pages} {021045} (\bibinfo {year}
  {2020})}\BibitemShut {NoStop}%
\bibitem [{\citenamefont {Palyulin}\ \emph
  {et~al.}(2014{\natexlab{b}})\citenamefont {Palyulin}, \citenamefont
  {Chechkin},\ and\ \citenamefont {Metzler}}]{PCM14b}%
  \BibitemOpen
  \bibfield  {author} {\bibinfo {author} {\bibfnamefont {V.~V.}\ \bibnamefont
  {Palyulin}}, \bibinfo {author} {\bibfnamefont {A.~V.}\ \bibnamefont
  {Chechkin}}, \ and\ \bibinfo {author} {\bibfnamefont {R.}~\bibnamefont
  {Metzler}},\ }\bibfield  {title} {\enquote {\bibinfo {title}
  {{Space-fractional Fokker–Planck equation and optimization of random search
  processes in the presence of an external bias}},}\ }\href@noop {} {\bibfield
  {journal} {\bibinfo  {journal} {J. Stat. Mech. Theory Exp.}\ }\textbf
  {\bibinfo {volume} {2014}},\ \bibinfo {pages} {11031} (\bibinfo {year}
  {2014}{\natexlab{b}})}\BibitemShut {NoStop}%
\bibitem [{\citenamefont {Palyulin}\ \emph {et~al.}(2016)\citenamefont
  {Palyulin}, \citenamefont {Chechkin}, \citenamefont {Klages},\ and\
  \citenamefont {Metzler}}]{PCKM16}%
  \BibitemOpen
  \bibfield  {author} {\bibinfo {author} {\bibfnamefont {V.~V.}\ \bibnamefont
  {Palyulin}}, \bibinfo {author} {\bibfnamefont {A.~V.}\ \bibnamefont
  {Chechkin}}, \bibinfo {author} {\bibfnamefont {R.}~\bibnamefont {Klages}}, \
  and\ \bibinfo {author} {\bibfnamefont {R.}~\bibnamefont {Metzler}},\
  }\bibfield  {title} {\enquote {\bibinfo {title} {{Search reliability and
  search efficiency of combined L{\'e}vy-{B}rownian motion: Long relocations
  mingled with thorough local exploration}},}\ }\href@noop {} {\bibfield
  {journal} {\bibinfo  {journal} {J. Phys. A Math. Theor.}\ }\textbf {\bibinfo
  {volume} {49}},\ \bibinfo {pages} {394002} (\bibinfo {year}
  {2016})}\BibitemShut {NoStop}%
\bibitem [{\citenamefont {Palyulin}\ \emph {et~al.}(2017)\citenamefont
  {Palyulin}, \citenamefont {Mantsevich}, \citenamefont {Klages}, \citenamefont
  {Metzler},\ and\ \citenamefont {Chechkin}}]{PMKMC17}%
  \BibitemOpen
  \bibfield  {author} {\bibinfo {author} {\bibfnamefont {V.~V.}\ \bibnamefont
  {Palyulin}}, \bibinfo {author} {\bibfnamefont {V.~N.}\ \bibnamefont
  {Mantsevich}}, \bibinfo {author} {\bibfnamefont {R.}~\bibnamefont {Klages}},
  \bibinfo {author} {\bibfnamefont {R.}~\bibnamefont {Metzler}}, \ and\
  \bibinfo {author} {\bibfnamefont {A.~V.}\ \bibnamefont {Chechkin}},\
  }\bibfield  {title} {\enquote {\bibinfo {title} {Comparison of pure and
  combined search strategies for single and multiple targets},}\ }\href@noop {}
  {\bibfield  {journal} {\bibinfo  {journal} {Eur. Phys. J. B}\ }\textbf
  {\bibinfo {volume} {90}},\ \bibinfo {pages} {170} (\bibinfo {year}
  {2017})}\BibitemShut {NoStop}%
\bibitem [{\citenamefont {Lomholt}\ \emph {et~al.}(2008)\citenamefont
  {Lomholt}, \citenamefont {Koren}, , \citenamefont {Metzler},\ and\
  \citenamefont {Klafter}}]{LKMK08}%
  \BibitemOpen
  \bibfield  {author} {\bibinfo {author} {\bibfnamefont {M.~A.}\ \bibnamefont
  {Lomholt}}, \bibinfo {author} {\bibfnamefont {T.}~\bibnamefont {Koren}}, ,
  \bibinfo {author} {\bibfnamefont {R.}~\bibnamefont {Metzler}}, \ and\
  \bibinfo {author} {\bibfnamefont {J.}~\bibnamefont {Klafter}},\ }\bibfield
  {title} {\enquote {\bibinfo {title} {L\'evy strategies in intermittent search
  processes are advantageous},}\ }\href@noop {} {\bibfield  {journal} {\bibinfo
   {journal} {Proc. Natl. Acad. Sci.}\ }\textbf {\bibinfo {volume} {105}},\
  \bibinfo {pages} {11055.} (\bibinfo {year} {2008})}\BibitemShut {NoStop}%
\bibitem [{\citenamefont {Bartumeus}\ \emph {et~al.}(2005)\citenamefont
  {Bartumeus}, \citenamefont {Da~Luz}, \citenamefont {Viswanathan},\ and\
  \citenamefont {Catalan}}]{bartu2}%
  \BibitemOpen
  \bibfield  {author} {\bibinfo {author} {\bibfnamefont {F.}~\bibnamefont
  {Bartumeus}}, \bibinfo {author} {\bibfnamefont {M.~G.~E.}\ \bibnamefont
  {Da~Luz}}, \bibinfo {author} {\bibfnamefont {G.~M.}\ \bibnamefont
  {Viswanathan}}, \ and\ \bibinfo {author} {\bibfnamefont {J.}~\bibnamefont
  {Catalan}},\ }\bibfield  {title} {\enquote {\bibinfo {title} {Animal search
  strategies: a quantitative random-walk analysis},}\ }\href@noop {} {\bibfield
   {journal} {\bibinfo  {journal} {Ecology}\ }\textbf {\bibinfo {volume}
  {86}},\ \bibinfo {pages} {3078} (\bibinfo {year} {2005})}\BibitemShut
  {NoStop}%
\bibitem [{\citenamefont {Bartumeus}\ and\ \citenamefont
  {Levin}(2008)}]{BaLe08}%
  \BibitemOpen
  \bibfield  {author} {\bibinfo {author} {\bibfnamefont {F.}~\bibnamefont
  {Bartumeus}}\ and\ \bibinfo {author} {\bibfnamefont {S.~A.}\ \bibnamefont
  {Levin}},\ }\bibfield  {title} {\enquote {\bibinfo {title} {Fractal
  reorientation clocks: Linking animal behavior to statistical patterns of
  search},}\ }\href@noop {} {\bibfield  {journal} {\bibinfo  {journal} {Proc.
  Natl. Acad. Sci.}\ }\textbf {\bibinfo {volume} {105}},\ \bibinfo {pages}
  {19072} (\bibinfo {year} {2008})}\BibitemShut {NoStop}%
\bibitem [{\citenamefont {Bartumeus}\ \emph {et~al.}(2014)\citenamefont
  {Bartumeus}, \citenamefont {Raposo}, \citenamefont {Viswanathan},\ and\
  \citenamefont {{Da Luz}}}]{BRVL14}%
  \BibitemOpen
  \bibfield  {author} {\bibinfo {author} {\bibfnamefont {F.}~\bibnamefont
  {Bartumeus}}, \bibinfo {author} {\bibfnamefont {E.~P.}\ \bibnamefont
  {Raposo}}, \bibinfo {author} {\bibfnamefont {G.~M.}\ \bibnamefont
  {Viswanathan}}, \ and\ \bibinfo {author} {\bibfnamefont {M.~G.~E.}\
  \bibnamefont {{Da Luz}}},\ }\bibfield  {title} {\enquote {\bibinfo {title}
  {Stochastic optimal foraging: Tuning intensive and extensive dynamics in
  random searches},}\ }\href@noop {} {\bibfield  {journal} {\bibinfo  {journal}
  {PLoS One}\ }\textbf {\bibinfo {volume} {9}},\ \bibinfo {pages} {106373}
  (\bibinfo {year} {2014})}\BibitemShut {NoStop}%
\bibitem [{\citenamefont {Klafter}\ and\ \citenamefont
  {Sokolov}(2011)}]{KlSo11}%
  \BibitemOpen
  \bibfield  {author} {\bibinfo {author} {\bibfnamefont {J.}~\bibnamefont
  {Klafter}}\ and\ \bibinfo {author} {\bibfnamefont {I.~M.}\ \bibnamefont
  {Sokolov}},\ }\href@noop {} {\emph {\bibinfo {title} {First Steps in Random
  Walks: From Tools to Applications}}}\ (\bibinfo  {publisher} {Oxford
  University Press},\ \bibinfo {address} {Oxford},\ \bibinfo {year}
  {2011})\BibitemShut {NoStop}%
\bibitem [{\citenamefont {Kolmogorov}(1940)}]{kolmogorov1940wiener}%
  \BibitemOpen
  \bibfield  {author} {\bibinfo {author} {\bibfnamefont {A.~N.}\ \bibnamefont
  {Kolmogorov}},\ }\bibfield  {title} {\enquote {\bibinfo {title} {The wiener
  spiral and some other interesting curves in hilbert space},}\ }\bibfield
  {booktitle} {\emph {\bibinfo {booktitle} {Dokl. Akad. Nauk SSSR}},\
  }\href@noop {} {\ \textbf {\bibinfo {volume} {26}},\ \bibinfo {pages} {115}
  (\bibinfo {year} {1940})}\BibitemShut {NoStop}%
\bibitem [{\citenamefont {Yaglom}(1958)}]{Yag58}%
  \BibitemOpen
  \bibfield  {author} {\bibinfo {author} {\bibfnamefont {A.M.}\ \bibnamefont
  {Yaglom}},\ }\bibfield  {title} {\enquote {\bibinfo {title} {Correlation
  theory of processes with random stationary nth increments},}\ }\href@noop {}
  {\bibfield  {journal} {\bibinfo  {journal} {Am. Math. Soc. Transl.}\ }\textbf
  {\bibinfo {volume} {8}},\ \bibinfo {pages} {87} (\bibinfo {year}
  {1958})}\BibitemShut {NoStop}%
\bibitem [{\citenamefont {Embrechts}(2009)}]{embrechts2009selfsimilar}%
  \BibitemOpen
  \bibfield  {author} {\bibinfo {author} {\bibfnamefont {P.}~\bibnamefont
  {Embrechts}},\ }\href@noop {} {\emph {\bibinfo {title} {Selfsimilar
  processes}}}\ (\bibinfo  {publisher} {Princeton University Press},\ \bibinfo
  {year} {2009})\BibitemShut {NoStop}%
\bibitem [{\citenamefont {Jeon}\ and\ \citenamefont {Metzler}(2010)}]{jeon1}%
  \BibitemOpen
  \bibfield  {author} {\bibinfo {author} {\bibfnamefont {J.-H.}\ \bibnamefont
  {Jeon}}\ and\ \bibinfo {author} {\bibfnamefont {R.}~\bibnamefont {Metzler}},\
  }\bibfield  {title} {\enquote {\bibinfo {title} {Fractional {B}rownian motion
  and motion governed by the fractional {L}angevin equation in confined
  geometries},}\ }\href@noop {} {\bibfield  {journal} {\bibinfo  {journal}
  {Phys. Rev. E}\ }\textbf {\bibinfo {volume} {81}},\ \bibinfo {pages} {021103}
  (\bibinfo {year} {2010})}\BibitemShut {NoStop}%
\bibitem [{\citenamefont {Burov}\ \emph {et~al.}(2011)\citenamefont {Burov},
  \citenamefont {Jeon}, \citenamefont {Metzler},\ and\ \citenamefont
  {Barkai}}]{burov2011single}%
  \BibitemOpen
  \bibfield  {author} {\bibinfo {author} {\bibfnamefont {S.}~\bibnamefont
  {Burov}}, \bibinfo {author} {\bibfnamefont {J.-H.}\ \bibnamefont {Jeon}},
  \bibinfo {author} {\bibfnamefont {R.}~\bibnamefont {Metzler}}, \ and\
  \bibinfo {author} {\bibfnamefont {E.}~\bibnamefont {Barkai}},\ }\bibfield
  {title} {\enquote {\bibinfo {title} {Single particle tracking in systems
  showing anomalous diffusion: the role of weak ergodicity breaking},}\
  }\href@noop {} {\bibfield  {journal} {\bibinfo  {journal} {Phys. Chem. Chem.
  Phys.}\ }\textbf {\bibinfo {volume} {13}},\ \bibinfo {pages} {1800} (\bibinfo
  {year} {2011})}\BibitemShut {NoStop}%
\bibitem [{\citenamefont {Fuli{\'n}ski}(2017)}]{fuli}%
  \BibitemOpen
  \bibfield  {author} {\bibinfo {author} {\bibfnamefont {A.}~\bibnamefont
  {Fuli{\'n}ski}},\ }\bibfield  {title} {\enquote {\bibinfo {title} {Fractional
  {B}rownian motions: memory, diffusion velocity, and correlation functions},}\
  }\href@noop {} {\bibfield  {journal} {\bibinfo  {journal} {J. Phys. A-Math.
  Theor.}\ }\textbf {\bibinfo {volume} {50}},\ \bibinfo {pages} {054002}
  (\bibinfo {year} {2017})}\BibitemShut {NoStop}%
\bibitem [{\citenamefont {Goychuk}(2012)}]{goy}%
  \BibitemOpen
  \bibfield  {author} {\bibinfo {author} {\bibfnamefont {I.}~\bibnamefont
  {Goychuk}},\ }\bibfield  {title} {\enquote {\bibinfo {title} {Viscoelastic
  subdiffusion: Generalized langevin equation approach},}\ }\href@noop {}
  {\bibfield  {journal} {\bibinfo  {journal} {Adv. Chem. Phys.}\ }\textbf
  {\bibinfo {volume} {150}},\ \bibinfo {pages} {187} (\bibinfo {year}
  {2012})}\BibitemShut {NoStop}%
\bibitem [{\citenamefont {Weiss}(2013)}]{weis1}%
  \BibitemOpen
  \bibfield  {author} {\bibinfo {author} {\bibfnamefont {M.}~\bibnamefont
  {Weiss}},\ }\bibfield  {title} {\enquote {\bibinfo {title} {Single-particle
  tracking data reveal anticorrelated fractional {B}rownian motion in crowded
  fluids},}\ }\href@noop {} {\bibfield  {journal} {\bibinfo  {journal} {Phys.
  Rev. E}\ }\textbf {\bibinfo {volume} {88}},\ \bibinfo {pages} {010101}
  (\bibinfo {year} {2013})}\BibitemShut {NoStop}%
\bibitem [{\citenamefont {Khadem}\ \emph {et~al.}(2016)\citenamefont {Khadem},
  \citenamefont {Hille}, \citenamefont {L{\"o}hmannsr{\"o}ben},\ and\
  \citenamefont {Sokolov}}]{khadem}%
  \BibitemOpen
  \bibfield  {author} {\bibinfo {author} {\bibfnamefont {S.~M..J.}\
  \bibnamefont {Khadem}}, \bibinfo {author} {\bibfnamefont {C.}~\bibnamefont
  {Hille}}, \bibinfo {author} {\bibfnamefont {H.-G.}\ \bibnamefont
  {L{\"o}hmannsr{\"o}ben}}, \ and\ \bibinfo {author} {\bibfnamefont {I.~M.}\
  \bibnamefont {Sokolov}},\ }\bibfield  {title} {\enquote {\bibinfo {title}
  {What information is contained in the fluorescence correlation spectroscopy
  curves, and where},}\ }\href@noop {} {\bibfield  {journal} {\bibinfo
  {journal} {Phys. Rev. E}\ }\textbf {\bibinfo {volume} {94}},\ \bibinfo
  {pages} {022407} (\bibinfo {year} {2016})}\BibitemShut {NoStop}%
\bibitem [{\citenamefont {Szymanski}\ and\ \citenamefont
  {Weiss}(2009)}]{szymanski2009elucidating}%
  \BibitemOpen
  \bibfield  {author} {\bibinfo {author} {\bibfnamefont {J.}~\bibnamefont
  {Szymanski}}\ and\ \bibinfo {author} {\bibfnamefont {M.}~\bibnamefont
  {Weiss}},\ }\bibfield  {title} {\enquote {\bibinfo {title} {Elucidating the
  origin of anomalous diffusion in crowded fluids},}\ }\href@noop {} {\bibfield
   {journal} {\bibinfo  {journal} {Phys. Rev. Lett.}\ }\textbf {\bibinfo
  {volume} {103}},\ \bibinfo {pages} {038102} (\bibinfo {year}
  {2009})}\BibitemShut {NoStop}%
\bibitem [{\citenamefont {Jeon}\ \emph
  {et~al.}(2011{\natexlab{a}})\citenamefont {Jeon}, \citenamefont {Tejedor},
  \citenamefont {Burov}, \citenamefont {Barkai}, \citenamefont
  {Selhuber-Unkel}, \citenamefont {Berg-S{\o}rensen}, \citenamefont
  {Oddershede},\ and\ \citenamefont {Metzler}}]{jeonvivi}%
  \BibitemOpen
  \bibfield  {author} {\bibinfo {author} {\bibfnamefont {J.-H.}\ \bibnamefont
  {Jeon}}, \bibinfo {author} {\bibfnamefont {V.}~\bibnamefont {Tejedor}},
  \bibinfo {author} {\bibfnamefont {S.}~\bibnamefont {Burov}}, \bibinfo
  {author} {\bibfnamefont {E.}~\bibnamefont {Barkai}}, \bibinfo {author}
  {\bibfnamefont {C.}~\bibnamefont {Selhuber-Unkel}}, \bibinfo {author}
  {\bibfnamefont {K.}~\bibnamefont {Berg-S{\o}rensen}}, \bibinfo {author}
  {\bibfnamefont {L.}~\bibnamefont {Oddershede}}, \ and\ \bibinfo {author}
  {\bibfnamefont {R.}~\bibnamefont {Metzler}},\ }\bibfield  {title} {\enquote
  {\bibinfo {title} {In vivo anomalous diffusion and weak ergodicity breaking
  of lipid granules},}\ }\href@noop {} {\bibfield  {journal} {\bibinfo
  {journal} {Phys. Rev. Lett.}\ }\textbf {\bibinfo {volume} {106}},\ \bibinfo
  {pages} {048103} (\bibinfo {year} {2011}{\natexlab{a}})}\BibitemShut
  {NoStop}%
\bibitem [{\citenamefont {Weiss}(2008)}]{weisp}%
  \BibitemOpen
  \bibfield  {author} {\bibinfo {author} {\bibfnamefont {M.}~\bibnamefont
  {Weiss}},\ }\bibfield  {title} {\enquote {\bibinfo {title} {Probing the
  interior of living cells with fluorescence correlation spectroscopy},}\
  }\href@noop {} {\bibfield  {journal} {\bibinfo  {journal} {Ann. NY. Acad.
  Sci}\ }\textbf {\bibinfo {volume} {1130}},\ \bibinfo {pages} {21} (\bibinfo
  {year} {2008})}\BibitemShut {NoStop}%
\bibitem [{\citenamefont {Weiss}\ \emph {et~al.}(2004)\citenamefont {Weiss},
  \citenamefont {Elsner}, \citenamefont {Kartberg},\ and\ \citenamefont
  {Nilsson}}]{weiss2004anomalous}%
  \BibitemOpen
  \bibfield  {author} {\bibinfo {author} {\bibfnamefont {M.}~\bibnamefont
  {Weiss}}, \bibinfo {author} {\bibfnamefont {M.}~\bibnamefont {Elsner}},
  \bibinfo {author} {\bibfnamefont {F.}~\bibnamefont {Kartberg}}, \ and\
  \bibinfo {author} {\bibfnamefont {T.}~\bibnamefont {Nilsson}},\ }\bibfield
  {title} {\enquote {\bibinfo {title} {Anomalous subdiffusion is a measure for
  cytoplasmic crowding in living cells},}\ }\href@noop {} {\bibfield  {journal}
  {\bibinfo  {journal} {Biophys. J.}\ }\textbf {\bibinfo {volume} {87}},\
  \bibinfo {pages} {3518} (\bibinfo {year} {2004})}\BibitemShut {NoStop}%
\bibitem [{\citenamefont {Wiese}(2019)}]{Wiese19}%
  \BibitemOpen
  \bibfield  {author} {\bibinfo {author} {\bibfnamefont {K.~J.}\ \bibnamefont
  {Wiese}},\ }\bibfield  {title} {\enquote {\bibinfo {title} {First passage in
  an interval for fractional {B}rownian motion},}\ }\href@noop {} {\bibfield
  {journal} {\bibinfo  {journal} {Phys. Rev. E}\ }\textbf {\bibinfo {volume}
  {99}},\ \bibinfo {pages} {032106} (\bibinfo {year} {2019})}\BibitemShut
  {NoStop}%
\bibitem [{\citenamefont {Reynolds}(2009)}]{Rey09}%
  \BibitemOpen
  \bibfield  {author} {\bibinfo {author} {\bibfnamefont {A.~M.}\ \bibnamefont
  {Reynolds}},\ }\bibfield  {title} {\enquote {\bibinfo {title} {Scale-free
  animal movement patterns: L{\'{e}}vy walks outperform fractional {B}rownian
  motions and fractional l{\'{e}}vy motions in random search scenarios},}\
  }\href@noop {} {\bibfield  {journal} {\bibinfo  {journal} {J. Phys. A: Math.
  Theor.}\ }\textbf {\bibinfo {volume} {42}},\ \bibinfo {pages} {434006}
  (\bibinfo {year} {2009})}\BibitemShut {NoStop}%
\bibitem [{\citenamefont {Nava-Sede{\~{n}}o}\ \emph {et~al.}(2017)\citenamefont
  {Nava-Sede{\~{n}}o}, \citenamefont {Hatzikirou}, \citenamefont {Klages},\
  and\ \citenamefont {Deutsch}}]{NHKD17}%
  \BibitemOpen
  \bibfield  {author} {\bibinfo {author} {\bibfnamefont {J.~M.}\ \bibnamefont
  {Nava-Sede{\~{n}}o}}, \bibinfo {author} {\bibfnamefont {H.}~\bibnamefont
  {Hatzikirou}}, \bibinfo {author} {\bibfnamefont {R.}~\bibnamefont {Klages}},
  \ and\ \bibinfo {author} {\bibfnamefont {A.}~\bibnamefont {Deutsch}},\
  }\bibfield  {title} {\enquote {\bibinfo {title} {Cellular automaton models
  for time-correlated random walks: Derivation and analysis},}\ }\href@noop {}
  {\bibfield  {journal} {\bibinfo  {journal} {Sci. Rep.}\ }\textbf {\bibinfo
  {volume} {7}},\ \bibinfo {pages} {16952} (\bibinfo {year}
  {2017})}\BibitemShut {NoStop}%
\bibitem [{\citenamefont {Estrada}(2020)}]{Estr20}%
  \BibitemOpen
  \bibfield  {author} {\bibinfo {author} {\bibfnamefont {E.}~\bibnamefont
  {Estrada}},\ }\bibfield  {title} {\enquote {\bibinfo {title} {Analyzing the
  impact of sars cov-2 on the human proteome},}\ }\href@noop {} {\  (\bibinfo
  {year} {2020})}\BibitemShut {NoStop}%
\bibitem [{\citenamefont {Molchan}(1999)}]{Molch99}%
  \BibitemOpen
  \bibfield  {author} {\bibinfo {author} {\bibfnamefont {G.~M.}\ \bibnamefont
  {Molchan}},\ }\bibfield  {title} {\enquote {\bibinfo {title} {{Maximum of a
  fractional {B}rownian motion: Probabilities of small values}},}\ }\href@noop
  {} {\bibfield  {journal} {\bibinfo  {journal} {Commun. Math. Phys.}\ }\textbf
  {\bibinfo {volume} {205}},\ \bibinfo {pages} {97} (\bibinfo {year}
  {1999})}\BibitemShut {NoStop}%
\bibitem [{\citenamefont {Guigas}\ and\ \citenamefont {Weiss}(2008)}]{GuWe08}%
  \BibitemOpen
  \bibfield  {author} {\bibinfo {author} {\bibfnamefont {G.}~\bibnamefont
  {Guigas}}\ and\ \bibinfo {author} {\bibfnamefont {M.}~\bibnamefont {Weiss}},\
  }\bibfield  {title} {\enquote {\bibinfo {title} {{Sampling the cell with
  anomalous diffusion - The discovery of slowness}},}\ }\href@noop {}
  {\bibfield  {journal} {\bibinfo  {journal} {Biophys. J.}\ }\textbf {\bibinfo
  {volume} {94}},\ \bibinfo {pages} {90} (\bibinfo {year} {2008})}\BibitemShut
  {NoStop}%
\bibitem [{\citenamefont {Jeon}\ \emph
  {et~al.}(2011{\natexlab{b}})\citenamefont {Jeon}, \citenamefont {Chechkin},\
  and\ \citenamefont {Metzler}}]{JCM11}%
  \BibitemOpen
  \bibfield  {author} {\bibinfo {author} {\bibfnamefont {J.~H.}\ \bibnamefont
  {Jeon}}, \bibinfo {author} {\bibfnamefont {A.~V.}\ \bibnamefont {Chechkin}},
  \ and\ \bibinfo {author} {\bibfnamefont {R.}~\bibnamefont {Metzler}},\
  }\bibfield  {title} {\enquote {\bibinfo {title} {{First passage behaviour of
  fractional {B}rownian motion in two-dimensional wedge domains}},}\
  }\href@noop {} {\bibfield  {journal} {\bibinfo  {journal} {Europhys. Lett.}\
  }\textbf {\bibinfo {volume} {94}},\ \bibinfo {pages} {20008} (\bibinfo {year}
  {2011}{\natexlab{b}})}\BibitemShut {NoStop}%
\bibitem [{\citenamefont {Sanders}\ and\ \citenamefont
  {Ambjörnsson}(2012)}]{SaAm12}%
  \BibitemOpen
  \bibfield  {author} {\bibinfo {author} {\bibfnamefont {L.~P.}\ \bibnamefont
  {Sanders}}\ and\ \bibinfo {author} {\bibfnamefont {T.}~\bibnamefont
  {Ambjörnsson}},\ }\bibfield  {title} {\enquote {\bibinfo {title} {First
  passage times for a tracer particle in single file diffusion and fractional
  {B}rownian motion},}\ }\href@noop {} {\bibfield  {journal} {\bibinfo
  {journal} {J. Chem. Phys.}\ }\textbf {\bibinfo {volume} {136}},\ \bibinfo
  {pages} {175103} (\bibinfo {year} {2012})}\BibitemShut {NoStop}%
\bibitem [{\citenamefont {Gu{\'e}rin}\ \emph {et~al.}(2016)\citenamefont
  {Gu{\'e}rin}, \citenamefont {Levernier}, \citenamefont {B{\'e}nichou},\ and\
  \citenamefont {Voituriez}}]{guerin}%
  \BibitemOpen
  \bibfield  {author} {\bibinfo {author} {\bibfnamefont {T.}~\bibnamefont
  {Gu{\'e}rin}}, \bibinfo {author} {\bibfnamefont {N.}~\bibnamefont
  {Levernier}}, \bibinfo {author} {\bibfnamefont {O.}~\bibnamefont
  {B{\'e}nichou}}, \ and\ \bibinfo {author} {\bibfnamefont {R.}~\bibnamefont
  {Voituriez}},\ }\bibfield  {title} {\enquote {\bibinfo {title} {Mean
  first-passage times of non-markovian random walkers in confinement},}\
  }\href@noop {} {\bibfield  {journal} {\bibinfo  {journal} {Nature}\ }\textbf
  {\bibinfo {volume} {534}},\ \bibinfo {pages} {356} (\bibinfo {year}
  {2016})}\BibitemShut {NoStop}%
\bibitem [{\citenamefont {Romanczuk}\ \emph {et~al.}(2012)\citenamefont
  {Romanczuk}, \citenamefont {B{\"a}r}, \citenamefont {Ebeling}, \citenamefont
  {Lindner},\ and\ \citenamefont {Schimansky-Geier}}]{RBELS12}%
  \BibitemOpen
  \bibfield  {author} {\bibinfo {author} {\bibfnamefont {P.}~\bibnamefont
  {Romanczuk}}, \bibinfo {author} {\bibfnamefont {M.}~\bibnamefont {B{\"a}r}},
  \bibinfo {author} {\bibfnamefont {W.}~\bibnamefont {Ebeling}}, \bibinfo
  {author} {\bibfnamefont {B.}~\bibnamefont {Lindner}}, \ and\ \bibinfo
  {author} {\bibfnamefont {L.}~\bibnamefont {Schimansky-Geier}},\ }\bibfield
  {title} {\enquote {\bibinfo {title} {Active {B}rownian particles},}\
  }\href@noop {} {\bibfield  {journal} {\bibinfo  {journal} {Eur. Phys. J.
  Spec. Top.}\ }\textbf {\bibinfo {volume} {202}},\ \bibinfo {pages} {1}
  (\bibinfo {year} {2012})}\BibitemShut {NoStop}%
\bibitem [{\citenamefont {Bechinger}\ \emph {et~al.}(2016)\citenamefont
  {Bechinger}, \citenamefont {Di~Leonardo}, \citenamefont {L\"owen},
  \citenamefont {Reichhardt}, \citenamefont {Volpe},\ and\ \citenamefont
  {Volpe}}]{BeDiL16}%
  \BibitemOpen
  \bibfield  {author} {\bibinfo {author} {\bibfnamefont {C.}~\bibnamefont
  {Bechinger}}, \bibinfo {author} {\bibfnamefont {R.}~\bibnamefont
  {Di~Leonardo}}, \bibinfo {author} {\bibfnamefont {H.}~\bibnamefont
  {L\"owen}}, \bibinfo {author} {\bibfnamefont {C.}~\bibnamefont {Reichhardt}},
  \bibinfo {author} {\bibfnamefont {G.}~\bibnamefont {Volpe}}, \ and\ \bibinfo
  {author} {\bibfnamefont {G.}~\bibnamefont {Volpe}},\ }\bibfield  {title}
  {\enquote {\bibinfo {title} {Active particles in complex and crowded
  environments},}\ }\href@noop {} {\bibfield  {journal} {\bibinfo  {journal}
  {Rev. Mod. Phys.}\ }\textbf {\bibinfo {volume} {88}},\ \bibinfo {pages}
  {045006} (\bibinfo {year} {2016})}\BibitemShut {NoStop}%
\bibitem [{\citenamefont {Duzgun}\ and\ \citenamefont
  {Selinger}(2018)}]{duzgun2018active}%
  \BibitemOpen
  \bibfield  {author} {\bibinfo {author} {\bibfnamefont {A.}~\bibnamefont
  {Duzgun}}\ and\ \bibinfo {author} {\bibfnamefont {J.~V.}\ \bibnamefont
  {Selinger}},\ }\bibfield  {title} {\enquote {\bibinfo {title} {Active
  {B}rownian particles near straight or curved walls: Pressure and boundary
  layers},}\ }\href@noop {} {\bibfield  {journal} {\bibinfo  {journal} {Phys.
  Rev. E}\ }\textbf {\bibinfo {volume} {97}},\ \bibinfo {pages} {032606}
  (\bibinfo {year} {2018})}\BibitemShut {NoStop}%
\bibitem [{\citenamefont {Das}\ \emph {et~al.}(2015)\citenamefont {Das},
  \citenamefont {Garg}, \citenamefont {Campbell}, \citenamefont {Howse},
  \citenamefont {Sen}, \citenamefont {Velegol}, \citenamefont {Golestanian},\
  and\ \citenamefont {Ebbens}}]{das2015boundaries}%
  \BibitemOpen
  \bibfield  {author} {\bibinfo {author} {\bibfnamefont {S.}~\bibnamefont
  {Das}}, \bibinfo {author} {\bibfnamefont {A.}~\bibnamefont {Garg}}, \bibinfo
  {author} {\bibfnamefont {A.~I.}\ \bibnamefont {Campbell}}, \bibinfo {author}
  {\bibfnamefont {J.}~\bibnamefont {Howse}}, \bibinfo {author} {\bibfnamefont
  {A.}~\bibnamefont {Sen}}, \bibinfo {author} {\bibfnamefont {D.}~\bibnamefont
  {Velegol}}, \bibinfo {author} {\bibfnamefont {R.}~\bibnamefont
  {Golestanian}}, \ and\ \bibinfo {author} {\bibfnamefont {S.~J.}\ \bibnamefont
  {Ebbens}},\ }\bibfield  {title} {\enquote {\bibinfo {title} {Boundaries can
  steer active {J}anus spheres},}\ }\href@noop {} {\bibfield  {journal}
  {\bibinfo  {journal} {Nat. Commun.}\ }\textbf {\bibinfo {volume} {6}},\
  \bibinfo {pages} {8999} (\bibinfo {year} {2015})}\BibitemShut {NoStop}%
\bibitem [{\citenamefont {Elgeti}\ and\ \citenamefont
  {Gompper}(2015)}]{elgeti2015run}%
  \BibitemOpen
  \bibfield  {author} {\bibinfo {author} {\bibfnamefont {J.}~\bibnamefont
  {Elgeti}}\ and\ \bibinfo {author} {\bibfnamefont {G.}~\bibnamefont
  {Gompper}},\ }\bibfield  {title} {\enquote {\bibinfo {title} {Run-and-tumble
  dynamics of self-propelled particles in confinement},}\ }\href@noop {}
  {\bibfield  {journal} {\bibinfo  {journal} {Europhys. Lett.}\ }\textbf
  {\bibinfo {volume} {109}},\ \bibinfo {pages} {58003} (\bibinfo {year}
  {2015})}\BibitemShut {NoStop}%
\bibitem [{\citenamefont {Kaiser}\ \emph {et~al.}(2012)\citenamefont {Kaiser},
  \citenamefont {Wensink},\ and\ \citenamefont
  {L{\"o}wen}}]{kaiser2012capture}%
  \BibitemOpen
  \bibfield  {author} {\bibinfo {author} {\bibfnamefont {A.}~\bibnamefont
  {Kaiser}}, \bibinfo {author} {\bibfnamefont {H.~H.}\ \bibnamefont {Wensink}},
  \ and\ \bibinfo {author} {\bibfnamefont {H.}~\bibnamefont {L{\"o}wen}},\
  }\bibfield  {title} {\enquote {\bibinfo {title} {How to capture active
  particles},}\ }\href@noop {} {\bibfield  {journal} {\bibinfo  {journal}
  {Phys. Rev. Lett.}\ }\textbf {\bibinfo {volume} {108}},\ \bibinfo {pages}
  {268307} (\bibinfo {year} {2012})}\BibitemShut {NoStop}%
\bibitem [{\citenamefont {Guggenberger}\ \emph {et~al.}(2019)\citenamefont
  {Guggenberger}, \citenamefont {Pagnini}, \citenamefont {Vojta},\ and\
  \citenamefont {Metzler}}]{gugg}%
  \BibitemOpen
  \bibfield  {author} {\bibinfo {author} {\bibfnamefont {T.}~\bibnamefont
  {Guggenberger}}, \bibinfo {author} {\bibfnamefont {G.}~\bibnamefont
  {Pagnini}}, \bibinfo {author} {\bibfnamefont {T.}~\bibnamefont {Vojta}}, \
  and\ \bibinfo {author} {\bibfnamefont {R.}~\bibnamefont {Metzler}},\
  }\bibfield  {title} {\enquote {\bibinfo {title} {Fractional {B}rownian motion
  in a finite interval: correlations effect depletion or accretion zones of
  particles near boundaries},}\ }\href@noop {} {\bibfield  {journal} {\bibinfo
  {journal} {New J. Phys.}\ }\textbf {\bibinfo {volume} {21}},\ \bibinfo
  {pages} {022002} (\bibinfo {year} {2019})}\BibitemShut {NoStop}%
\bibitem [{\citenamefont {Vojta}\ \emph {et~al.}(2020)\citenamefont {Vojta},
  \citenamefont {Halladay}, \citenamefont {Skinner}, \citenamefont
  {Janu\ifmmode~\check{s}\else \v{s}\fi{}onis}, \citenamefont {Guggenberger},\
  and\ \citenamefont {Metzler}}]{VHSJGM20}%
  \BibitemOpen
  \bibfield  {author} {\bibinfo {author} {\bibfnamefont {T.}~\bibnamefont
  {Vojta}}, \bibinfo {author} {\bibfnamefont {S.}~\bibnamefont {Halladay}},
  \bibinfo {author} {\bibfnamefont {S.}~\bibnamefont {Skinner}}, \bibinfo
  {author} {\bibfnamefont {Sk.}\ \bibnamefont {Janu\ifmmode~\check{s}\else
  \v{s}\fi{}onis}}, \bibinfo {author} {\bibfnamefont {T.}~\bibnamefont
  {Guggenberger}}, \ and\ \bibinfo {author} {\bibfnamefont {R.}~\bibnamefont
  {Metzler}},\ }\bibfield  {title} {\enquote {\bibinfo {title} {Reflected
  fractional {B}rownian motion in one and higher dimensions},}\ }\href@noop {}
  {\bibfield  {journal} {\bibinfo  {journal} {Phys. Rev. E}\ }\textbf {\bibinfo
  {volume} {102}},\ \bibinfo {pages} {032108} (\bibinfo {year}
  {2020})}\BibitemShut {NoStop}%
\bibitem [{\citenamefont {Zaburdaev}\ \emph {et~al.}(2016)\citenamefont
  {Zaburdaev}, \citenamefont {Fouxon}, \citenamefont {Denisov},\ and\
  \citenamefont {Barkai}}]{ZFDB16}%
  \BibitemOpen
  \bibfield  {author} {\bibinfo {author} {\bibfnamefont {V.}~\bibnamefont
  {Zaburdaev}}, \bibinfo {author} {\bibfnamefont {I.}~\bibnamefont {Fouxon}},
  \bibinfo {author} {\bibfnamefont {S.}~\bibnamefont {Denisov}}, \ and\
  \bibinfo {author} {\bibfnamefont {E.}~\bibnamefont {Barkai}},\ }\bibfield
  {title} {\enquote {\bibinfo {title} {Superdiffusive dispersals impart the
  geometry of underlying random walks},}\ }\href@noop {} {\bibfield  {journal}
  {\bibinfo  {journal} {Phys. Rev. Lett.}\ }\textbf {\bibinfo {volume} {117}},\
  \bibinfo {pages} {270601} (\bibinfo {year} {2016})}\BibitemShut {NoStop}%
\bibitem [{\citenamefont {Rangarajan}\ \emph {et~al.}(2003)\citenamefont
  {Rangarajan}, \citenamefont {Ding},\ and\ \citenamefont
  {Ding}}]{rangarajan2003processes}%
  \BibitemOpen
  \bibfield  {author} {\bibinfo {author} {\bibfnamefont {G.}~\bibnamefont
  {Rangarajan}}, \bibinfo {author} {\bibfnamefont {M.}~\bibnamefont {Ding}}, \
  and\ \bibinfo {author} {\bibfnamefont {M.}~\bibnamefont {Ding}},\ }\href@noop
  {} {\emph {\bibinfo {title} {Processes with long-range correlations: Theory
  and applications}}}\ (\bibinfo  {publisher} {Springer Science \& Business
  Media},\ \bibinfo {year} {2003})\BibitemShut {NoStop}%
\bibitem [{\citenamefont {Fodor}\ \emph {et~al.}(2016)\citenamefont {Fodor},
  \citenamefont {Nardini}, \citenamefont {Cates}, \citenamefont {Tailleur},
  \citenamefont {Visco},\ and\ \citenamefont {van Wijland}}]{FNCTVW16}%
  \BibitemOpen
  \bibfield  {author} {\bibinfo {author} {\bibfnamefont {{\'E}.}~\bibnamefont
  {Fodor}}, \bibinfo {author} {\bibfnamefont {C.}~\bibnamefont {Nardini}},
  \bibinfo {author} {\bibfnamefont {M.E.}\ \bibnamefont {Cates}}, \bibinfo
  {author} {\bibfnamefont {J.}~\bibnamefont {Tailleur}}, \bibinfo {author}
  {\bibfnamefont {P.}~\bibnamefont {Visco}}, \ and\ \bibinfo {author}
  {\bibfnamefont {F.}~\bibnamefont {van Wijland}},\ }\bibfield  {title}
  {\enquote {\bibinfo {title} {How far from equilibrium is active matter?}}\
  }\href@noop {} {\bibfield  {journal} {\bibinfo  {journal} {Phys. Rev. Lett.}\
  }\textbf {\bibinfo {volume} {117}},\ \bibinfo {pages} {038103} (\bibinfo
  {year} {2016})}\BibitemShut {NoStop}%
\bibitem [{\citenamefont {Um}\ \emph {et~al.}(2019)\citenamefont {Um},
  \citenamefont {Song},\ and\ \citenamefont {Jeon}}]{USJ19}%
  \BibitemOpen
  \bibfield  {author} {\bibinfo {author} {\bibfnamefont {J.}~\bibnamefont
  {Um}}, \bibinfo {author} {\bibfnamefont {T.}~\bibnamefont {Song}}, \ and\
  \bibinfo {author} {\bibfnamefont {J.-H.}\ \bibnamefont {Jeon}},\ }\bibfield
  {title} {\enquote {\bibinfo {title} {Langevin dynamics driven by a
  telegraphic active noise},}\ }\href@noop {} {\bibfield  {journal} {\bibinfo
  {journal} {Front. Phys.}\ }\textbf {\bibinfo {volume} {7}},\ \bibinfo {pages}
  {143} (\bibinfo {year} {2019})}\BibitemShut {NoStop}%
\bibitem [{\citenamefont {Chechkin}\ \emph {et~al.}(2012)\citenamefont
  {Chechkin}, \citenamefont {Lenz},\ and\ \citenamefont {Klages}}]{ChKl12}%
  \BibitemOpen
  \bibfield  {author} {\bibinfo {author} {\bibfnamefont {A.V.}\ \bibnamefont
  {Chechkin}}, \bibinfo {author} {\bibfnamefont {F.}~\bibnamefont {Lenz}}, \
  and\ \bibinfo {author} {\bibfnamefont {R.}~\bibnamefont {Klages}},\
  }\bibfield  {title} {\enquote {\bibinfo {title} {Normal and anomalous
  fluctuation relations for {G}aussian stochastic dynamics},}\ }\href@noop {}
  {\bibfield  {journal} {\bibinfo  {journal} {J. Stat. Mech.: Theor. Exp.}\
  }\textbf {\bibinfo {volume} {2012}},\ \bibinfo {pages} {L11001} (\bibinfo
  {year} {2012})}\BibitemShut {NoStop}%
\bibitem [{\citenamefont {Hosking}(1984)}]{hosking}%
  \BibitemOpen
  \bibfield  {author} {\bibinfo {author} {\bibfnamefont {J.~R.~M.}\
  \bibnamefont {Hosking}},\ }\bibfield  {title} {\enquote {\bibinfo {title}
  {Modeling persistence in hydrological time series using fractional
  differencing},}\ }\href@noop {} {\bibfield  {journal} {\bibinfo  {journal}
  {Water Resour. Res.}\ }\textbf {\bibinfo {volume} {20}},\ \bibinfo {pages}
  {1898} (\bibinfo {year} {1984})}\BibitemShut {NoStop}%
\bibitem [{\citenamefont {Asmussen}(1998)}]{asmus}%
  \BibitemOpen
  \bibfield  {author} {\bibinfo {author} {\bibfnamefont {S.}~\bibnamefont
  {Asmussen}},\ }\href@noop {} {\emph {\bibinfo {title} {Stochastic simulation
  with a view towards stochastic processes}}}\ (\bibinfo  {publisher}
  {University of Aarhus, Centre for Mathematical Physics and Stochastics},\
  \bibinfo {year} {1998})\BibitemShut {NoStop}%
\bibitem [{\citenamefont {Davies}\ and\ \citenamefont {Harte}(1987)}]{davi}%
  \BibitemOpen
  \bibfield  {author} {\bibinfo {author} {\bibfnamefont {R.~B.}\ \bibnamefont
  {Davies}}\ and\ \bibinfo {author} {\bibfnamefont {D.~S.}\ \bibnamefont
  {Harte}},\ }\bibfield  {title} {\enquote {\bibinfo {title} {Tests for {H}urst
  effect},}\ }\href@noop {} {\bibfield  {journal} {\bibinfo  {journal}
  {Biometrika}\ }\textbf {\bibinfo {volume} {74}},\ \bibinfo {pages} {95}
  (\bibinfo {year} {1987})}\BibitemShut {NoStop}%
\bibitem [{\citenamefont {Willinger}\ \emph {et~al.}(1997)\citenamefont
  {Willinger}, \citenamefont {Taqqu}, \citenamefont {Sherman},\ and\
  \citenamefont {Wilson}}]{willi}%
  \BibitemOpen
  \bibfield  {author} {\bibinfo {author} {\bibfnamefont {W.}~\bibnamefont
  {Willinger}}, \bibinfo {author} {\bibfnamefont {M.~S.}\ \bibnamefont
  {Taqqu}}, \bibinfo {author} {\bibfnamefont {R.}~\bibnamefont {Sherman}}, \
  and\ \bibinfo {author} {\bibfnamefont {D.~V.}\ \bibnamefont {Wilson}},\
  }\bibfield  {title} {\enquote {\bibinfo {title} {Self-similarity through
  high-variability: statistical analysis of ethernet lan traffic at the source
  level},}\ }\href@noop {} {\bibfield  {journal} {\bibinfo  {journal} {IEEE
  Acm. T. Network.}\ }\textbf {\bibinfo {volume} {5}},\ \bibinfo {pages} {71}
  (\bibinfo {year} {1997})}\BibitemShut {NoStop}%
\bibitem [{\citenamefont {Norros}\ \emph {et~al.}(1999)\citenamefont {Norros},
  \citenamefont {Mannersalo},\ and\ \citenamefont {Wang}}]{noros}%
  \BibitemOpen
  \bibfield  {author} {\bibinfo {author} {\bibfnamefont {I.}~\bibnamefont
  {Norros}}, \bibinfo {author} {\bibfnamefont {P.}~\bibnamefont {Mannersalo}},
  \ and\ \bibinfo {author} {\bibfnamefont {J.~L.}\ \bibnamefont {Wang}},\
  }\bibfield  {title} {\enquote {\bibinfo {title} {Simulation of fractional
  {B}rownian motion with conditionalized random midpoint displacement},}\
  }\href@noop {} {\bibfield  {journal} {\bibinfo  {journal} {Adv. Perform.
  Anal.}\ }\textbf {\bibinfo {volume} {2}},\ \bibinfo {pages} {77} (\bibinfo
  {year} {1999})}\BibitemShut {NoStop}%
\bibitem [{\citenamefont {Dieker}\ and\ \citenamefont {Mandjes}(2003)}]{diek}%
  \BibitemOpen
  \bibfield  {author} {\bibinfo {author} {\bibfnamefont {A.~B.}\ \bibnamefont
  {Dieker}}\ and\ \bibinfo {author} {\bibfnamefont {M.}~\bibnamefont
  {Mandjes}},\ }\bibfield  {title} {\enquote {\bibinfo {title} {On spectral
  simulation of fractional {B}rownian motion},}\ }\href@noop {} {\bibfield
  {journal} {\bibinfo  {journal} {Probab. Eng. Inf. Sci.}\ }\textbf {\bibinfo
  {volume} {17}},\ \bibinfo {pages} {417} (\bibinfo {year} {2003})}\BibitemShut
  {NoStop}%
\bibitem [{\citenamefont {Levernier}\ \emph {et~al.}(2018)\citenamefont
  {Levernier}, \citenamefont {B\'enichou}, \citenamefont {Gu\'erin},\ and\
  \citenamefont {Voituriez}}]{LBGV18}%
  \BibitemOpen
  \bibfield  {author} {\bibinfo {author} {\bibfnamefont {N.}~\bibnamefont
  {Levernier}}, \bibinfo {author} {\bibfnamefont {O.}~\bibnamefont
  {B\'enichou}}, \bibinfo {author} {\bibfnamefont {T.}~\bibnamefont
  {Gu\'erin}}, \ and\ \bibinfo {author} {\bibfnamefont {R.}~\bibnamefont
  {Voituriez}},\ }\bibfield  {title} {\enquote {\bibinfo {title} {Universal
  first-passage statistics in aging media},}\ }\href@noop {} {\bibfield
  {journal} {\bibinfo  {journal} {Phys. Rev. E}\ }\textbf {\bibinfo {volume}
  {98}},\ \bibinfo {pages} {022125} (\bibinfo {year} {2018})}\BibitemShut
  {NoStop}%
\bibitem [{\citenamefont {B\'enichou}\ \emph {et~al.}(2010)\citenamefont
  {B\'enichou}, \citenamefont {Chevalier},\ and\ \citenamefont
  {Klafter}}]{BCK10}%
  \BibitemOpen
  \bibfield  {author} {\bibinfo {author} {\bibfnamefont {O.}~\bibnamefont
  {B\'enichou}}, \bibinfo {author} {\bibfnamefont {C.}~\bibnamefont
  {Chevalier}}, \ and\ \bibinfo {author} {\bibfnamefont {J.}~\bibnamefont
      {Klafter}},\ }\bibfield  {title} {\enquote {\bibinfo {title} {Geometry-controlled kinetics},}\ }\href@noop {} {\bibfield
  {journal} {\bibinfo  {journal} {Nature Chem.}\ }\textbf {\bibinfo {volume}
  {2}},\ \bibinfo {pages} {472} (\bibinfo {year} {2010})}\BibitemShut {NoStop}%
\bibitem [{\citenamefont {Bénichou}\ and\ \citenamefont
  {Voituriez}(2014)}]{BeVo14}%
  \BibitemOpen
  \bibfield  {author} {\bibinfo {author} {\bibfnamefont {O.}~\bibnamefont
  {Bénichou}}\ and\ \bibinfo {author} {\bibfnamefont {R.}~\bibnamefont
  {Voituriez}},\ }\bibfield  {title} {\enquote {\bibinfo {title} {From
  first-passage times of random walks in confinement to geometry-controlled
  kinetics},}\ }\href@noop {} {\bibfield  {journal} {\bibinfo  {journal} {Phys.
  Rep.}\ }\textbf {\bibinfo {volume} {539}},\ \bibinfo {pages} {225} (\bibinfo
  {year} {2014})}\BibitemShut {NoStop}%
\bibitem [{\citenamefont {Sarvaharman}\ \emph {et~al.}(2019)\citenamefont
  {Sarvaharman}, \citenamefont {Heiblum~Robles},\ and\ \citenamefont
  {Giuggioli}}]{SHG19}%
  \BibitemOpen
  \bibfield  {author} {\bibinfo {author} {\bibfnamefont {S.}~\bibnamefont
  {Sarvaharman}}, \bibinfo {author} {\bibfnamefont {A.}~\bibnamefont
  {Heiblum~Robles}}, \ and\ \bibinfo {author} {\bibfnamefont {L.}~\bibnamefont
  {Giuggioli}},\ }\bibfield  {title} {\enquote {\bibinfo {title} {From
  micro-to-macro: How the movement statistics of individual walkers affect the
  formation of segregated territories in the territorial random walk model},}\
  }\href@noop {} {\bibfield  {journal} {\bibinfo  {journal} {Front. Phys.}\
  }\textbf {\bibinfo {volume} {7}},\ \bibinfo {pages} {129} (\bibinfo {year}
  {2019})}\BibitemShut {NoStop}%
\bibitem [{\citenamefont {Dimidov~C.}(2016)}]{DOT16}%
  \BibitemOpen
  \bibfield  {author} {\bibinfo {author} {\bibfnamefont {V.~Trianni}\
  \bibnamefont {C.~Dimidov}, \bibfnamefont {G.~Oriolo}},\ }\bibfield  {title}
  {\enquote {\bibinfo {title} {Random walks in swarm robotics: An experiment
  with kilobots},}\ } \href@noop {} {\ \bibinfo {series} {Lecture Notes in
  Computer Science},\ \textbf {\bibinfo {volume} {9882}},\ \bibinfo {pages}
  {185} (\bibinfo {year} {2016})}\BibitemShut {NoStop}%
\end{thebibliography}

%

\end{document}